%% file: After_submission_Jigsaw_full.tex
\documentclass[letterpaper,twocolumn,10pt]{article}
\usepackage{usenix}

\makeatother
\usepackage{algpseudocode}
\usepackage[ruled,linesnumbered,boxed,commentsnumbered]{algorithm2e}
\usepackage{multirow} 

\usepackage{amssymb}
\usepackage{graphicx}
\usepackage{subfigure}
\usepackage{makecell}
\usepackage{threeparttable}
\usepackage{tikz}
\usepackage{amsmath}
\usepackage{appendix}
\usepackage{float}
\usepackage{diagbox}
\usepackage{soul}
\usepackage{xcolor}
\usepackage{caption}
\usepackage{authblk}
\usepackage{marvosym}

\usepackage{authblk}

\newcommand*\emptycirc[1][0.8ex]{\tikz\draw (0,0) circle (#1);} 
\newcommand*\halfcirc[1][0.8ex]{%
	\begin{tikzpicture}
	\draw[fill] (0,0)-- (90:#1) arc (90:270:#1) -- cycle ;
	\draw (0,0) circle (#1);
	\end{tikzpicture}}
\newcommand*\fullcirc[1][0.8ex]{\tikz\fill (0,0) circle (#1);} 

\newcounter{Clactr}
\newenvironment{observation}[1]{\stepcounter{Clactr}\normalfont\bfseries\par\noindent{Observation \arabic{Clactr}.}\space#1}{}

\begin{document}

\date{}

\title{\Large \bf Query Recovery from Easy to Hard: Jigsaw Attack against SSE}

\author[1]{Hao Nie}
\author[1,\Letter]{Wei Wang}
\author[1,2,3,4,\Letter]{Peng Xu}
\author[1]{Xianglong Zhang}
\author[1,5]{Laurence T. Yang}
\author[6]{Kaitai Liang}
\affil[1]{\footnotesize\textit{Huazhong University of Science and Technology}}
\affil[2]{\footnotesize\textit{Hubei Key Laboratory of Distributed System Security, School of Cyber Science and Engineering}}
\affil[3]{\footnotesize\textit{JinYinHu Laboratory}}
\affil[4]{\footnotesize\textit{State Key Laboratory of Cryptology}}
\affil[5]{\footnotesize\textit{St. Francis Xavier University}}
\affil[6]{\footnotesize\textit{Delft University of Technology}}
\affil[\Letter]{\small\textit{Corresponding authors: viviawangwei@hust.edu.cn, xupeng@hust.edu.cn}}

\maketitle

\begin{abstract}
Searchable symmetric encryption schemes often unintentionally disclose certain sensitive information, such as access, volume, and search patterns.  
Attackers can exploit such leakages and other available knowledge related to the user's database to recover queries.
We find that the effectiveness of query recovery attacks depends on the volume/frequency distribution of keywords.  
Queries containing keywords with high volumes/frequencies are more susceptible to recovery, even when countermeasures are implemented.
Attackers can also effectively leverage these ``special'' queries to recover all others. 

By exploiting the above finding, we propose a Jigsaw attack that begins by accurately identifying and recovering those distinctive queries. 
Leveraging the volume, frequency, and co-occurrence information, our attack achieves $90\%$ accuracy in three tested datasets, which is comparable to previous attacks (Oya et al., USENIX' 22 and Damie et al., USENIX' 21). 
With the same runtime, our attack demonstrates an advantage over the attack proposed by Oya et al (approximately $15\%$ more accuracy when the keyword universe size is 15k). 
Furthermore, our proposed attack outperforms existing attacks against widely studied countermeasures, achieving roughly $60\%$ and $85\%$ accuracy against the padding and the obfuscation, respectively. 
In this context, with a large keyword universe ($\geq$3k), it surpasses current state-of-the-art attacks by more than $20\%$.

\end{abstract}

\section{Introduction}

\input{table_related_work.tex}

\label{sec:introduction}
Searchable Symmetric Encryption (SSE) \cite{song2000practical,DBLP:conf/ccs/CurtmolaGKO06,DBLP:conf/eurocrypt/CashT14,DBLP:conf/ccs/ChamaniPPJ18,DBLP:conf/eurocrypt/KamaraM17,DBLP:conf/esorics/PatelPY18,DBLP:conf/ccs/SunYLSSVN18,bost2016ovarphiovarsigma,bost2017forward,xu2022rose} enables users to securely search encrypted databases stored on remote servers.  
An SSE scheme typically consists of setup and search protocols. 
In the setup, the user sends encrypted indexes of the documents to the server. 
In the search, the user generates a search token and sends it to the server who then returns the matched documents. 
The search process does not reveal any confidential information about the documents or the user's search query, except for the volume pattern (also known as the response length) and the access pattern, which reveal the number of matched documents and their identities, respectively.
Additionally, the server may also know the search pattern, which indicates whether two queries are identical by comparing their search tokens or access patterns.

Passive attacks on SSE can exploit the above leakages and some prior knowledge to recover queries. 
According to the prior knowledge given to the attacker, we categorize two main attacks in Table \ref{tab:tab_passive_attacks}:  
1) \textbf{known-data attacks}\cite{ning2021leap,blackstone2020revisiting,islam2012access,cash2015leakage,pouliot2016shadow,VALESORICS,xu2021interpreting}, which assume that the attacker has access to partial/full plain texts of the documents in the user's dataset; and  
2) \textbf{similar-data attacks}\cite{oya2021hiding,oya2021ihop,damie2021highly,pouliot2016shadow,liu2014search}, which enable the attacker to obtain a similar document set or estimations on the users' query distribution. 
Unlike known-data attacks, that require the plain texts, similar data attacks can recover queries by exploiting statistical information from a similar dataset, such as query frequency and the probability of two keywords appearing in the same document.
Without relying on the ``strong assumption'' that the attacker must be provided plain texts of documents, similar-data attacks are relatively practical to deploy and bypass countermeasures. Existing works \cite{oya2021hiding,oya2021ihop} have shown that similar-data attacks can bypass some defenses employing pattern randomization techniques\cite{DBLP:conf/ndss/ShangOPK21,DBLP:conf/infocom/ChenLRZ18}.   
We note that SSE is also vulnerable to active attacks \cite{DBLP:conf/uss/ZhangKP16,DBLP:conf/eurosp/PoddarWLP20,blackstone2020revisiting,zhang2023high} leveraging file injection to recover queries, which is orthogonal to this work.

Damie et al. \cite{damie2021highly} explored an intriguing phenomenon across various query distributions, indicating the correlation between accuracy and query volume (i.e., the number of documents containing a particular keyword).  
Oya et al. \cite{oya2021hiding} demonstrated that high-frequency queries (i.e., the frequency with which the user queries a specific keyword) give a greater probability of being successfully recovered in their proposed attack.
We note that a similar notice was given in \cite{blackstone2020revisiting} that the effectiveness of known-data attacks is also influenced by the volume of queries, wherein high-volume queries are easily recoverable. 
Building upon the aforementioned interesting hints, we conduct experiments that confirm the influence of volume and/or frequency on the performance of attacks. 
Moreover, we discover that leveraging this knowledge enables us to enhance the effectiveness of similar-data attacks. 
We provide two crucial observations regarding the query recovery. 

\begin{observation}
\mdseries\itshape Queries containing keywords with a high volume/frequency are much easier to recover than others. 
\label{obs:observation1}
\end{observation}

In a database, the volume of keywords follows Zipf's law\cite{zipf2016human}, which states that the volume of a keyword ranks $n$th in a sorted list (sorted by volume) is inversely proportional to $n$. 
We also observe that the frequency of keywords follows almost the same law. 
We confirm the above phenomenons by showing the concrete results in three datasets (See Section \ref{sec:keyword distribution} and Appendix \ref{Appendix:simpleattack} for more details). 
Keywords with higher volume or frequency display larger disparities, which consequently makes it easier for attackers to recover those queries.

\begin{observation}
\mdseries\itshape By revealing the queries from observation 1, the attacker can gain advantage to retrieve further queries (even all queries).
\label{obs:observation2}
\end{observation}

In \cite{damie2021highly}, Damie et al. proposed an efficient similar-data attack (i.e. refined score attack, RSA) that achieves around $85\%$ accuracy {in recovering all queries} by utilizing only $10$ known queries. 
They also show that when utilizing known queries with a higher volume, the attack's accuracy increases and becomes more stable.

\noindent \textbf{Challenges.} Observation 1 does not explicitly facilitate a way to identify and recover those distinctive queries as they consistently intermingle with others. 
To the best of our knowledge, there is no attack that first focuses on filtering those distinctive queries, thereby allowing for the recovery of queries from easy to hard. 
Setting a start with immediately recovering all queries based on known or recovered queries is not trivial and could be defended against by countermeasures. 
For example, previous work \cite{damie2021highly} cannot work effectively at the outset without any proper known query set. 
This also implicitly explains why it achieves low accuracy under padding\cite{cash2015leakage} and obfuscation\cite{DBLP:conf/infocom/ChenLRZ18}.

\noindent \textbf{Contributions.}
We propose a new effective similar-data attack called Jigsaw providing a ``granular and incremental'' strategy, which comprises three core modules.
1) The first module uses the keyword's volume and frequency information to locate and recover the most distinctive queries. 
2) The second module further refines the recovered queries by matching the queries with the keywords according to the co-occurrence matrix. 
This module eliminates those incorrect query recoveries from the first module and tries to achieve near-perfect accuracy.
3) The last module is to recover the remaining queries using the outputs from the second module. 
We generate scores for query-keyword combinations using the co-occurrence matrix, volume, and frequency information. We optimize the score for queries to obtain matches between queries and keywords. 
We also provide comprehensive evaluations of Jigsaw. 
Concretely, our contributions are outlined as follows.
\smallskip
\\
$\bullet$ \textit{Localization and recovery of distinctive queries.} 
We measure the distinguishability of each query and use the volume and frequency to recover the most distinguishable queries (the first module of Jigsaw). 
For queries with a high volume and frequency (e.g., the top $10\%$ of queries in volume and frequency in Enron\cite{Enron}, Lucene\cite{Lucene}, and Wikipedia\cite{Wikipedia}), we can obtain an accuracy $>70\%$. 
\smallskip
\\
$\bullet$ \textit{Precise verification of recovered queries.} 
We make a further refinement by filtering out those queries that do not align well with the co-occurrence information (the second module). 
We here obtain nearly $100\%$ accuracy at the expense of recovering a smaller number of queries (about a dozen queries in Enron and 50 queries in Lucene and Wikipedia). 
\smallskip
\\
$\bullet$ \textit{Accurate recovery of all queries.} 
We at last utilize the recovered queries to recover all queries with about $95\%$ accuracy (the last module). 
Even if the frequency information is not given, we can still capture roughly $90\%$ accuracy.  
We state that Jigsaw exhibits durability, as it can hold its effectiveness (dropping $<5\%$ of accuracy) in the future period even by exploiting the auxiliary frequency information that was leaked long ago (e.g., $30$ months in Wikipedia and $150$ weeks in Enron and Lucene).  
\smallskip
\\
$\bullet$ \textit{Comprehensive evaluations and comparisons.} 
We present empirical experiments and comparisons with the state of art similar-data attacks (including Graphm \cite{pouliot2016shadow}, SAP \cite{oya2021hiding}, RSA \cite{damie2021highly}, and IHOP \cite{oya2021ihop}) to highlight the performance of Jigsaw. 
Our attack provides $>90\%$ accuracy surpassing the Graphm and SAP attacks, similar to the RSA and IHOP.
Within the same runtime, Jigsaw exhibits about $15\%$ more accuracy than IHOP when the keyword universe is 15k. 
Also, Jigsaw outperforms them when countering the defenses (padding\cite{cash2015leakage} and obfuscation\cite{DBLP:conf/infocom/ChenLRZ18}). 
It maintains $>60\%$ and $>85\%$ accuracy against the padding (in most cases) and obfuscation and takes the lead in accuracy in most cases.

\section{Related Work}

Except for the SSE schemes\cite{DBLP:conf/uss/DemertzisPPS20,DBLP:conf/crypto/GargMP16,OBImultipathORAM} based on expensive primitives, such as ORAM and PIR, most SSE schemes leak the access pattern, search, volume, and response size pattern (i.e., the size of each document).
With some prior knowledge, passive attacks abuse the above leakages to recover users' queries. These attacks can be categorized as similar-data attacks\cite{oya2021hiding,oya2021ihop,damie2021highly,pouliot2016shadow,liu2014search} and known-data attacks\cite{ning2021leap,blackstone2020revisiting,islam2012access,cash2015leakage,pouliot2016shadow,VALESORICS,xu2021interpreting}. 

\noindent{\textbf{Similar-data attacks}}. Liu et al.\cite{liu2014search} proposed the Freq attack that exploits the search pattern and the query frequency information to recover users' queries.
Recently, Oya et al.\cite{oya2021hiding} proposed the SAP attack, which utilizes the search and volume pattern to get the frequency and volume of each query.
However, Freq and SAP strongly rely on the frequency information and achieve a relatively low accuracy. 
Attacks that abuse the access pattern have higher accuracy.
Pouliot et al.\cite{pouliot2016shadow} proposed the GraphM attack that formalizes the query recovery as a weighted graph match problem and solves it by PATH\cite{PATH} or Umeyama\cite{Umeyama} algorithm. 
Oya et al.\cite{oya2021ihop} proposed the IHOP attack, which uses a co-occurrence matrix of queries and keywords, along with query frequency, to launch its attack.
IHOP supposes that queries are correlated and follow a Markov process, allowing it even to threaten frequency-smoothing defenses such as PANCAKE\cite{DBLP:conf/uss/GrubbsKLBL0R20}.
Damie et al.\cite{damie2021highly} proposed the RSA, which starts with some known queries. 
Though providing an accuracy of about $85\%$, RSA still requires some known queries as a prerequisite. 
Without those, the attack will not work effectively.

\noindent{\textbf{Known-data attacks}}. Islam et al.\cite{islam2012access} proposed the first known-data attack (IKK) with all documents and partial queries to recover queries.
Cash et al.\cite{cash2015leakage} proposed the Count attack that can recover most queries without known queries.
In \cite{blackstone2020revisiting}, Blackstone et al. proposed an attack that performs perfectly (accuracy approaching $100\%$) with fully known documents, but poorly (less than $10\%$) with a small portion of known documents.
Ning et al.\cite{ning2021leap} proposed LEAP, which can recover half of all queries with $100\%$ accuracy with only $1\%$ of documents.
These attacks are all dependent on known data information. However, the known data are hard to obtain, and the attacks are easy to prevent with countermeasures. 

\noindent{\textbf{Countermeasures}}. To defend against leakage abuse attacks, many countermeasures \cite{cash2015leakage,DBLP:conf/uss/DemertzisPPS20,DBLP:journals/corr/shielddb,DBLP:journals/iacr/BostF17,xu2019hardening,DBLP:conf/infocom/ChenLRZ18,islam2012access} have been proposed. 
Among those, padding is one of the most commonly used methods. 
Cash et al.\cite{cash2015leakage} first presented the padding strategy. 
The volume of each query is padded to the nearest multiple of an integer $k$. 
After that, Demertzis et al.\cite{DBLP:conf/uss/DemertzisPPS20} proposed SEAL, which pads the volume of each query to the nearest power of an integer $x$. 
The padded documents would add noise to the volume and access patterns, hampering attacks abusing those leakages. 
Several works also consider the keywords clustering \cite{DBLP:journals/corr/shielddb,DBLP:journals/iacr/BostF17}. 
Each cluster contains no less than $\alpha$ keywords, and then each keyword is padded to the largest volume in the cluster.

Another countermeasure is obfuscation\cite{DBLP:conf/infocom/ChenLRZ18}. 
When querying for a keyword, if a document contains the keyword, the document will be returned with probability $p$ (the true positive rate, TPR); otherwise, each document will be returned with probability $q$ (the false positive rate, FPR). 
Shang et al.\cite{DBLP:conf/ndss/ShangOPK21} proposed OSSE, which provides the same response effect and produces fresh obfuscation in each query.

\section{SSE Scheme and Attack Model}
We revisit the standard SSE\cite{DBLP:conf/ccs/CurtmolaGKO06}, define the leakage function of queries, and describe the attacker's prior knowledge as the prerequisite of our attack. 
Note we put the summary of notations in Table \ref{tab:sum_of_nota}, Appendix \ref{AppenA}.

\subsection{SSE}
An SSE scheme\cite{DBLP:conf/ccs/CurtmolaGKO06} facilitates keyword searches over encrypted data, denoted as $ED$, while maintaining the confidentiality of the data and the keywords. 
The typical components of an SSE scheme encompass the setup, update, and query processes.
Initially, the user possesses a dataset $D$, which comprises a set of documents $d$ identified by $id(d)$. 
Each document contains a list of keywords $k$.
During the setup phase, the user can construct and encrypt an index, and then upload it along with the encrypted document set $ED$ to the server.
In the update, the user can dynamically update the index stored on the server.
During the query process, to search a keyword $k$, the user generates a trapdoor $td(k)$ for the server. 
Eventually, the user retrieves the list $D(k)$, which is a list of $id(d)$ satisfying that $k$ appears in $d$. 
We also denote the list $D(k)$ as $D(td(k))$ for convenience. 
We note that the user also needs to retrieve the encrypted documents according to the $id(d)$ and decrypt them to complete the search.

\subsection{Leakages}

An efficient SSE scheme typically leaks the volume pattern, the access pattern, and the search pattern of queries to the server and potential eavesdroppers. 
For a sequence of $s$ queries $Td^{s}=[ td(x_1),td(x_2),\ldots,td(x_s)]$, the leakages often used in attacks are summarized as follows.

$\bullet$ \textbf{Access pattern} is the family of functions $ap:ED\times Td^s \to AP^s$ where $AP$ is a list $[ D(x_1),D(x_2),\ldots,D(x_s) ]$. 
For each query, the scheme leaks the identifiers of corresponding encrypted documents. 
This leakage happens in most SSE schemes\cite{bost2017forward,bost2016ovarphiovarsigma,DBLP:conf/crypto/CashJJKRS13,DBLP:conf/acns/ChangM05,DBLP:conf/ccs/CurtmolaGKO06,DBLP:conf/ccs/KamaraPR12,DBLP:conf/sp/NaveedPG14,DBLP:conf/ndss/StefanovPS14,xu2022rose} when the user retrieves the encrypted document. 
Some schemes use primitives such as ORAM\cite{DBLP:journals/jacm/GoldreichO96} or PIR\cite{DBLP:conf/focs/ChorGKS95} to hide access pattern, but those primitives lead to expensive costs.  

$\bullet$ \textbf{Volume pattern} is the family of functions $vp:ED\times Td^s \to VP^s$, where $VP$ is a list $[ |D(x_1)|,|D(x_2)|,\ldots,|D(x_s)| ]$. 
For each query, the scheme leaks the number of documents returned by the server. 
 
$\bullet$ \textbf{Search pattern} is the family of functions $sp:ED\times Td^s \to M^{s\times s}$, where $M^{s\times s}$ is a $s\times s$ binary matrix such that $M^{s\times s}[i,j]=1$, if the underlying keywords of $td_i$ and $td_j$ are the same and otherwise $M^{s \times s}[i,j]=0$. For any two queries $td(x_i),td(x_j)\in Td,i\neq j$, the attacker knows whether $x_i$ equals $x_j$. 
While schemes may not directly reveal the search pattern from queries, repeated querying of the same keyword leads to the exposure of the same access pattern. Attackers can utilize this access pattern to infer whether two queries correspond to the same keyword\cite{oya2021hiding}.

\subsection{Attackers}

We respectively consider two kinds of attackers targeting the user's queries: honest but curious servers and eavesdroppers:

$\bullet$ An \textit{honest but curious server} follows the SSE protocols but attempts to recover the users' queries by utilizing the leakage pattern and other prior knowledge, such as known data or similar data. The server also has access to all the encrypted documents.

$\bullet$ An \textit{eavesdropper} who intercepts the traffic between the server and the user can observe the encrypted documents returned in each query and possess the same knowledge of leakages as the server, except for all the encrypted documents. {With a similar dataset, the eavesdropper can also launch similar-data attacks.}

Different from known-data attacks, the server could utilize outdated (obtained from the past) or leaked documents which are not necessarily included in the current user's dataset to launch similar-data attacks.
Note the eavesdropper could also know some similar data in several scenarios by acting as a legal user. 
For instance, the eavesdropper may share the same email system (producing similar email data) with his colleagues.
We state that both attackers can employ our Jigsaw attack by utilizing the following knowledge derived from leakage and similar data.

\noindent{\textbf{Attackers' knowledge derived from leakages}}.
We use the leakages to derive the frequency, volume, and co-occurrence of queries.
We assume the user issues $s$ queries, denoted as $Td^{s}$, from which the attacker identifies $l$ different queries by the search pattern. We denote the query list identified by the attacker as $Td_{r}=[ td_1,td_2,\ldots,td_l]$, which does not contain repeated queries. 
The attacker can also observe the returned documents $[D(td_1), D(td_2),\ldots, D(td_l)]$. 
The server possesses knowledge of the total number of documents, denoted as $|D|$. While the eavesdropper, who can only observe the query traffic, does not know the total number of documents. We also use $|D|$ to represent the size of the document set observed by the eavesdropper. 
Then for each query $td$, we normalize the volume pattern of query denoted $v_{td}={|D(td)|}/{|D|}$. 
For the query list $Td_{r}$, $V_{r}=[ v_{td_1},v_{td_2},\ldots,v_{td_l} ] $ is the vector of volume of all queries in $Td_{r}$. 
With the search pattern, the attacker acquires knowledge of the frequency at which a $td$ appears in $Td^s$. 
We denote the frequency of $td$ as $f_{td} = {Count(td)}/{|Td^s|}$, where $Count(td)$ computes the number of $td$ in $Td^s$. For the query list $Td_{r}$, $F_{r} = [ f_{td_1},f_{td_2},\ldots,f_{td_l}]$ is the vector of frequency of all queries in $Td_{r}$. 
Based on the access pattern, the attacker can construct a $l\times{|D|}$ matrix $ID_{r}$.  $ID_{r}[i,j]=1$, if the search result for $td_i$ contains $d_j$, $0$ otherwise. 
Then, the attacker can construct the $l\times l$ co-occurrence matrix $C_{r}= ID_{r}ID_{r}^\top / {|D|}$.

\noindent{\textbf{Attackers' knowledge derived from similar data}}.
We assume the attacker knows a list of similar documents $D_{s}=[ d_1,d_2,\ldots,d_k]$ and employs the same algorithm to extract keywords as the user. 
Actually, the attacker can easily obtain a similar keyword universe as the user's database with a small similar data\cite{zhang2023high}. 
We denote the keyword universe extracted by the attacker as $W_{s}=[w_1,w_2,\ldots,w_m]$.
Then, the attacker can construct the volume $V_{s}=[ v_{w_1},v_{w_2},\ldots,v_{w_m}]$, where $v_{w_i}={|D_{s}(w_i)|}/|D_s|$ and the $D_s(w_i)$ is the documents in $D_s$ which contain keyword $w_i$. 
The attacker also constructs $ID_{s}$ from $W_{s}$ and $D_{s}$ in a similar manner as the construction of $ID_r$. 
Based on the $ID_{s}$, the attacker calculates the co-occurrence matrix $C_{s}= ID_{s}ID_{s}^\top /|D_s|$.
The attacker can also obtain a similar query frequency $F_{s}=[ f_{w_1},f_{w_2},\ldots,f_{w_m}]$ for $W_s$ by public information such as Google Trend\cite{GoogleTrends} or outdated frequency information.

\section{Jigsaw Attack}
\label{sec:our attack}
{We first demonstrate that Observation 1 does apply to all tested datasets although it does not directly provide an approach to recover those distinctive queries. 
We introduce the differential distance in the first module of Jigsaw to establish a way to distinguish these queries. 
We then define a distance between queries and keywords, enabling us to pair the distinctive queries with their nearest keywords.
We accomplish the refinement in the second module by testing the queries' compatibility with co-occurrence relationships.
Building on Observation 2, in the third module, we systematically recover all remaining queries by using those obtained in the second module.  
We define a specified ``score'' for each query-keyword pair and maximize it to get matches. 
The recovery process is iterative, with previously recovered queries being utilized in the recovery of subsequent queries.}

\subsection{Distribution of Keywords}

\label{sec:keyword distribution}

\begin{figure}[!t]
    	\centering
    	\includegraphics[width=0.7\linewidth]{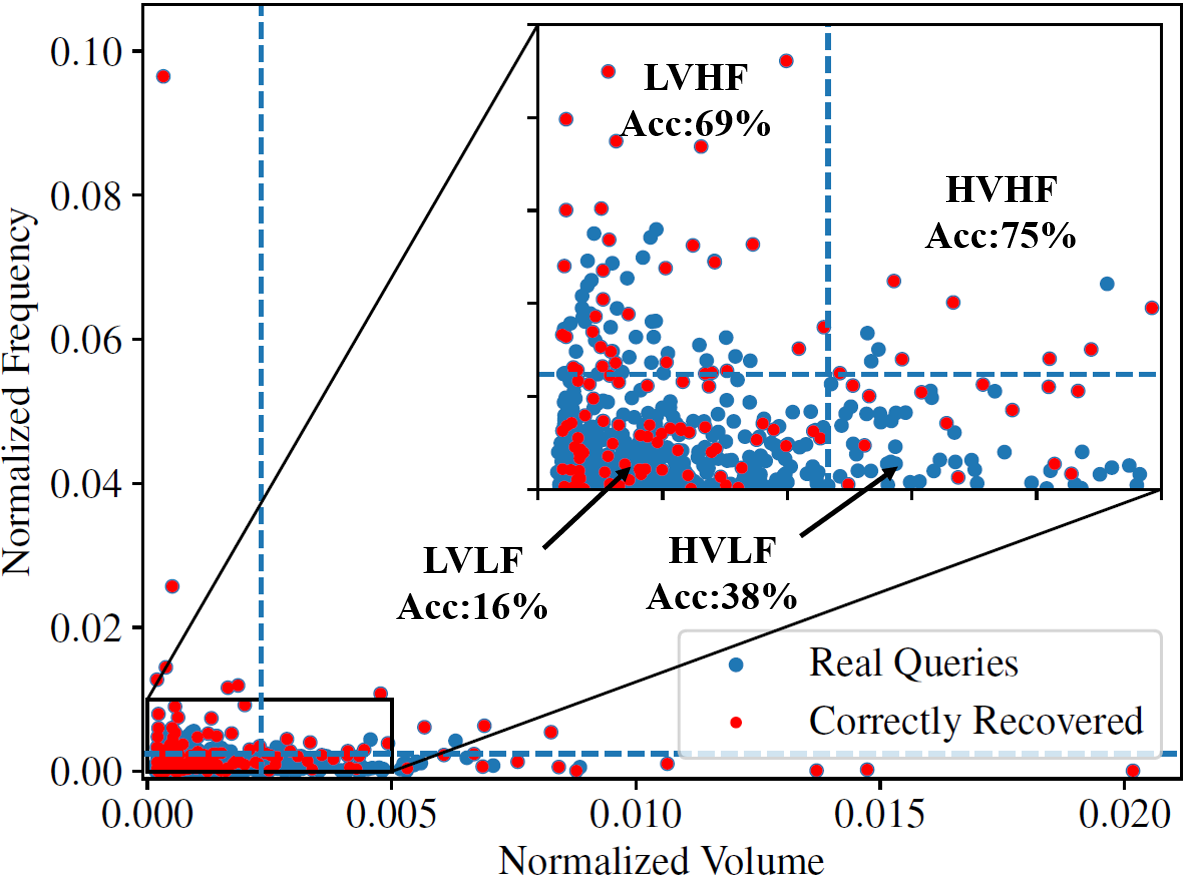}
    	\caption{The distribution of queries on Enron.
     The horizontal dashed line divides the top $10\%$ queries on volume from other queries; the vertical dashed line divides the top $10\%$ queries on frequency from other queries. The blue dots denote the real queries issued by the user; the red dots denote the queries successfully recovered by the simple attack presented in Appendix \ref{Appendix:simpleattack}.
     }
    	\label{fig:distribution}
    \end{figure}

Before proceeding to the Jigsaw attack, we present a comprehensive exposition of volume and frequency distribution (commonly found) in typical databases. 
We also showcase how the distribution impacts the effectiveness of attacks. 

According to the Zipf's law\cite{zipf2016human}, the volume of keywords in a database follows the Zipfian distribution, illustrating that \textit{a small portion of keywords associates with a high volume.} 
As a result, these keywords exhibit significant disparities among each other due to their limited quantity but extensive volume range. 
A similar phenomenon emerges concerning query frequency, which indicates that a small subset of keywords could tend to be more frequently queried by users.

We present a simple similar-data attack to show that queries with high volume and frequency are relatively easier to recover. 
We employ this attack solely to demonstrate the distribution of keywords and its influence on the recovery. 
In this attack, the attacker generates the volume and frequency of queries and pairs them with keywords that exhibit the closest similarity w.r.t. volume and frequency. 
We simulate the attack on Enron\cite{Enron}, Lucene\cite{Lucene}, and Wikipedia\cite{Wikipedia} and provide the results for Enron in Figure \ref{fig:distribution} (See Appendix \ref{Appendix:simpleattack} for details regarding the attack and its results on other datasets).

Figure \ref{fig:distribution} shows the volume and frequency correlated distribution of queries and the effectiveness of the attack (on recovery). 
{We select the top 1, 000 keywords on volume except for the stop words as the
keyword universe and categorize the corresponding queries} into four quadrants: HVHF, HVLF, LVHF, and LVLF, where ``L'' and ``H'' represent ``low'' and ``high'', and the ``V'' and ``F'' denote ``volume'' and ``frequency'', respectively. {High-volume queries refer to the top $10\%$ of queries based on volume, while low-volume queries are the remaining $90\%$.  
The same classification applies to high-frequency and low-frequency queries.}
The lower-left corner of the figure is magnified to highlight that keywords with low volume and low frequency exhibit dense packing, which makes it difficult to differentiate these keywords.
In contrast, the HVHF quadrant consists of sparser queries that are more distinguishable from each other. 
Using the similarity in volume and frequency, the attacker can easily recover these queries. 
Experimental results confirm this, with the simple attack achieving an accuracy of nearly $80\%$ in the HVHF quadrant ($>90\%$ in Lucene and Wikipedia datasets).
In the LVLF quadrant, however, the accuracy drops to $10\%$. 
We also achieve a moderately high accuracy of $62\%$ and $34\%$ in the LVHF and HVLF quadrants. 
Similar findings in \cite{blackstone2020revisiting,damie2021highly,oya2021hiding} also present the correlation between frequency, volume, and accuracy.

\subsection{Locating and Recovering the Most Distinctive Queries}

\label{sec:our attack 1}
As previously described, some queries are more distinguishable and easier to recover. 
Our first module aims to locate and recover these queries.
We outline its details in Algorithm \ref{alg1}, which takes the $Td_r$, $V_r$, and $F_r$ derived from query leakages in SSE and $W_s$, $V_s$, and $F_s$ from similar data as input and outputs $BaseRec$ predictions $Pred$. 
The $BaseRec$ determines the number of recovered queries in this module. 
{As we recover queries from high to low distinctiveness, a larger value of $BaseRec$ can yield more predictions of queries lacking distinctiveness, consequently leading to a decline in accuracy.}

Concretely, we first identify the most distinctive queries from all the queries by evaluating the \textit{differential distance}. 
We define the {differential distance} $d_{td_i}$ of a query $td_i$ as follows: 
\begin{equation}
    d_{td_i}=\min_{{td_j\in Td_r\land{j\ne i}}}\alpha\cdot|v_{td_i}-v_{td_j}|+(1-\alpha)|f_{td_i}-f_{td_j}|.
    \label{equ:d}
\end{equation}
Note that in the measurement (line \ref{A1Caldiss}-\ref{A1Caldise}), $\alpha$ is the weight of the volume, and $(1-\alpha)$ is the weight of the frequency. 
The differential distance $d_{td}$ can assess the sparsity around the query $td$ and thus the level of distinctiveness of the query $td$. 
Then, we sort the queries in descending order by $d_{td}$ (line \ref{A1sortdis}). 
We regard the top $BaseRec$ queries in $d_{td}$ as the most distinctive queries and the attack target in the first module. 

Finally, we recover top $BaseRec$ queries (line \ref{A1recs}-\ref{A1rece}). 
Given a query $td_j$, we calculate the distance $s(td_i,w_j)$ between the query $td_i$ in real data and any keyword $w_j$ in similar data. 
We define the $s(td_i,w_j)$ as
\begin{equation}
    s(td_i,w_j) = \alpha\cdot|v_{td_i}-v_{w_j}|+(1-\alpha)|f_{td_i}-f_{w_j}|.\label{equ:score}
\end{equation}
We then recover the query $td_i$ as 
\begin{equation}
    Pred(td_i)=\arg\min_{w_j\in W_s} s(td_i,w_j)
\end{equation}

By properly adjusting the weight of volume and frequency information, this module can recover the distinctive queries with high accuracy (e.g., averagely $77\%$ in Enron when $BaseRec=100$, see Section \ref{sec:test_alg1_and_alg2}).
{We use the L1 norm here. 
It's worth noting that we also tested other norms and found that the L1 norm yields the best performance.} 
We note that Zipf's law could not be applicable to certain datasets, such as those containing randomly generated texts or artificially padded datasets. 
In this case, the module's performance could be negatively affected. 
Nonetheless, it is uncommon for real-world datasets to deviate significantly from Zipf's law.
Furthermore, padding a dataset to the extent that the volume of keywords diverges the Zipfian distribution would result in an inflated storage cost.
Though, we can still achieve relatively high accuracy as demonstrated in Section \ref{sec:Against Countermeasures}.

\subsection{Adjustment by Query Co-occurrence}

In the second module (see Algorithm \ref{alg2}), we utilize the co-occurrence matrix to further refine the recovered queries output by Algorithm \ref{alg1}. 
Note that the attack accuracy of all queries relies on the precise recovery of the distinctive queries.
Any incorrect recovery could significantly impact the overall accuracy.

Before launching the second module, we set the following input parameters: 
\\
$\bullet$ The recovered queries $Pred$ and its cardinality $BaseRec$ from the first module. 
\\
$\bullet$ The co-occurrence matrix $C_r'$ of queries and $C_s'$ of keywords in $Pred$. We construct the two co-occurrence matrices by first extracting the columns and rows in the co-occurrence matrix $C_r$ and $C_s$ when the corresponding queries and keywords appear in $Pred$. Then each row of $C_r'$ and $C_s'$ is normalized by dividing the sum of that row.
\\
$\bullet$ The parameter $ConfRec$ ($\leq{BaseRec}$), which reflects the number of recovered queries after refinement. 
Similar to the $BaseRec$, a smaller value of $ConfRec$ results in higher accuracy and a reduced number of recovered queries. 
In some restricted scenarios where we are only given a little prior knowledge or the leakage patterns have been ``noised'' by countermeasures, we should set the $ConfRec$ to a smaller value to capture high accuracy. 

With the above input, the module can verify the recovered queries in $Pred$ and output the predictions of $ConfRec$ queries with higher accuracy through the following process.

If most of the predictions in $Pred$ are accurate, then for a correct prediction $(td_i, w_i)$ in $Pred$, $C'_{r}[i]$ should be similar to $C'_{s}[i]$; otherwise, $td_i$ and $w_i$ are only similar in terms of volume and frequency and the relevant in co-occurrence matrix is not significant. 
Such similarity and deviation of $Pred[i]$ can be captured by calculating the Euclidean norm of $C'_{r}[i]-C'_{s}[i]$. 
We define $revconf$ as the reversed confidence of a prediction $(td_i, w_i)$ as
\begin{equation}
    revconf=||C'_{r}[i]-C'_{s}[i]||.
\end{equation}

If a prediction provides a smaller value of $revconf$, then it is considered more confident. 

Based on the $revconf$, we calculate $Revconf$ containing $(td,revconf)$ for all $td\in Td'$ (line \ref{A2calconfs}-\ref{A2calconfe}). 
Then, we sort $Revconf$ in descent order (line \ref{A2sort}). 
To provide $ConfRec$ verified predictions, we remove the top $BaseRec-ConfRec$ queries from $Pred$ and return the remaining predictions (line \ref{A2res}-\ref{A2ree}).

At the expense of recovering a smaller number of queries, Algorithm \ref{alg2} can reach almost perfect accuracy. 
We show in Section \ref{sec:test_alg1_and_alg2} that when the $BaseRec$ is set to 100, and the $ConfRec$ is 20, we obtain $96.9\%$ and $100\%$ accuracy in Enron and Wikipedia, respectively.   

\input{alg1}
\input{alg2.tex}

\subsection{Dynamic Recovery for All Queries}
\input{alg3.tex}
The prior modules provide predictions for a subset of queries. 
In Algorithm \ref{alg3}, we present the last module of our attack that leverages the relation between the recovered distinctive queries and the remaining queries. 
This module recovers queries through an iterative approach, where the recovered queries by the second module serve as known queries.

The module takes the following information as input and outputs the predictions for all queries. 
\\
$\bullet$ The predictions $Pred$ from the second module. 
\\
$\bullet$ The co-occurrence matrices $C_{r}$ and the query list $Td_r$ from the leakages.
\\
$\bullet$ The co-occurrence matrices $C_{s}$ and the keyword universe $W_s$ from the similar data.
\\
$\bullet$ The parameter $RefSpeed$, which controls the number of recovered queries in each iteration.

We denote the $unknownTd$ as currently un-recovered queries and the $unpairedW$ as currently unpaired keywords. 
We denote $C_r^s$ and $C_s^s$ as sub-matrices of $C_r$ and $C_s$, which represent the co-occurrence matrix between un-recovered and recovered queries, and between unpaired and paired keywords, respectively.
We normalize each row of $C_r^s$ and $C_s^s$ by dividing the sum of that
row at each time the set of recovered queries changes.

We use the matrix $C_r^s$, $C_s^s$, and the distance $s$ to evaluate the score between an un-recovered query $td$ and an unpaired keyword $w$. { 
If a row of $C_s^s$ is similar to one of $C_r^s$, it might indicate a correct prediction for the corresponding keyword and query.}
The score contains two parts, the L2-norm of $C_r^s[td]-C_s^s[w]$ and the distance $s(td,w)$ (calculated in Equation \ref{equ:score}), which are summed with weight $\beta$ and $(1-\beta)$. 
The score of a prediction $(td_i,w_j)$ is defined as:
\begin{equation}
    score(td_i,w_j)=-\ln( \beta||C^s_{r}[td_i]-C^s_{s}[w_j]||+(1-\beta)s(td_i,w_j)).
\end{equation}
If a prediction $(td,w)$ is correct, the $s(td,w)$ and the $||C^s_{r}[td]-C^s_{s}[w]||$ will be small, which results in a high score.

Inspired by the RSA\cite{damie2021highly}, we use a similar concept - $certainty$ - to measure the level of assurance in the prediction for a query. Given a query $td$, the prediction $(td,w_i)$ is considered certain if $score(td,w_i)$ is much higher than the score of any other predictions for $td$. 
The $certainty$ of a prediction $(td,w_i)$ is defined as:
\begin{equation}
    certainty(td,w_i)=score(td,w_i)-\max\limits_{j\neq i} score(td,w_j).
    \label{equ:certainty}
\end{equation}
{For example, if an un-recovered query has scores of 2, 3, and 7 with all three unpaired keywords, then the certainty of this query with unpaired keywords is $-5$, $-4$, and $4$, respectively.}  
In each iteration, we exclusively recover the queries with the highest certainty predictions.

This module runs through multiple iterations, each consisting of three main processes: 
\begin{enumerate}
  \item For all un-recovered queries $unknownTd$, calculate the $score$ of all predictions between $unknownTd$ and $unpairedW$. {Then, based on the $score$, calculate the highest $certainty$ along with the prediction of each query, and add them to $Temp\_Pred$. (line \ref{A3calcers}-\ref{A3calcere})}
  \item If the number of un-recovered queries is less than $RefSpeed$, then add all the predictions in $Temp\_Pred$ to $Final\_Pred$, else add $RefSpeed$ predictions with largest $certainty$ in $Temp\_Pred$ to $Final\_Pred$ (line \ref{A3addpres}-\ref{A3addpree}). 
  \item Update the $unknownTd$ and $unpariedW$. 
  Update and normalize the $C_r^s$ and $C_s^s$ accordingly. (line \ref{A3update})
\end{enumerate}

{
In the initial iterative process, the recovered queries have high accuracy due to their elevated level of $certainty$. 
As the process further operates, subsequently recovered queries are also recovered with high precision, primarily because of the augmented correlation between these queries and those that have already been recovered in previous iterations.}

In this module, we use a similar method as the one introduced in RSA~\cite{damie2021highly}. 
However, our approach provides several crucial differences. 
First, while the RSA algorithm exclusively utilizes the co-occurrence matrix to calculate the $score$, we incorporate both volume and frequency information in our $score$ calculation.  
{What's more, the performance of RSA is constrained to the ``pre-set'' known queries of high volume from the outset \cite{damie2021highly}.  
In Jigsaw, we use the second module to actively collect and recover the high-volume queries and further feed them into the third module.  
Their high volume provides Jigsaw with an advantage in query recovery.} 
Moreover, for a query $td$, the RSA calculates the $score$ of all keywords, potentially resulting in matching the query to a keyword that is already paired with another query. 
Our approach only calculates the $score$ of unpaired keywords. 
Furthermore, we normalize the co-occurrence matrix in each iteration, which differs from the RSA algorithm. 
These differences collectively contribute to a more robust and accurate outcome for our algorithm, particularly when encountering defenses (for example, under the obfuscation in CLRZ\cite{DBLP:conf/infocom/ChenLRZ18} in Enron, our attack achieves $>80\%$ accuracy while the RSA only captures $<40\%$.). 
We provide a detailed analysis of the advantages of our approach in Section \ref{sec:comparison_with_other_attacks} and Section \ref{sec:Against Countermeasures}.

\section{Evaluations}
\label{sec:evaluations}
We evaluate our attack under various metrics in real-world datasets to  show its effectiveness. 
We use Python 3.95 to simulate and run codes in Ubuntu 22.04.1 with 16 cores of an Intel(R) Xeon(R) Gold 5120 CPU (2.20GHz) and 64 GB RAM. {Our code is publicly available in \url{https://github.com/JigsawAttack/JigsawAttack.git}.}

\input{figures_in_experiments_distribution_enron_only}

\subsection{Experimental Setup}
\label{sec:Setup}
\noindent \textbf{Datasets.} 
We utilize three datasets, Enron, Lucene, and Wikipedia, for our experiments.
The Enron email corpus\cite{Enron} was collected between 2000-2002, consisting of 30,109 emails, which is a widely used dataset in previous research.
Lucene mailing list was formed between 2001-2020, with 66,491 emails from Apache Foundation\cite{Lucene}. 
For both Enron and Lucene datasets, we utilize pre-processed versions available in \cite{oya2021hiding}.
We use Wikipedia dataset\cite{Wikipedia} in 2020 and extract a subset of 1,000,000 documents by the algorithm in \cite{PlainTextWikipedia}. 
We employ the NLTK package\cite{NLTK} in Python to obtain all English words in datasets except the stop words for keyword extraction. 
In the experiments, we assume that the attacker obtains the same keyword universe as the user.

\noindent \textbf{Frequency information.} For the tests in Enron and Lucene datasets, we adopt the Google Trend\cite{GoogleTrends}, which contains 260 weeks of search trends in Google between October 2016 and October 2021, to generate query frequency for each keyword. 
Specifically, we calculate the sum of each query frequency in $1$ to $50$ weeks as the attacker's auxiliary knowledge. 
We normalize each keyword's frequency by dividing the frequency sum of all keywords as $F_{s}$. 
We also generate the user's queries according to the summed frequency in $1+\tau$ to $50+\tau$ weeks (denote as $F$) where $\tau$ is the time offset between the attacker's knowledge and the observation.
For tests in the Wikipedia dataset, we use the Pageviews Analysis\cite{Pageviews}, which contains 75 months of page views from July 2015 to September 2021. We use the sum of each query frequency in $1$ to $30$ months as the attacker's auxiliary knowledge and the frequency of $1+\tau$ to $30+\tau$ months to generate queries. 

\noindent \textbf{Attacker's knowledge.} 
We randomly divide all documents into two disjoint subsets of equal size. 
We use one subset as the user's encrypted database (i.e., the real data $D_{r}$) and another as the similar data $D_{s}$. 
Then, the user generates single-keyword queries according to the frequency $F$.
The attacker generates $W_{s}$, $V_{s}$, and $C_s$ from similar data and observes all the user's queries to obtain $Td_{r}$, $V_{r}$, $F_{r}$, and $C_r$.
We perform 30 independent simulations. In each simulation, we randomly select half of the documents as similar data and generate queries according to $F$.

\noindent \textbf{Accuracy definition.}
We use the terms $\textit{accuracy}$ and $\textit{recovery rate}$ to evaluate the attack performance. 
The recovery rate refers to the proportion of recovered queries in all observed queries (i.e., $|Recovered(Td^s)|/|Td^s|$). The accuracy denotes the correctly recovered queries out of recovered queries (i.e., $|CorrectRec(Td^s)|/|Recovered(Td^s)|$).

{\subsection{Performance of Algorithm \ref{alg1} and \ref{alg2}}}
\label{sec:test_alg1_and_alg2}
\input{table_Rec_Ver_Num.tex}
{We here provide evaluations for Algorithm \ref{alg1} and \ref{alg2} (the results of the entire Jigsaw are in Section \ref{sec:test_step3} and after).}
We first demonstrate the results of Algorithm \ref{alg1} in four quadrants (including HVHF, HVLF, LVHF, and LVLF).
For our experiments, we extract the top $1,000$ keywords based on their volume and generate $100,000$ queries with $\tau=0$.

To evaluate the recovery in different quadrants, we sort the queries in $Td_{r}$ according to their volume in descending order. 
We treat the first $rv\cdot l$ queries as high-volume queries ($l=|Td_r|$), while the remaining are low-volume queries. 
Similarly, we divide the queries into high and low-frequency queries, containing $rf\cdot l$ and $(1-rf)\cdot l$ queries, respectively. 
Using the above division, we categorize the queries into the HVHF, HVLF, LVHF, and LVLF quadrants.
We set $\alpha$ to $0.5$ and $BaseRec$ to $l$ to recover all queries and test the accuracy in four quadrants by varying $rv$ and $rf$. 
Figure \ref{fig:quadrants_enron_only} demonstrates the accuracy of Algorithm \ref{alg1} on Enron.
Detailed results for Lucene and Wikipedia are given in Appendix \ref{appendix:Extra results on Lucene and Wikipedia}.

Figure \ref{HVHF_3D_Enron} depicts the results of the HVHF quadrant in  Enron. When $rv<0.2$ or $rf<0.1$, the accuracy is approximately $70\%$. 
As the increase of $rf$ or $rv$ indicates a greater proportion of queries with lower volume and frequency within the quadrant, the accuracy falls. 
This indirectly proves that queries with high frequency or high volume are easier recoverable. 
Similar trends can be observed in the performance of Lucene and Wikipedia. 
In contrast, the accuracy in the LVLF quadrant is only $<0.3$ (see Figure \ref{LVLF_3D_Enron}). 
From Figure \ref{fig:distribution}, we can see that the queries in this quadrant are much denser as compared to  other quadrants. 
There is a lack of distinguishability based on volume and frequency, resulting in such a low accuracy.
Figure \ref{HVLF_3D_Enron} exhibits the accuracy of queries confined to the HVLF quadrant, representing the top $rv\cdot l$ and bottom $(1-rf)\cdot l$ queries w.r.t. volume and frequency, respectively. 
The recovery of queries  mainly relies on volume, and the accuracy reaches the summit when $rv < 0.1$ and $rf<0.8$, which implies that queries with high volume and low frequency can be recovered with high accuracy.
The LVHF quadrant delivers a similar result to the HVLF quadrant, see 
Figure \ref{LVHF_3D_Enron}. 
When $rf$ is $<0.2$, the recovery provides high accuracy, approaching $70\%$.

We also investigate the impact of the parameter $\alpha$ on the accuracy in the four quadrants.
We set $rv=0.1$, $rf=0.1$, and $BaseRec=l$ to recover all queries, and the results are shown in Figure \ref{fig:quadrants_test_alpha_enron_only}.
The accuracy decreases when $\alpha$ is either 0 or 1, indicating that relying solely on frequency or volume for query recovery leads to poor accuracy.
When $\alpha$ is appropriately configured, the accuracy $>50\%$ in the HVHF, HVLF, and LVHF quadrants, while it remains relatively low in the LVLF quadrant.
We also observe  that the $\alpha$ displays a distinct impact on the accuracy in the quadrants. 
For example, in the HVLF quadrant, when $\alpha = 0.05$, we achieve the highest accuracy, while the LVHF quadrant's best performance is when $\alpha$ is about $0.3$. 
In the HVLF quadrant, we achieve the highest accuracy when $\alpha=0.05$, while in the LVHF quadrant, the best performance is obtained with $\alpha$ around $0.3$.
This suggests that selecting an appropriate value of $\alpha$ accordingly can lead to higher accuracy in different scenarios.

Table \ref{tab:tab_Rec_Ver_Num} presents the recovery results of Algorithm \ref{alg1} and \ref{alg2} when we consider $BaseRec\in\{25, 100, 400\}$ and $ConfRec/BaseRec\in\{100\%, 50\%, 20\%\}$. 
Overall, the accuracy achieved for various parameter combinations surpasses 50\%. 
As $BaseRec$ and $ConfRec/BaseRec$ decrease, indicating a reduction in the number of recovered queries, the recovery rate decreases while the accuracy exhibits an increase. 
For instance, when we set $BaseRec=25$ and $ConfRec/BaseRe=20\%$, Algorithm \ref{alg2} achieves around $1\%$ recovery rate but $100\%$ accuracy on all the datasets. {We also observe that with the same parameters, the results in Wikipedia are better than those in Enron and Lucene, indicating that Wikipedia contains more distinctive queries.} 

Despite a decline in the recovery rate, when Algorithm \ref{alg2} is employed, its accuracy remains remarkably high with a small $BaseRec$ and $ConfRec$.
In comparison to RSA\cite{damie2021highly} relying on a dozen known queries and their relations (with other queries) to recover all queries accurately, the refinement on queries proposed by Algorithm \ref{alg2} is sufficient to pose a significant threat to all the user's queries. We demonstrate this in detail in the experiments of the next subsection.

\subsection{Results of the Jigsaw Attack}

\label{sec:test_step3}

\input{figures_in_test_beta}

We show the results of the Jigsaw attack and demonstrate how $\beta$ influences the accuracy here. 
We use the same setting as in Section \ref{sec:test_alg1_and_alg2}. Besides, we set $BaseRec$ to $100$, $ConfRec$ to $50$,  $\alpha$ to $0.3$, and $RefSpeed$ to $15$.

The experimental results when varying the values of $\beta$ in Algorithm \ref{alg3} are depicted in the first column of Figure \ref{fig:test_beta}.
When $\beta=0$, meaning the recovery only relies on the frequency and volume, the accuracy reaches around $30\%$ in Enron and Wikipedia and $60\%$ in Lucene. 
As $\beta$ increases, we see the rise in accuracy, ultimately reaching the peak of over 95\% accuracy for Enron and over 98\% for Lucene and Wikipedia.

As demonstrated above, using similar documents and knowledge of query frequency, Jigsaw achieves $>95\%$ accuracy.
However, there are some cases where the attacker may not have access to query frequency, such as when dealing with newly established databases.
To examine the effectiveness of Jigsaw under such circumstances, we perform additional evaluations without utilizing frequency information.
In the absence of frequency information, the attacker should discern fewer distinctive queries, yielding a negative influence on the performance of Algorithm \ref{alg1} and Algorithm \ref{alg2}. 
However, by adjusting the parameter $ConfRec$ to recover a smaller number of queries, the first two modules can still achieve high accuracy in identifying distinctive queries.
We note that with this smaller yet highly accurate set of recovered queries, our attack's accuracy still remains stable and reaches $>90\%$ when $\beta$ is set to $0.8$ and $1.0$ (See in the second column of Figure \ref{fig:test_beta}).

\subsection{Durability}

\input{table_durability.tex}

In the experiment, we use an ``outdated'' query frequency obtained from the past as auxiliary information to enhance accuracy.
We here introduce the concept of \textit{durability} to measure the effect of the time offset between the outdated and target queries on the attack's recovery.
The time offset indicates how ``old'' the query frequency information is. 
{An outdated piece of frequency information might deviate significantly from the actual query frequency, possibly leading to the failure of attacks.}
We consider an attack to be durable if it can maintain its accuracy even as the time offset increases.

We conduct experiments to evaluate the durability of our attacks in Table \ref{tab:table_durability}. 
For Enron and Lucene, we use the frequency of the first 50 weeks in Google Trend as the attacker's auxiliary information, while the target queries are generated using the frequency during $\tau$ and $50+\tau$ weeks. 
For Wikipedia, we use the first 30 months' query frequency in Pageviews Analysis as the auxiliary information, and the queries are generated according to the frequency during $\tau$ and $30+\tau$ months.
Note that $\tau$ is the corresponding time offset. 
In the time offset between $10$ and $150$ weeks, the drop of the attack accuracy is rough $>0.05$ in Enron and Lucene.   
In Wikipedia, the accuracy decreases about $3.5\%$ as the time offset increases from $2$ to $30$ months. 
These results suggest that a leaked query frequency continues to have an impact on our attack even after several years have passed.   
\subsection{Summary of Evaluations}
The results clearly illustrate that the first module of Jigsaw  successfully recovers queries with a high level of accuracy. 
We also confirm that the second module can obtain nearly $100\%$ accuracy for dozens of queries, and the last module is able to recover all queries with about $95\%$ accuracy. 
Even without the frequency information, the accuracy does not experience a significant decline. 
At last, Jigsaw demonstrates its durability by maintaining consistent accuracy, even as the time offset increases from several weeks to years.

\section{Comparisons with Other Attacks}
\label{sec:comparison_with_other_attacks}

We compare the performance of our attack with Graph Match attack\cite{pouliot2016shadow} (Graphm), Sap attack\cite{oya2021hiding} (Sap), Refine Score attack\cite{damie2021highly} (RSA) and IHOP attack\cite{oya2021ihop} (IHOP). 
We do not test the Freq attack\cite{liu2014search} here as the Sap dominates its results.

{\subsection{Settings and Parameters}}

{We extract the top $|W|$ keywords based on their volume as keyword universe.} 
In Enron and Lucene, we evaluate the above attacks for different values of $|W|$, namely $500$, $1000$, and $2000$. 
But for Graphm, we omit $|W| = 1000, 2000$ due to the extended time required for the test. For example, the running time for Graphm with $|W|=1000$ exceeds $10,000$ seconds. 
We suppose the attacker knows the frequency of each keyword in the keyword universe in the first $50$ weeks and generate $\eta$ queries each week for a duration spanning from $50$ to $100$ weeks ($\tau=50$) using the frequency obtained from Google trend. 
We set $\eta$ to $100$, $500$, and $2500$. 
In Wikipedia, we test all the attacks except the Graphm, and we set the $|W|$ to $1000$, $3000$, and $5000$. 
We assume the attacker knows the initial $30$ months' query frequency from Pageviews Analysis, and the user generates $\eta$ queries based on the frequency during the period of $10$ and $40$ months ($\tau=10$). 
We set the $\eta$ to $1000$, $5000$, and $10000$.

{\textbf{Parameters for Jigsaw.}} 
{Recall that the selections of $\alpha$, $\beta$, $BaseRec$, and $ConfRec$ can significantly influence Jigsaw's accuracy (see Section \ref{sec:evaluations}).  
We briefly introduce the reasons behind these parameters selection. 
\\
$\bullet$
As for $\alpha$, we use it to control the weight of volume and frequency information. 
Based on Figure \ref{fig:quadrants_test_alpha_enron_only}, it is recommended to select the parameter from 0.05 to 0.4. 
If the volume information is not accurate (i.e. being noised by certain countermeasures), a smaller $\alpha$ is recommended, and vice versa. Here, we set the $\alpha$ to $0.3$. 
\\
$\bullet$ For $\beta$, one may choose $\beta$ from $0.8$ to $1.0$ as illustrated in Figure \ref{fig:test_beta}.
Note that as a larger $\beta$ indicates assigning more weight to co-occurrence rather than volume and frequency, the co-occurrence information appears to play an important role in recovering queries. 
Similar to the case of $\alpha$, if the co-occurrence is affected by noise, one can opt for a relatively small $\beta$. We set the $\beta$ to $0.9$ here.
\\
$\bullet$ For $BaseRec$ and $ConfRec$, we found that the $ConfRec$ should be set to at least 5 for the third module to initiate, and the $BaseRec$ should range from $1.2\times ConfRec$ to $2\times ConfRec$. 
In normal SSE settings, $BaseRec$ and $ConfRec$ should be sufficiently large to output more accurately recovered queries, whereas they are set to small when observed information is noised to ensure the accuracy of the second module. The $BaseRec$ and $ConfRec$ are set to $45$ and $35$, respectively.
\\
$\bullet$ When it comes to $Refspeed$, it controls how many queries to recover in each iteration. 
One may use a large $RefSpeed$ to optimize the runtime of Jigsaw (such as one-tenth of $|W|$), but this could harm attack accuracy. 
A gradually increased $RefSpeed$ is recommended when dealing with countermeasures, as it can yield both practical runtime and accuracy. 
The $RefSpeed$ is set to $10$ when the keyword universe is small ($<=2,000$) and $50$ when the universe is large ($>2,000$). As in the time-limited settings, we set the $Refspeed$ to $|W|/10$.

}

{\textbf{Parameters for other attacks.}} We use the implementation of PATH algorithm\cite{PATH} available in the package\footnote{http://projects.cbio.mines-paristech.fr/graphm/} to solve the graph matching problem in Graphm (aiming to produce the best performance).  
We set the $\alpha$ in Graphm to $0$ because we find that Graphm can perform its best when $\alpha=0$ in our settings. 
Note we conduct tests ranging from $\alpha=0$ to $\alpha=1$, with increments of $0.1$.  
Recall that RSA requires some known queries in the setup. 
We randomly choose $10$ queries and reveal the true keywords for RSA. We also include the results of RSA with varying numbers of known queries in the Appendix \ref{sec:comparison_which_rsa_with_different_number_of_known_queries}. 
The refine speed in RSA is set to the same as Jigsaw.
We set the $\alpha$ in Sap to $0.5$.
For IHOP, we set the $p_{free}$ to $0.25$ and the $n_{iters}$ to $500$.

{\subsection{Comparison Results}}

\begin{figure}[!ht]

	\subfigure[Enron]
	{
 \label{fig:comparison_enron}
		\begin{minipage}{.88\linewidth}
			\centering
                \includegraphics[width=\linewidth]
                {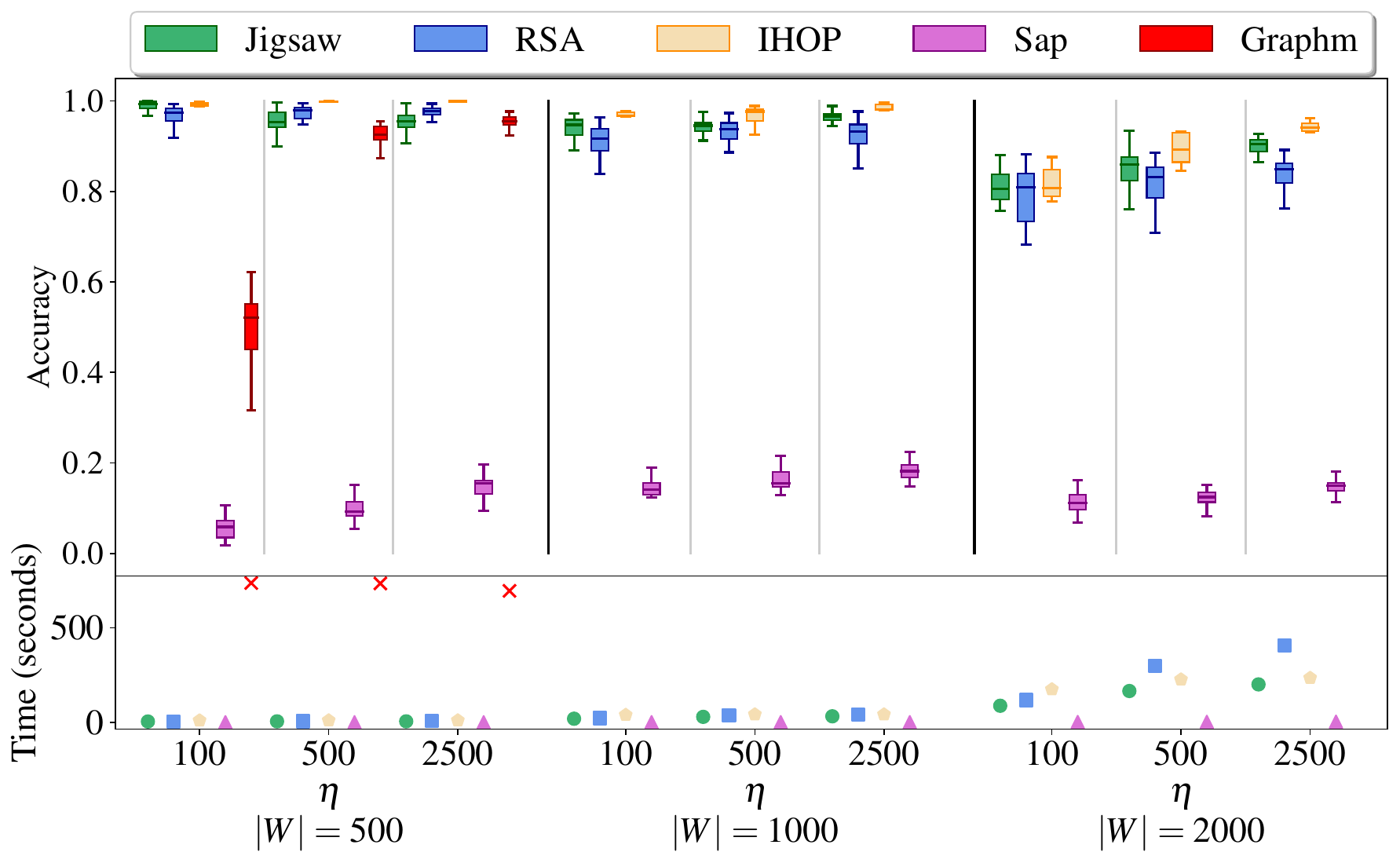}
		\end{minipage}
	}
	\subfigure[Lucene]
	{
 \label{fig:comparison_lucene}
		\begin{minipage}{.88\linewidth}
			\centering
			\includegraphics[width=\linewidth]{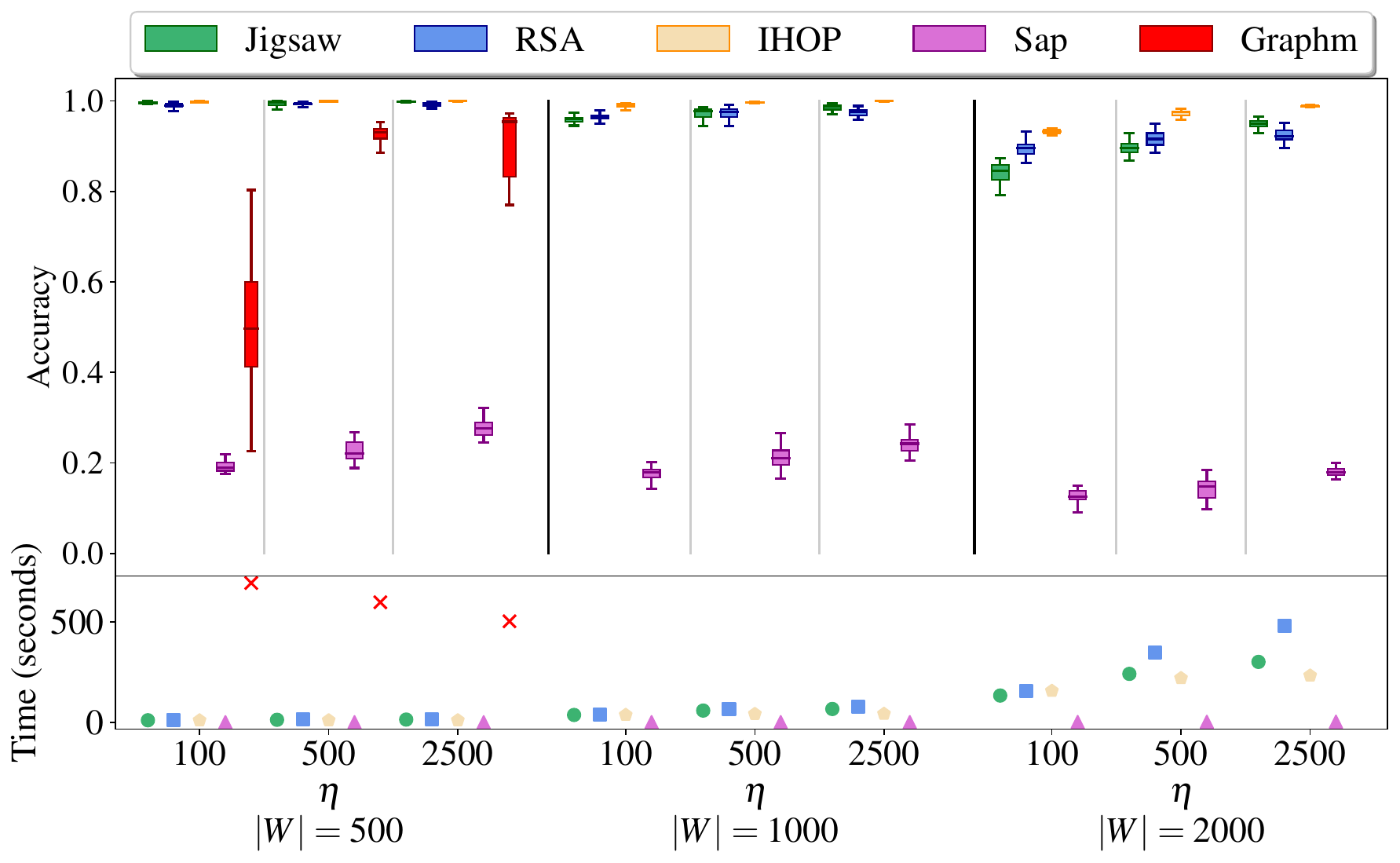}
		\end{minipage}
	}
        \subfigure[Wikipedia]
	{
 \label{fig:comparison_wiki}
		\begin{minipage}{.88\linewidth}
			\centering
			\includegraphics[width=\linewidth]{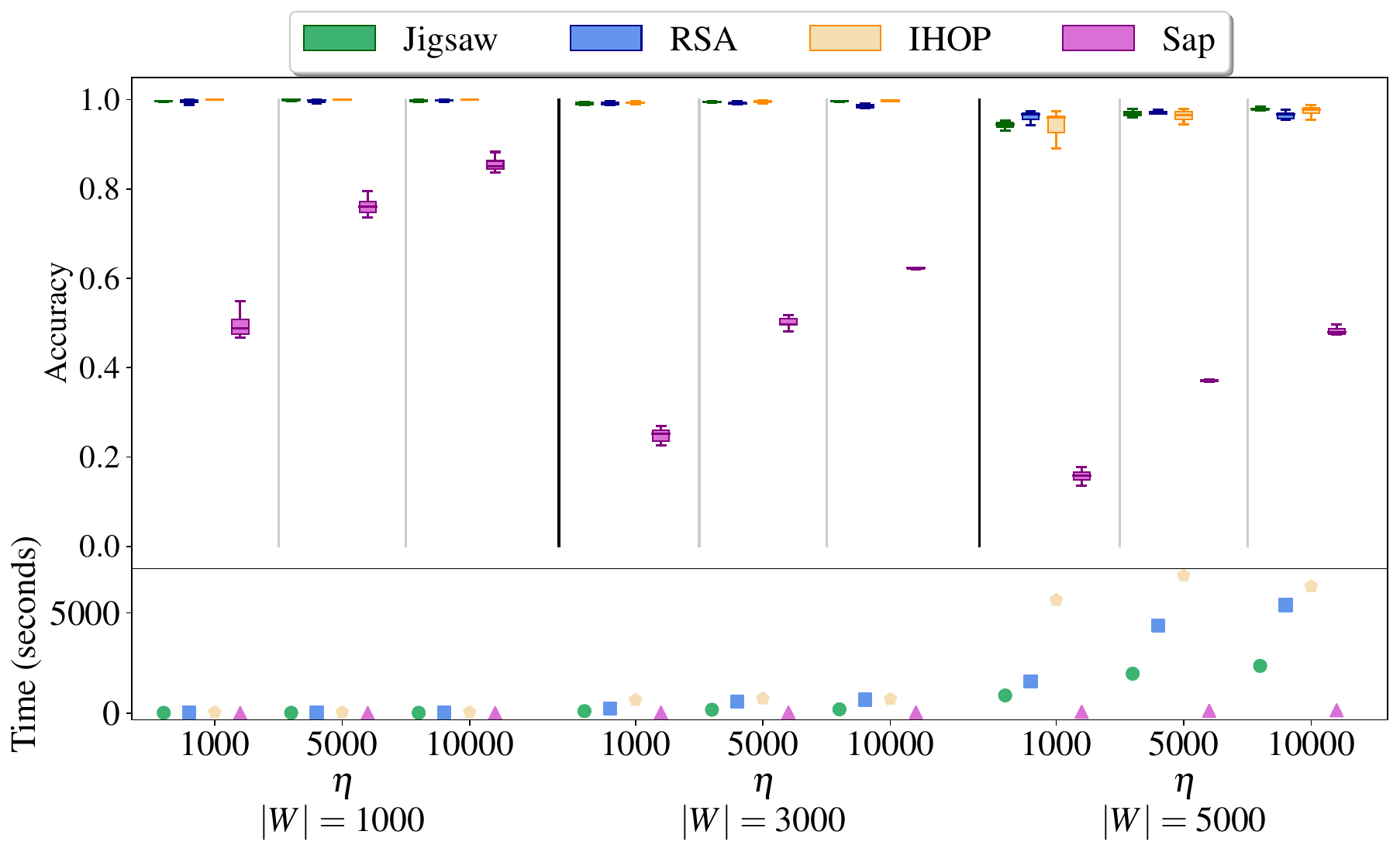}
		\end{minipage}
	}
	\caption{Accuracy \& Time comparisons in Enron, Lucene, and Wikipedia.}
	\label{fig:comparison}
	
\end{figure}

\begin{figure} [!ht]
    \centering
    \includegraphics[width=0.7\linewidth]{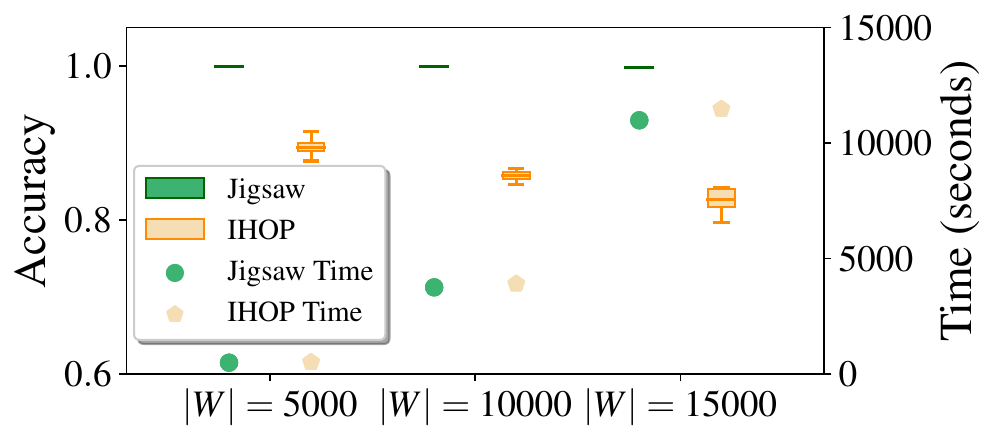}

    \caption{Accuracy comparisons with IHOP within similar runtime.}
    \label{fig:limited runtime}
\end{figure}

{\textbf{Comparisons with all tested attacks.}} 
Here, we demonstrate the results of all tested attacks in Figure \ref{fig:comparison}. 
Our attack provides comparable accuracy to RSA and IHOP while showing a significant advantage over Sap and Graphm.
Graphm's accuracy is low upon $\eta=100$ but improves if more queries are observed. 
It is argued that Graphm requires observation of almost all possible queries to achieve high accuracy~\cite{oya2021hiding}, and it also takes longer matching time between queries and keywords. 
On the other hand, Sap solely utilizes the frequency and volume information, resulting in relatively lower accuracy that increases as it observes more queries.  
{We also observe that increasing $|W|$ results in a decrease in all attacks' accuracy. 
This is because a larger value of $|W|$ introduces a greater number of low-volume keywords. 

}

In the context where a large number of queries are observed, our attack can outperform RSA. 
As the increase of this number, our attack delivers a boost in accuracy.
But RSA cannot gain advantages from more queries because it does not leverage the frequency information. 
Our performance is similar to that of IHOP, as both of the attacks exploit the frequency and co-occurrence of queries. 
We observe that IHOP achieves slightly higher accuracy than our attack.
This can be attributed to its random fixing and free strategy, which enhances the matching of a portion of the queries in each iteration.
This strategy consumes a significant amount of time, especially when the keyword universe is large. 
For example, when $|W|$ is $5000$ in Wikipedia, IHOP takes approximately twice as long as Jigsaw to complete the attack.
We are going to show in Section \ref{sec:Against Countermeasures} that the strategy used by IHOP is not robust once certain countermeasures introduce noise to the leaked information.

{\noindent \textbf{Comparisons with IHOP in large keyword universe under the same time limitation.} We also present the performance of Jigsaw and IHOP within the same time limitation when the keyword universe is large. 
Before the evaluations, we adjust the parameters by setting $RefSpeed$ to $|W|/10$ for Jigsaw and restricting $n_{iters}$ for IHOP to keep both runtimes at a similar pace. 
Note that without any adjusting, IHOP could take approx. 24,000 seconds for 100 iterations, and nearly five times that for 500 iterations, when $|W|=10,000$. 
We illustrate the results under Wikipedia in Figure \ref{fig:limited runtime}. 
As the keyword universe increases, from 5,000 to 15,000, the runtime costs of both attacks jump from $<1,000$ seconds to nearly 10,000 seconds.
IHOP demonstrates a continuous fall in accuracy, from roughly $89\%$ to $85\%$, while Jigsaw's accuracy stands at the same level, $10-15\%$ higher than that of IHOP.    

}

\section{Against Countermeasures}
\label{sec:Against Countermeasures}

We evaluate the attacks against the padding in CGPR\cite{cash2015leakage} and obfuscation\cite{DBLP:conf/infocom/ChenLRZ18}. 
We also show the results against the padding in SEAL\cite{DBLP:conf/uss/DemertzisPPS20} and the cluster-based padding\cite{DBLP:journals/corr/shielddb,DBLP:journals/iacr/BostF17} in Appendix \ref{appendix:Results of Against Other Padding Strategy}.
{We specifically compare the attacks against the padding strategy employed in SEAL, rather than the ``entire'' SEAL (i.e., padding + ORAM)\footnote{We note that ORAM, another crucial component within SEAL, can be used to hide the access and search pattern, and thus could probably counter all the attacks in Table  \ref{tab:tab_passive_attacks}.} 
.}

To counter obfuscation, Oya et al. \cite{oya2021ihop} proposed an adaptation for IHOP, which modifies the co-occurrence matrix of keywords in similar data. 
We apply the same philosophy to Jigsaw and RSA. 
Note that we also design adaptations on the compared attacks against the padding. 
These adaptations can effectively minimize the difference between the user's data after the noise injection and the similar data. 
We highlight that the accuracy of RSA and IHOP increases significantly in most situations after applying the adaptations (such as $30\%$ improvement against the padding in CGPR). 
We present a comprehensive overview of the adaptations and the performance of RSA and IHOP with/without the adaptations in Appendix \ref{appendix:Modifications of the Similar Data}. { We emphasize that there have been no systematic and proper studies for optimal adaptions in existing attacks, rendering this as an interesting open problem. 

}

We conduct a comparative analysis of our attack, RSA\cite{damie2021highly}, and IHOP\cite{oya2021ihop} w.r.t. the aforementioned countermeasures, in Enron, Lucene, and Wikipedia.
We do not test the SAP and Graphm in this section, as they exhibit relatively poor performance or require excessive computational time for the attacks. 
{We also introduce an $\alpha$ term into the objective function of IHOP to balance the weight between the frequency and volume terms against countermeasures (noted as IHOP-$\alpha$). 
We test the $\alpha$ from 0 to 1 with a step length of 0.1, and in most cases, $\alpha=0.1$ brings the best results for IHOP-$\alpha$. 
We fix $\alpha=0.1$ for IHOP-$\alpha$ in the following presentations.}

For Enron and Lucene, we set $|W|$ to $1000$ and set $\eta$ to $500$. 
The attacker is allowed to observe $\eta$ queries per week over a duration of $50$ weeks. 
In Wikipedia, we set the $|W|$ to {$1000$, $3000$ and $5000$} and $\eta$ to $5000$, and the attacker can observe $\eta$ queries per month for a total duration of $30$ months. 
The time offset $\tau$ is set to $0$. 
For Jigsaw, to account for the injection of noise to the volume (by the countermeasures), we set $\alpha$ to a relatively small value, $0.2$. 
And we set $\beta$ to $0.9$. 
The $BaseRec$ and $ConfRec$ are also set to relatively small, $15$ and $10$, to ensure that the second module of Jigsaw can produce correct recoveries. 
{We set the $RefSpeed$ of our attack and RSA to $5$ in Enron and Lucene. 
In Wikipedia, we use a gradually increased value to shorten the runtimes of Jigsaw and RSA.  For each iteration of Jigsaw's third module and RSA, the value increases by $10\%$.} 
The known query number is set to $15$ in RSA. 
The $n_{iters}$ and $p_{free}$ are set to $500$ and $0.25$ in IHOP {and IHOP-$\alpha$}.

\subsection{Against the Padding in CGPR}

\label{sec:against padding in CGPR}

\begin{figure}[!h]
	\centering
	\subfigure[Enron]
	{
 \label{fig:padding_CGPR_enron}
		\begin{minipage}{.45\linewidth}
			\centering
                \includegraphics[width=\linewidth]
                {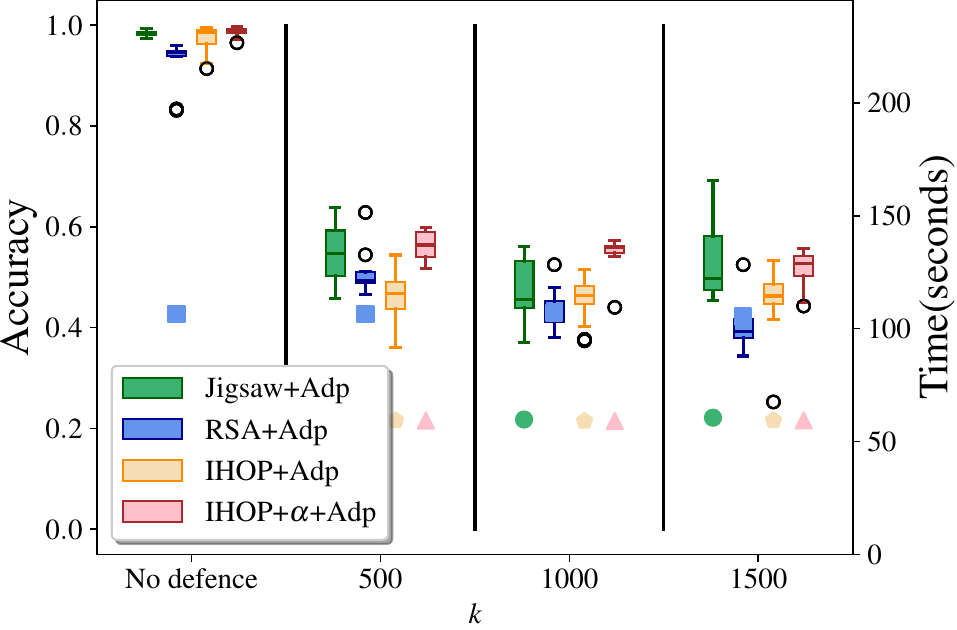}
		\end{minipage}
	}
	\subfigure[Lucene]
	{
 \label{fig:padding_CGPR_lucene}
		\begin{minipage}{.45\linewidth}
			\centering
			\includegraphics[width=\linewidth]{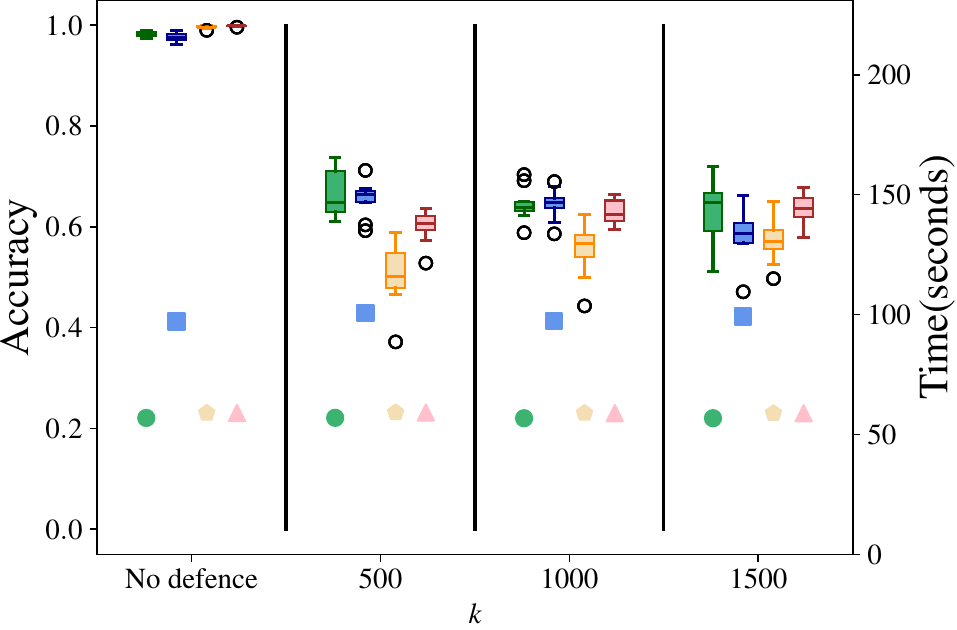}
		\end{minipage}
	}
	\caption{Comparisons with RSA and IHOP against the padding in CGPR\cite{cash2015leakage},  in Enron and Lucene.}
	\label{fig:against_CGPR_enron_lucene}
\end{figure}

\begin{figure}[!h]
	\centering
	\subfigure[Enron]
	{
 \label{fig:obfuscation_enron}
		\begin{minipage}{.45\linewidth}
			\centering
                \includegraphics[width=\linewidth]
                {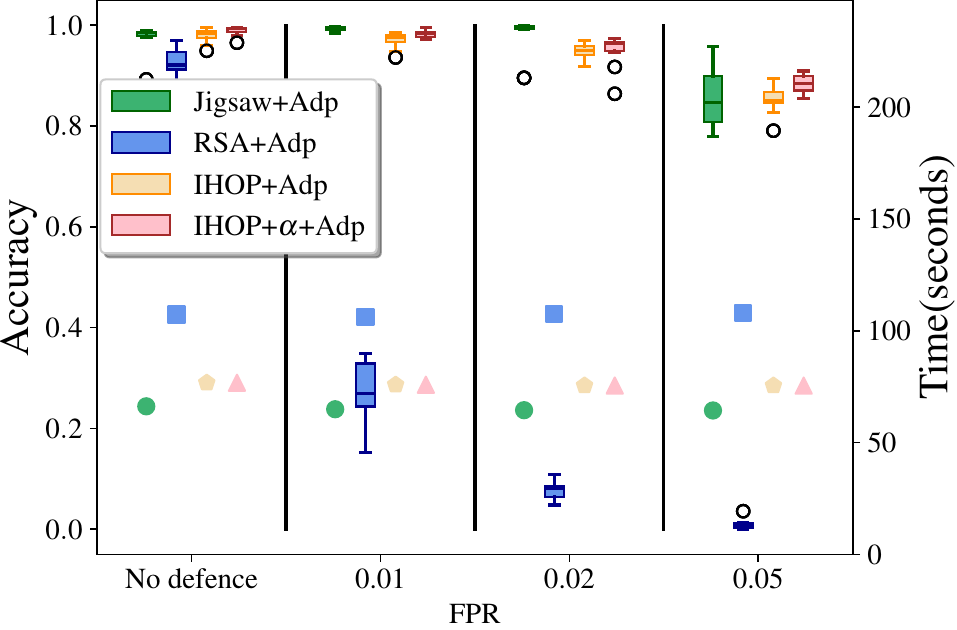}
		\end{minipage}
	}
	\subfigure[Lucene]
	{
 \label{fig:obfuscation_lucene}
		\begin{minipage}{.45\linewidth}
			\centering
			\includegraphics[width=\linewidth]{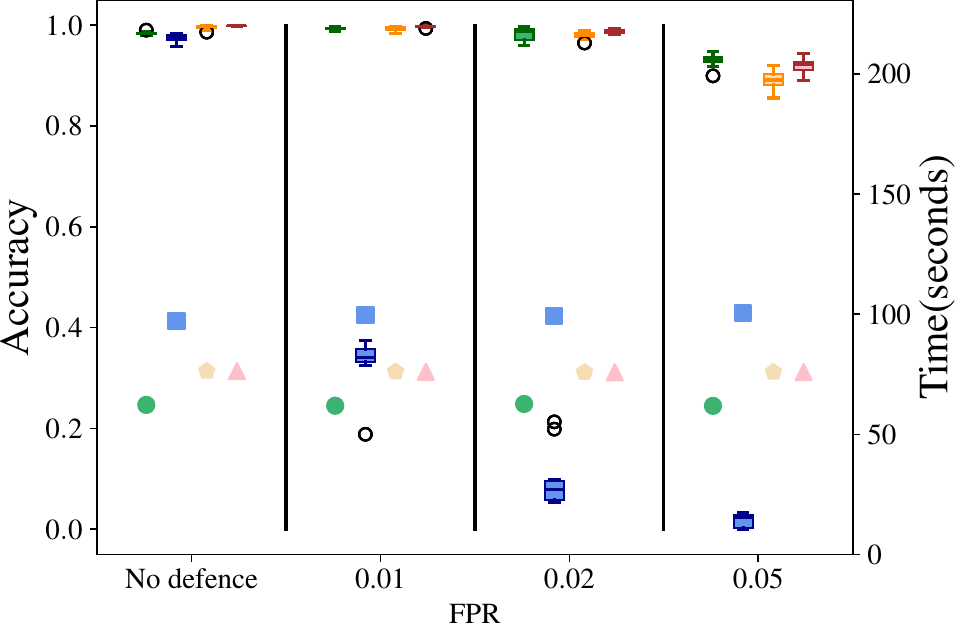}
		\end{minipage}
	}
	\caption{{Comparisons with RSA and IHOP against the obfuscation in CLRZ\cite{DBLP:conf/infocom/ChenLRZ18}, in Enron and Lucene.}}
	\label{fig:against_obfuscation_enron_lucene}
\end{figure}

\begin{figure*}[!h]
	\centering
	\subfigure[$|W|=1000$]
	{
 \label{fig:padding_CGPR_wiki_1000}
		\begin{minipage}{.30\linewidth}
			\centering
                \includegraphics[width=\linewidth]
                {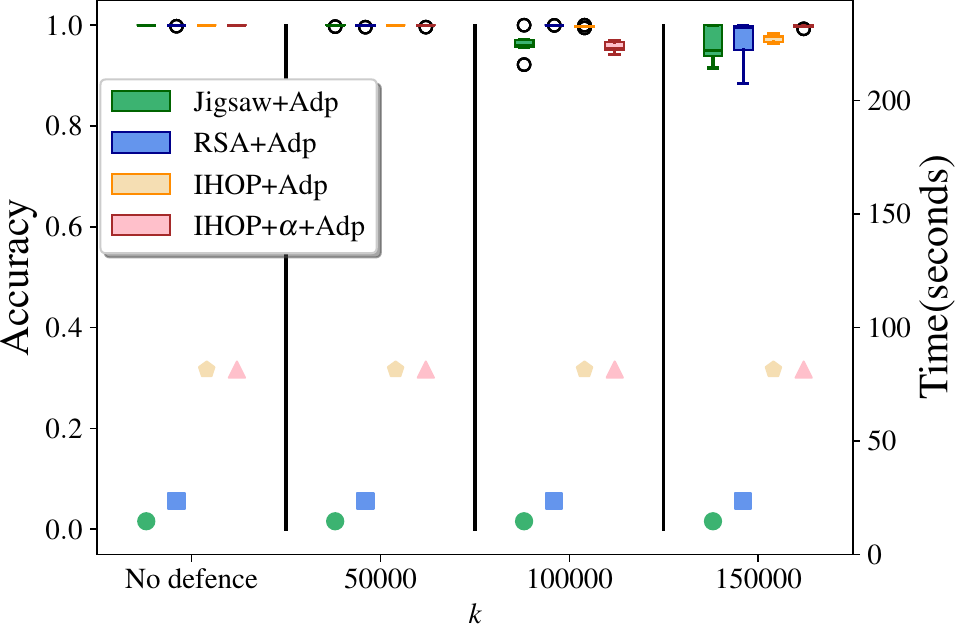}
		\end{minipage}
	}
	\subfigure[$|W|=3000$]
	{
 \label{fig:padding_CGPR_wiki_3000}
		\begin{minipage}{.30\linewidth}
			\centering
			\includegraphics[width=\linewidth]{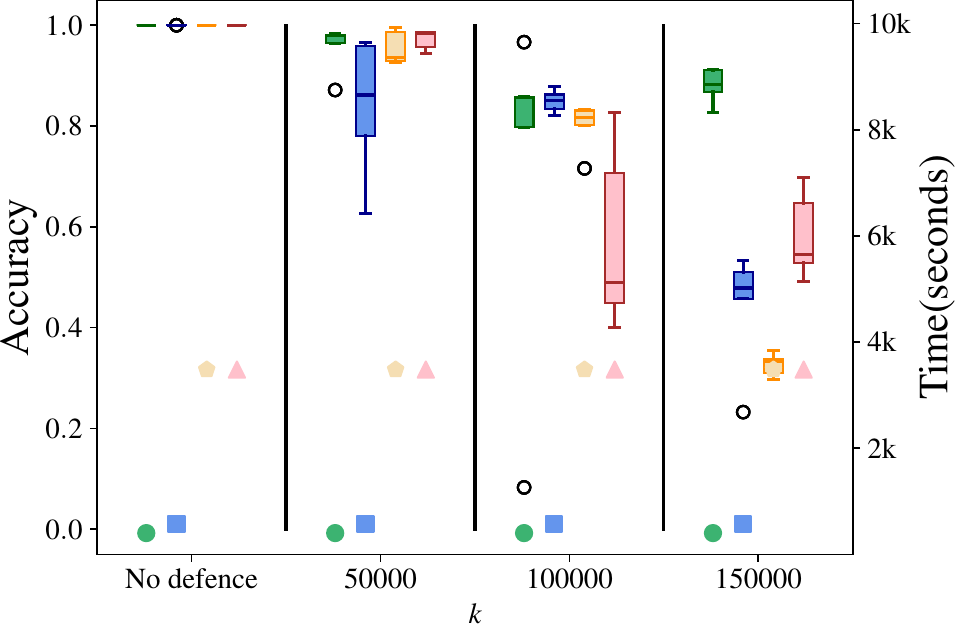}
		\end{minipage}
	}
        \subfigure[$|W|=5000$]
	{
 \label{fig:padding_CGPR_wiki_5000}
		\begin{minipage}{.30\linewidth}
			\centering
			\includegraphics[width=\linewidth]{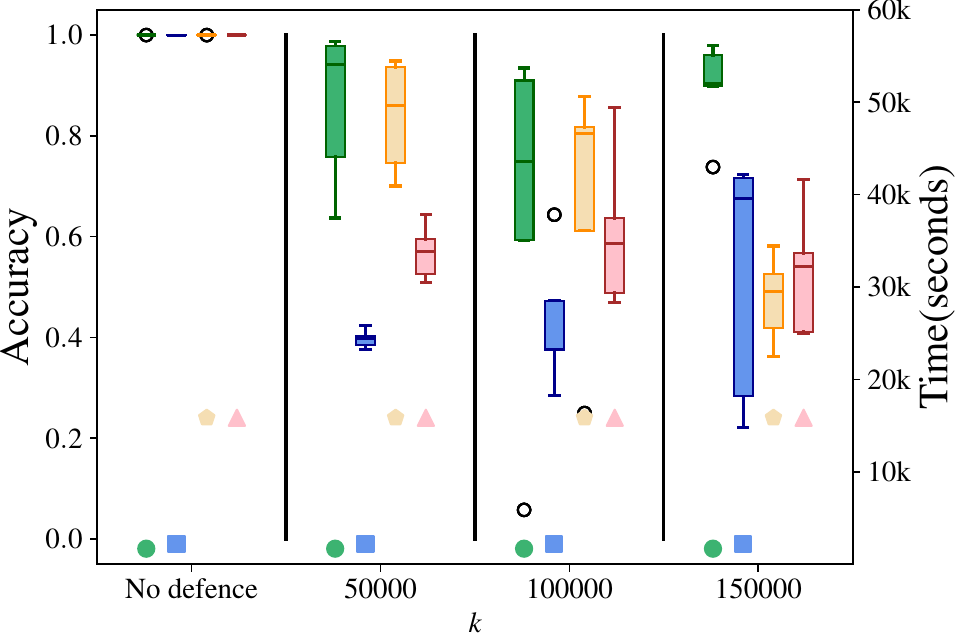}
		\end{minipage}
	}
	\caption{Comparisons with RSA and IHOP against the padding in CGPR\cite{cash2015leakage}, in Wikipedia.}
	\label{fig:against_CGPR_wiki}
\end{figure*}

\begin{figure*}[!h]
	\centering
	\subfigure[$|W|=1000$]
	{
 \label{fig:obfuscation_wiki_1000}
		\begin{minipage}{.30\linewidth}
			\centering
                \includegraphics[width=\linewidth]
                {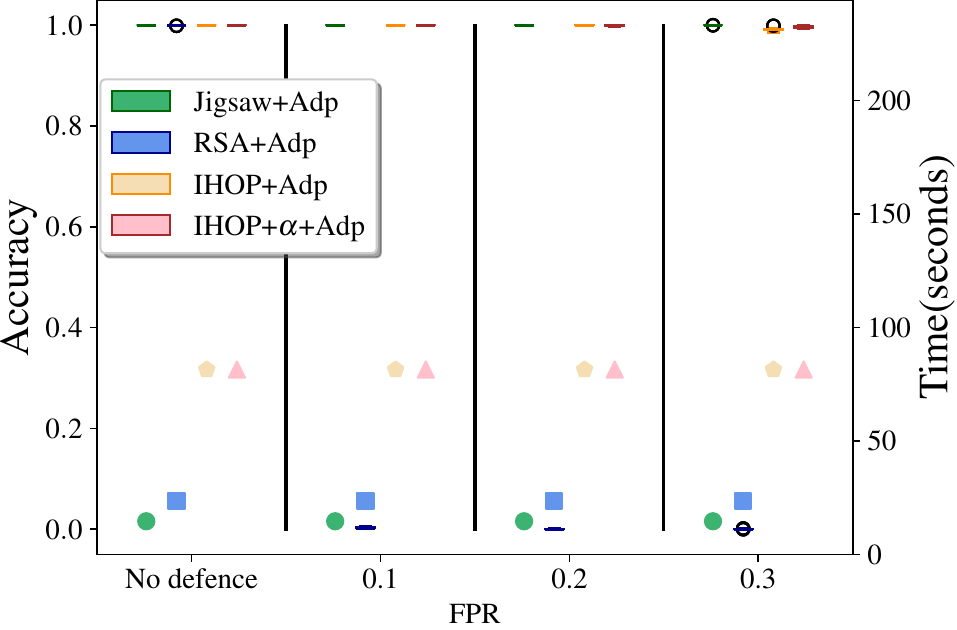}
		\end{minipage}
	}
	\subfigure[$|W|=3000$]
	{
 \label{fig:obfuscation_wiki_3000}
		\begin{minipage}{.30\linewidth}
			\centering
			\includegraphics[width=\linewidth]{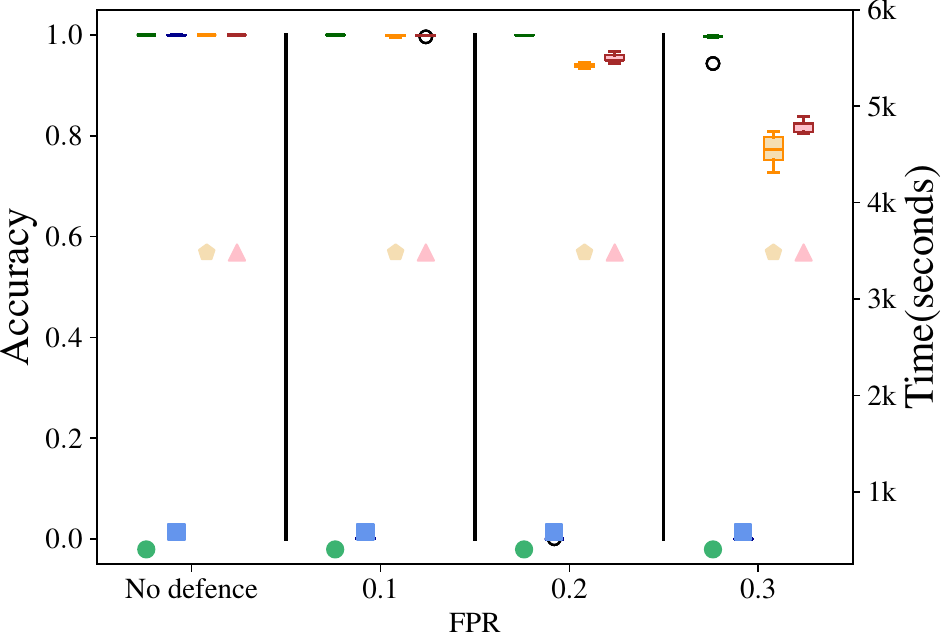}
		\end{minipage}
	}
        \subfigure[$|W|=5000$]
	{
 \label{fig:obfuscation_wiki_5000}
		\begin{minipage}{.30\linewidth}
			\centering
			\includegraphics[width=\linewidth]{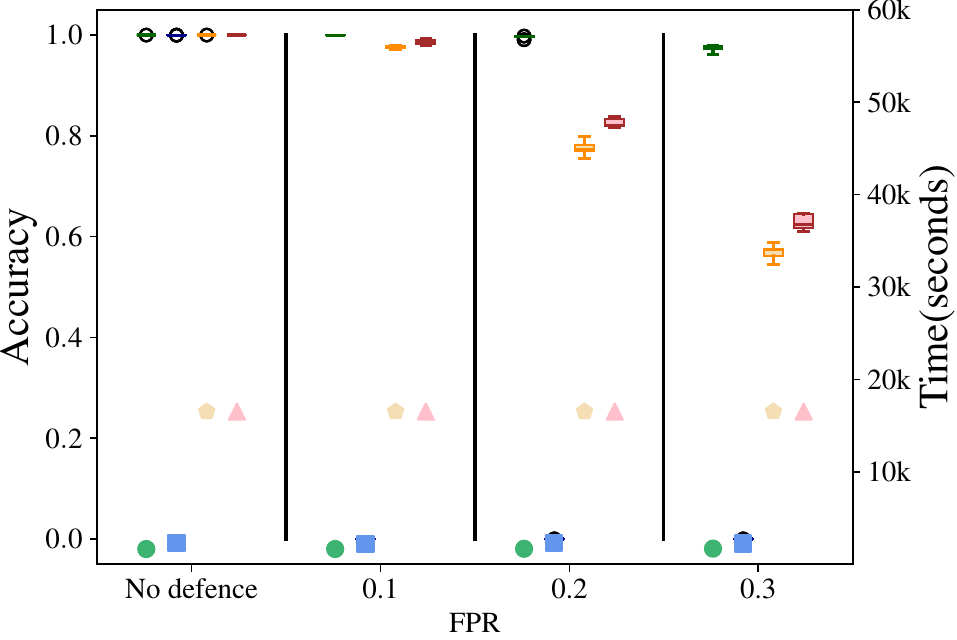}
		\end{minipage}
	}
	\caption{{Comparisons with RSA and IHOP against the obfuscation in CLRZ\cite{DBLP:conf/infocom/ChenLRZ18}, in Wikipedia.}}
	\label{fig:against_obfuscation_wiki}
\end{figure*}

Padding in CGPR \cite{cash2015leakage} injects fake documents to increase the query volume to the nearest multiple of $k$.
{This strategy adds noises to the volume and access pattern and at the same time increases the communication and storage costs (see Appendix \ref{Appendix:overhead} for experimental results).}
For queries with low volume, the padding can substantially change the access and volume patterns, and the volume is more likely to be ``expanded" to the same ``length''. 
But the high-volume queries, on the other hand, are poorly protected. 
This is so because, for those queries (already with a ``large'' volume), the (on-top) padding volume could be relatively small. 
In this case, the impact on the leakage patterns of high-volume queries is minimal. 
We state that since the first two modules of Jigsaw concentrate on recovering high-volume and high-frequency queries that are not significantly affected by the padding, they can keep producing highly accurate predictions. 
By leveraging these accurate recoveries, we can gain ``more'' pre-knowledge to pose a severe threat to low-volume queries.  
{In Figure \ref{fig:against_CGPR_enron_lucene}, we have $k\in\{500,1000,1500\}$ for Enron and Lucene; while in Figure \ref{fig:against_CGPR_wiki}, $k$ is set much larger, $\in\{50000,100000,150000\}$. 
This is because Wikipedia contains a significantly higher number of documents than other datasets.

All the tested attacks exhibit similar accuracy, about $50\%$ in Enron and $>60\%$ in Lucene. 
In Wikipedia, the gap in accuracy is more noticeable, and Jigsaw demonstrates a significant advantage over others when $|W|\geq 3000$.
For example, when $k=150,000$ and $|W|=3000$, Jigsaw provides nearly $90\%$ accuracy, whereas RSA and IHOP only obtain $<60\%$.
IHOP-$\alpha$ performs slightly better than IHOP, with a $60\%$ accuracy. 
This above result is attributed to the fact that the padding cannot protect the distinctive queries in Wikipedia, allowing Jigsaw to recognize and further recover them, which gives it an advantage in recovery.  
We clearly observe that Jigsaw consumes significantly less runtime than IHOP and IHOP-$\alpha$ in Wikipedia due to a gradually increased $RefSpeed$. 
}

\subsection{Against the Obfuscation in CLRZ}

We showcase the experimental results of attacks against the obfuscation in CLRZ\cite{DBLP:conf/infocom/ChenLRZ18}, which works by indexing a keyword to documents that do not contain the keyword with probability FPR and removing the index of documents that do contain the keyword with probability TPR. 
{Since the obfuscation does not involve padding, it does not affect storage costs.  
However, the communication costs will increase greatly due to a larger number of unrelated documents being retrieved (see Appendix \ref{Appendix:overhead}).}
{
In Figure \ref{fig:against_obfuscation_enron_lucene} and \ref{fig:against_obfuscation_wiki}, we set $TPR=0.999$ and $FPR\in\{0.01,0.02,0.05\}$ in Enron and Lucene and $FPR \in \{0.1,0.2,0.3\}$ in Wikipedia. 

Under the obfuscation, the accuracy of RSA drops abruptly to below $20\%$.
Jigsaw just experiences a minor decrease as the $FPR$ increases and, in most cases, maintains accuracy $>85\%$ in all tested datasets. 
IHOP and IHOP-$\alpha$ perform similarly to Jigsaw in Enron and Lucene. 
But in Wikipedia, their accuracy drops significantly with a large $|W|$.  
When $FPR=0.3$ and $|W|=3000$, the accuracy only reaches about $80\%$, dropping to $60\%$ when $|W|=5000$. 
In contrast, Jigsaw remains an accuracy above $95\%$ under the same settings. 
}

\subsection{Discussion}
{
Under all tested countermeasures, it is evident that Jigsaw (after the adaptations) achieves the highest accuracy, $>70\%$, in most cases. 
On average, RSA and IHOP could closely follow Jigsaw's performance. 
But they have some pitfalls. 
RSA is vulnerable to the obfuscation in CLRZ, resulting in $<20\%$ accuracy. 
{
IHOP and IHOP-$\alpha$ also experience low accuracy against the countermeasures with a large $|W|$. 
Their accuracy is approx. $50\%$ against the padding in CGPR (with $k=150,000$) and about $60\%$ against the obfuscation in CLRZ (with $TPR=0.3$) on Wikipedia with $|W|=5,000$.  
}
While one may have the option to apply the defenses to mitigate RSA and IHOP, Jigsaw proves to be a more ``robust'' attack that remains effective. 

{There are several countermeasures that might defend against Jigsaw. 
ORAM~\cite{OBImultipathORAM}, a popular solution to SSE attacks, conceals the access pattern and the derived co-occurrence matrix, reducing the effectiveness of Jigsaw's second and third modules. 
But this comes with an $\Omega(\log N)$ amortized blowup of communication cost for databases of size $N$. 
Providing similar efficacy on the access pattern, PIR\cite{henzinger2023one} is another potential option. 
However, it requires heavy server-side computation and does not support private updates by the client. 
Apart from completely hiding the access pattern, strong padding techniques may infuse noise into the volume pattern of the high-volume queries, making Jigsaw's first module unable to produce sufficient correct recoveries - resulting in low accuracy.  
A drawback of this solution is the necessity to pad a considerable amount of files, especially for high-volume queries. 
It remains an intriguing challenge to develop a padding technique that is both efficient and secure. 
}


\section{Conclusion}

{We propose the Jigsaw, a new similar-data attack against SSE, which works by first recovering the most distinctive queries and utilizing them to recover all queries further.
We test Jigsaw in different datasets and showcase the stable accuracy of around $95\%$ in query recovery. 
Moreover, our attack can provide an accuracy of about $60\%$ and $85\%$ against padding\cite{cash2015leakage} and obfuscation\cite{DBLP:conf/infocom/ChenLRZ18}, respectively, outperforming existing works \cite{liu2014search,oya2021hiding,damie2021highly,oya2021ihop}. The proposed attack exposes the vulnerabilities of existing SSE schemes. 
Developing secure and practical SSE schemes that are resistant to such attacks is an open problem.}

\bibliographystyle{plain}
\bibliography{sample}

\begin{appendices}
\section{Summary of Notations and Concepts}\label{AppenA}
We denote $[n]$ as a list of integers $[1,\ldots,n]$. For a list $L$, we use $|L|$ to represent its length and $L[i]$ to denote its $i$th element. For a set $S$, we use $|S|$ to represent its cardinality. For a matrix $M$, we use $M[i]$ to represent the $i$th row of the matrix and $M[i,j]$ to represent its element in the $i$th row and $j$th column. We use the $||\cdot||$ to denote the Euclidean norm of a vector or the Frobenius norm of a matrix. See the frequently used notations in Table \ref{tab:sum_of_nota}.

\section{Query Distribution and Simple Attack}
\label{Appendix:simpleattack}

We present a simple attack and its evaluation to showcase the relationship between query distribution and query recovery. 
The attack employs knowledge of frequency and volume information to pair queries with the keywords having the most similar frequency and volume. 
We assume the attacker knows a similar dataset $D_s$ and generates a keyword universe $W_s=[w_1,w_2,\ldots,w_m]$. 
Then, it generates the corresponding volume $V_s=[v_{w_1},v_{w_2},\ldots,v_{w_m}]$ of each keyword from $D_s$. 
It also knows a historical query frequency $F_s=[f_{w_1},f_{w_2},\ldots,f_{w_m}]$ of $W_s$.
Also, the attacker can observe the volume and search pattern of queries the user issues and generate the frequency $F_{r}$ and volume $V_{r}$ of them. 
After normalizing the $F_s$, $V_s$, $F_{r}$, and $V_{r}$, it pairs each query $td_i$ with keyword $w_j$ which has the smallest value of $|v_{w_j}-v_{td_i}|+|f_{w_j}-f_{td_i}|$.

\input{table_notations}

\input{figures_in_distribution}

We test the attack in Enron, Lucene, and Wikipedia. 
In Enron and Lucene, we use half of the dataset as the attacker's similar dataset and the other half as the user's dataset. 
We use the summed query frequency of the initial $50$ weeks in Google Trend as the attacker's prior knowledge and the subsequent $50$ weeks to generate the user's queries. 
In Wikipedia, we use 1,000,000 and 30,000 documents as the user's dataset and the attacker's similar dataset, respectively.  
We use the summed query frequency of the first $30$ months in Pageviews Analysis as the attacker's prior knowledge and the following $30$ months to produce the user's queries. 
We select the top $1,000$ keywords on volume except for the stop words as the keyword universe. 

The results are shown in Figure \ref{fig:distribution_a} and \ref{fig:distribution_b}. In Figure \ref{fig:distribution_a}, we normalize the volume and frequency of each query.
The horizontal dashed line divides the top $10\%$ queries on volume from other queries, and the vertical dashed line divides the top $10\%$ queries on frequency from other queries. The two lines divide the queries into four quadrants, i.e., the HVHF, HVLF, LVHF, and LVLF. 
The blue dots denote the queries, and the red dots denote the queries successfully recovered by the simple attack. 
In all tested datasets, the queries in the HVHF quadrant are sparse, and the attack has a high accuracy there. 
On the other hand, in the LVLF quadrant, the queries are nearly indistinguishable and hard to be recovered. 
We can zoom out the left low corner to show this more clearly.
We rank the queries according to their volume and frequency and show the queries according to their rank in Figure \ref{fig:distribution_b}. As the queries have a higher rank in volume or frequency, the red dot is denser, showing a higher accuracy in recovery.

{In Section \ref{sec:our attack}, we provide the definition of \textit{differential distance} $d_{td}$ of a query. 
The distinctiveness of a query increases as its differential distance becomes larger. 
Based on $d_{td}$, we define the number $K$ of 
{distinctive} queries in a dataset as: 
\begin{equation}
K=|\{i:\frac{d_{td_i}}{\sum_{td_j\in W} d_{td_j}/|W|}>\lambda \}|.
\end{equation}
In Enron, Lucene, and Wikipedia, setting $\lambda=5$, we have that the number is roughly $18$, $20$, and $33$, {which can concur with the results in Section \ref{sec:evaluations}, \ref{sec:comparison_with_other_attacks}, and \ref{sec:Against Countermeasures}}.}
 
\input{figures_in_experiments_distribution_lucene_and_wiki}

\section{Results on Lucene and Wikipedia}
\label{appendix:Extra results on Lucene and Wikipedia}

We present the results of Algorithm \ref{alg1} on Lucene and Wikipedia. 
For Lucene, we extract $1,000$ keywords with the largest volume and generate $100,000$ queries. 
And we use $3,000$ keywords and $150,000$ queries in Wikipedia. 
Figure \ref{fig:quadrants_lucene_wiki} shows the results on the four quadrants with different combinations of $rv$ and $rf$. Figure \ref{fig:quadrants_test_alpha_lucene_wiki} shows the results with different $\alpha$. 

The results on Lucene and Wikipedia are very similar to those on Enron.
The accuracy within each quadrant on Wikipedia tends to be lower than that on Enron and Lucene. 
This discrepancy is due to the larger keyword universe in the former dataset. 
However, as demonstrated in Section \ref{sec:test_alg1_and_alg2}, the number and precision of recovered distinctive queries in Wikipedia are not inferior to that in Enron and Lucene.

\input{table_RSA_with_different_query_numbers_wiki}
\section{V.s. RSA with Different Numbers of Known Queries}
\label{sec:comparison_which_rsa_with_different_number_of_known_queries}

As RSA requires some known queries to initiate the attack, we also evaluate the number of known queries for RSA to obtain high accuracy.
We conduct experiments by setting $\eta$ to $500$ and the keyword universe size to $1000$. 
We generate the keyword universe for both the user and the attacker in two ways: 1) selecting 1000 keywords randomly from the top 3000 keywords based on volume, and 2) selecting the top 1000 keywords based on volume. 
We vary the number of known queries as $5,10,25$ and $50$ corresponding to $0.5\%,1\%,2.5\%$ and $5\%$ of the total keywords in the universe. 
Since our attack does not rely on known queries, the known query percentage for Jigsaw is set to $0\%$. 

The results are shown in Table \ref{tab:table_RSA_kqn}. 
When selecting the keywords with the highest volume, the accuracy is higher compared to randomly selected keywords.
In this sense, our attack achieves roughly $9\%$ and $12\%$ higher accuracy in Enron and Lucene. 
RSA shows a similar pattern in the results. 
The difference in accuracy is due to the fact that randomly selected keywords contain many low-volume keywords, making them harder to recover. 
As the number of known queries increases, the accuracy of RSA also increases and reaches its peak around given $25$ known queries. 
Beyond that point, the performance of RSA remains relatively stable.  
In Table \ref{tab:tab_Rec_Ver_Num}, the second module of our attack can recover $25$ queries with near $100\%$ accuracy. 
Recall that our attack leverages the volume and frequency, combined with co-occurrence information to recover queries,  while RSA only considers the co-occurrence information. 
This makes Jigsaw more accurate than RSA even if RSA employs a $5\%$ known query, compared to our $0\%$.

\input{figures_bef_and_after_adp}

\section{Adaptations to Similar Data}\label{appendix:Modifications of the Similar Data}

It seems that the countermeasures such as padding and obfuscation do not consider the protection of the parameters. 
If gaining access to the parameters, the attacker will be able to make adaptations to similar data to weaken the countermeasures. 
For example, in the case of padding in CGPR, the attacker can utilize the parameter $k$ to pad the similar data, thereby minimizing the disparity between the similar data and the padded data. 
We say that this effectively mitigates the adverse effects of padding on query recovery. 
Specifically, our adaptations applied to Jigsaw, RSA, and IHOP are as follows. 
\\
$\bullet$ \textit{Padding in CGPR}\cite{cash2015leakage}. Recall that the user pads the query volume to the nearest multiple of $k$. 
We here employ the same padding approach with a different parameter, $k_{sim}$, on similar data. 
We calculate $k_{sim}$ as $k$ multiplied by the ratio of the sizes of the similar dataset ($D_s$) and the original dataset ($D$), i.e., $k_{sim}=k\cdot{|D_s|/|D|}$. 
Accordingly, this adjustment modifies $ID_s$, which subsequently affects the $C_s$ and $V_s$ parameters in the attacks. 
We also apply the same strategy for similar data in RSA and IHOP.
\\
$\bullet$ \textit{Obfuscation}\cite{DBLP:conf/infocom/ChenLRZ18}. 
We have two phases for this adaptation. Firstly, we apply the co-occurrence matrix (in  Equation \ref{equ:cs}) in \cite{oya2021ihop} to adapt the influence of obfuscation in similar data. 

\begin{figure}[!t]
	\centering
	\subfigure[Enron]
	{
 \label{fig:padding_cluster_enron}
		\begin{minipage}{.45\linewidth}
			\centering
                \includegraphics[width=\linewidth]
                {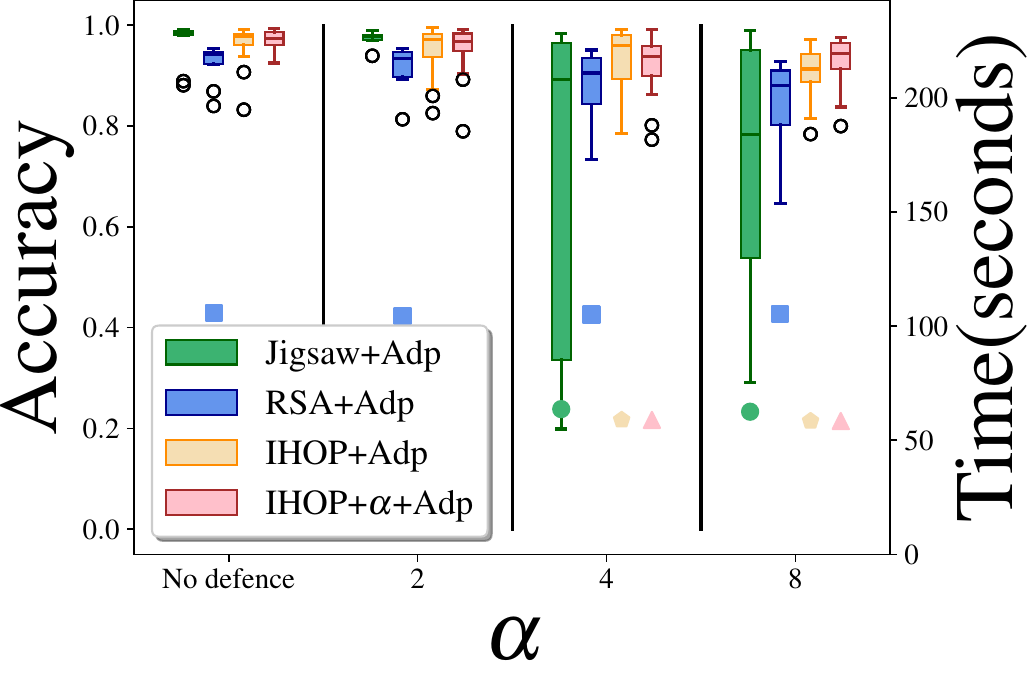}
		\end{minipage}
	}
	\subfigure[Lucene]
	{
 \label{fig:padding_cluster_lucene}
		\begin{minipage}{.45\linewidth}
			\centering
			\includegraphics[width=\linewidth]{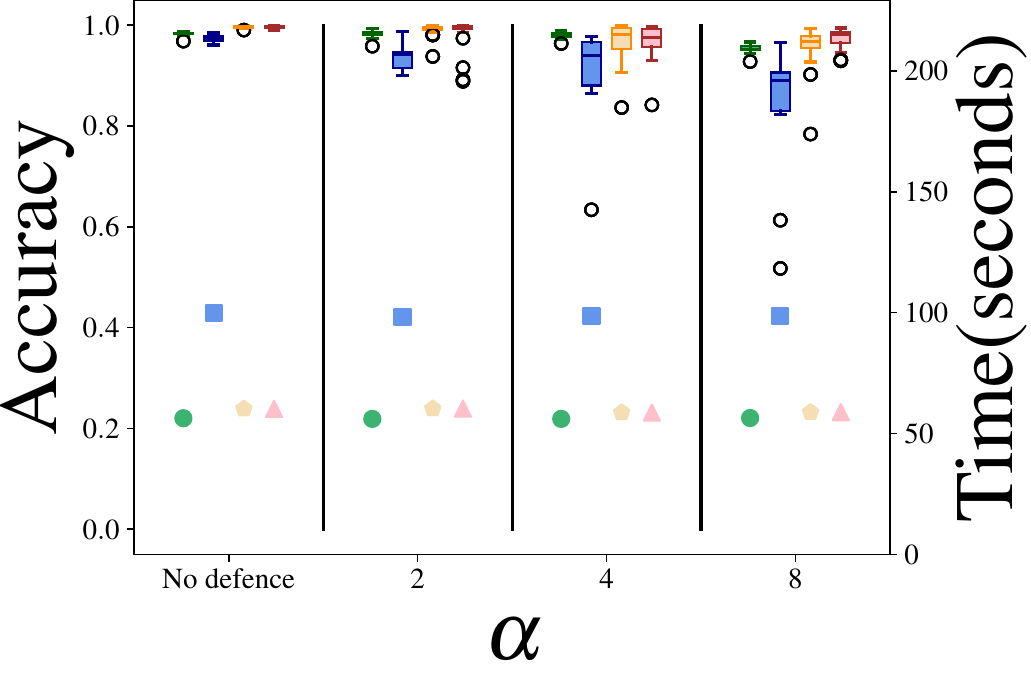}
		\end{minipage}
	}
	\caption{{Comparisons with RSA and IHOP against the cluster-based padding\cite{DBLP:journals/corr/shielddb,DBLP:journals/iacr/BostF17}, in Enron and Lucene.}}
	\label{fig:against_padding_cluster_enron_lucene}
\end{figure}

\begin{equation}
C_s^{obf}[i,j] = \left\{
    \begin{aligned}
     TPR^2 \cdot C_s[i,j] + FPR^2\cdot C_s^{not}[i,j]+\\
     TPR \cdot FPR \cdot(1-C_s[i,j]-C_s^{not}[i,j]),\ & i\neq j; \\
    TPR\cdot C_s[i,j] + FPR\cdot C_s^{not}[i,j],\ & i=j.
    \end{aligned}
    \right.
\label{equ:cs}
\end{equation}
where $C_s^{not}=(1-ID_s)(1-ID_s)^\top/|D_s|$. 
Secondly, we revise $V_s$ as $V_s^{obf}[i]=TPR\cdot V_s[i] + FPR\cdot(1-V_s[i])$ for further adaptation.
\\
$\bullet$ \textit{Padding in SEAL}\cite{DBLP:conf/uss/DemertzisPPS20}. 
{
In SEAL, the volume distribution of the padded queries is closely related to the size of the dataset. 
The varying sizes of similar data to the user's data result in different volume distributions after padding. 
To adapt Jigsaw against the SEAL's padding, we generate a new
similar data $D_s'$ with the size $|D|$ (aligning with the size of the user's data $D$) by expanding $D_s$ with its own data ``copies" in order to keep the volume distribution of $D_s$.} 
We then adopt the same padding strategy on $D_s'$ and replace the $D_s$ with the padded $D_s'$ in attacks.
\\
$\bullet$ \textit{Cluster-based padding}\cite{DBLP:journals/corr/shielddb,DBLP:journals/iacr/BostF17}. We generate a new similar dataset $D_s'$ by padding $D_s$ with the same parameter and replace $D_s$ with $D_s'$ in the attacks.

{We test Jigsaw, RSA\cite{damie2021highly}, and IHOP\cite{oya2021ihop} with/without the adaptations on Enron against the CGPR's padding\cite{cash2015leakage} ($k=1500$), the obfuscation\cite{DBLP:conf/infocom/ChenLRZ18} (TPR$=0.999$, FPR$=0.05$), the cluster-based padding\cite{DBLP:journals/corr/shielddb} ($\alpha=8$), and the SEAL's padding\cite{DBLP:conf/uss/DemertzisPPS20} ($x=4$). 
The parameters are the same as in Section \ref{sec:Against Countermeasures}. 
The results are presented in Figure \ref{fig:before and after adp}. 
For obfuscation, RSA with/without the adaption performs poorly ($<10\%$ accuracy). 
On the other hand, under three padding strategies, the proposed adaptations optimize the accuracy of RSA and IHOP significantly. 
Noticeably our Jigsaw attack with adaptations still takes the lead in most cases, see Section \ref{sec:Against Countermeasures} for comparison details.
}

\begin{figure*}[!t]
	\centering
	\subfigure[$|W|=1000$]
	{
 \label{fig:padding_cluster_wiki_1000}
		\begin{minipage}{.30\linewidth}
			\centering
                \includegraphics[width=\linewidth]
                {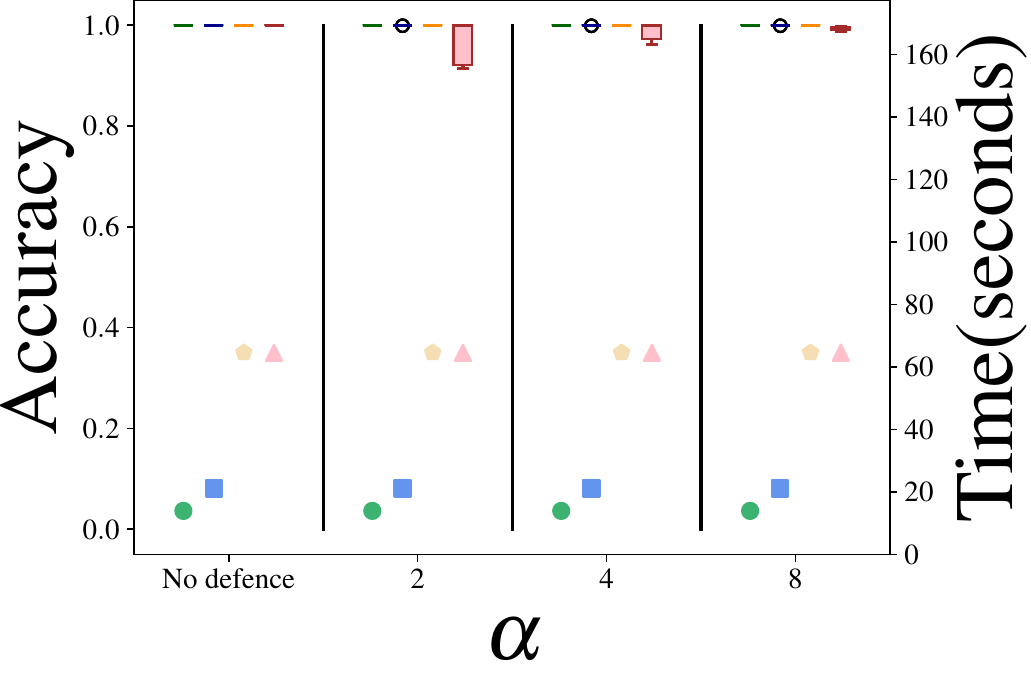}
		\end{minipage}
	}
	\subfigure[$|W|=3000$]
	{
 \label{fig:padding_cluster_wiki_3000}
		\begin{minipage}{.30\linewidth}
			\centering
			\includegraphics[width=\linewidth]{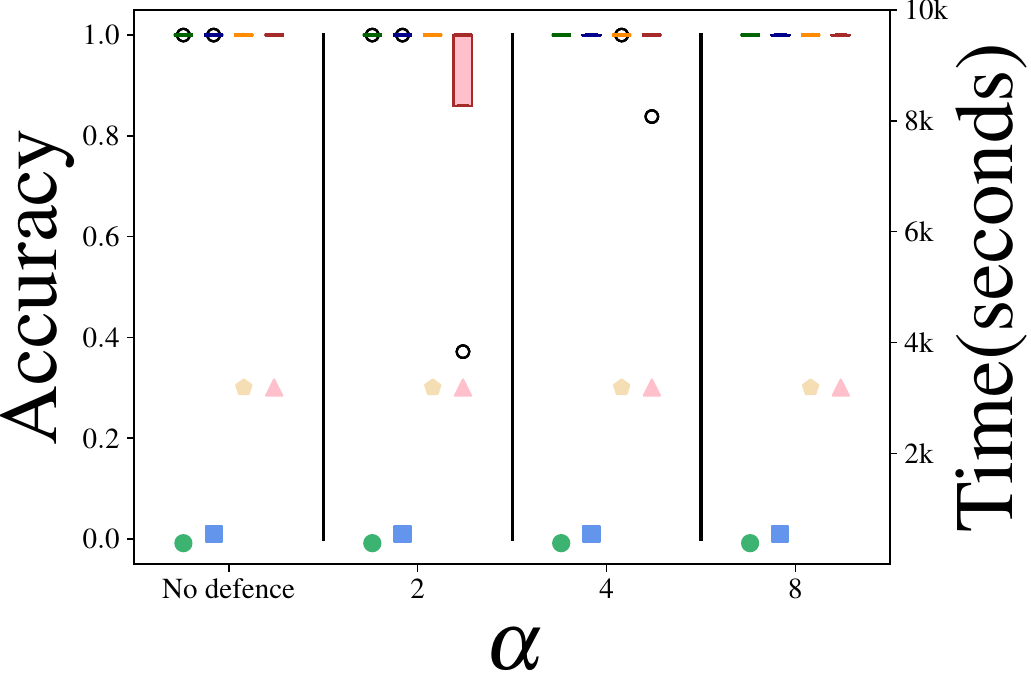}
		\end{minipage}
	}
        \subfigure[$|W|=5000$]
	{
 \label{fig:padding_cluster_wiki_5000}
		\begin{minipage}{.30\linewidth}
			\centering
			\includegraphics[width=\linewidth]{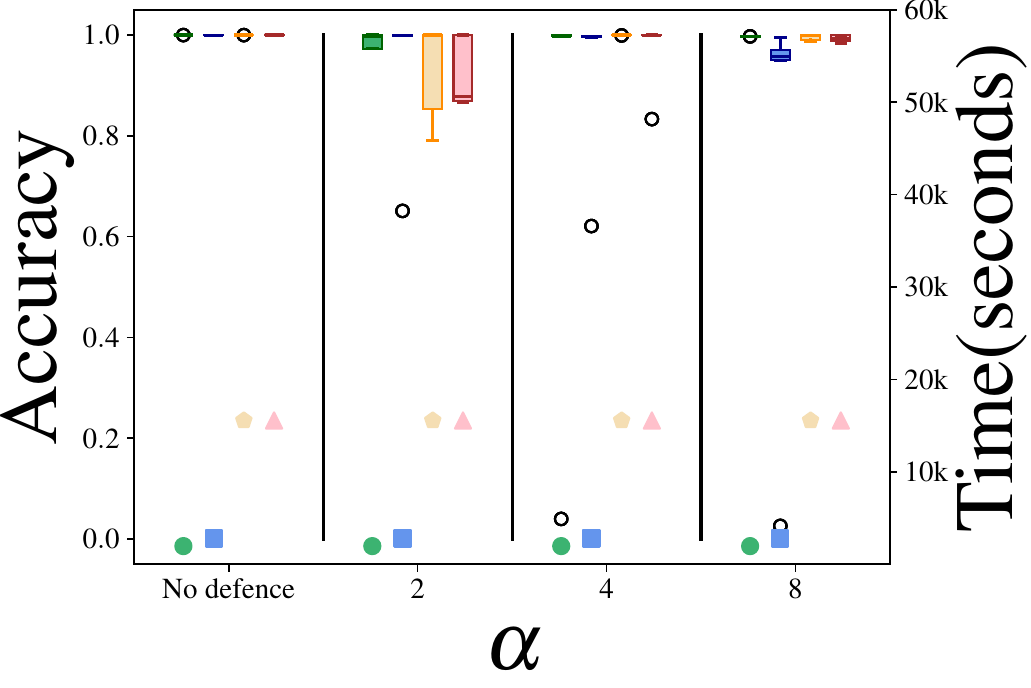}
		\end{minipage}
	}
	\caption{{Comparisons with RSA and IHOP against the cluster-based padding\cite{DBLP:journals/corr/shielddb,DBLP:journals/iacr/BostF17}, in Wikipedia.}}
	\label{fig:against_padding_cluster_wiki}
\end{figure*}

\begin{figure*}[!t]
	\centering
	\subfigure[$|W|=1000$]
	{
 \label{fig:padding_seal_wiki_1000}
		\begin{minipage}{.30\linewidth}
			\centering
                \includegraphics[width=\linewidth]
                {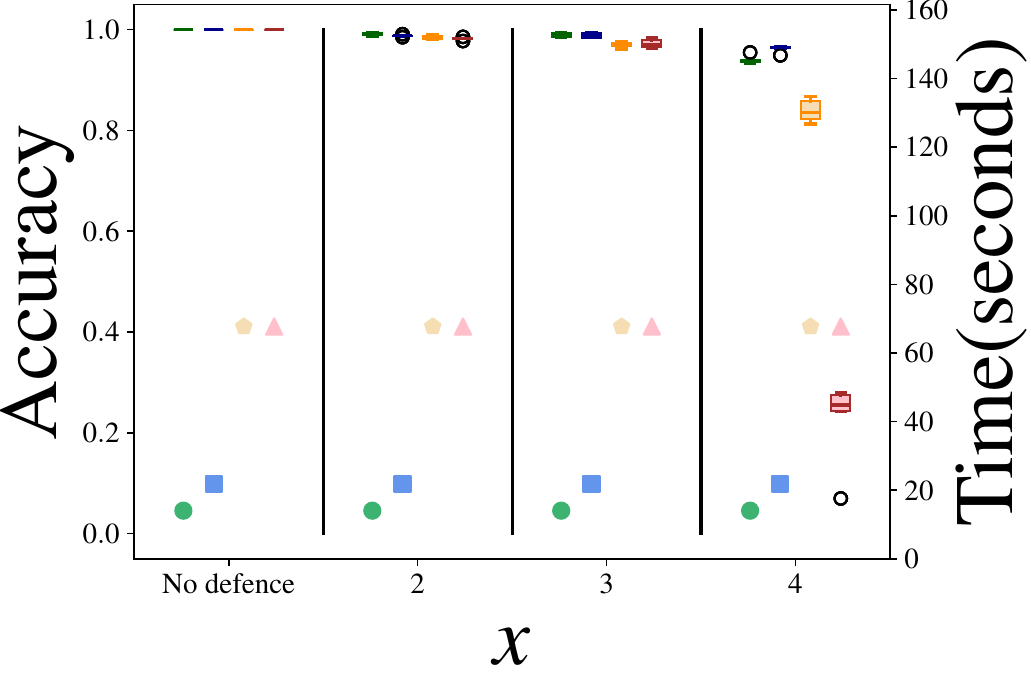}
		\end{minipage}
	}
	\subfigure[$|W|=3000$]
	{
 \label{fig:padding_seal_wiki_3000}
		\begin{minipage}{.30\linewidth}
			\centering
			\includegraphics[width=\linewidth]{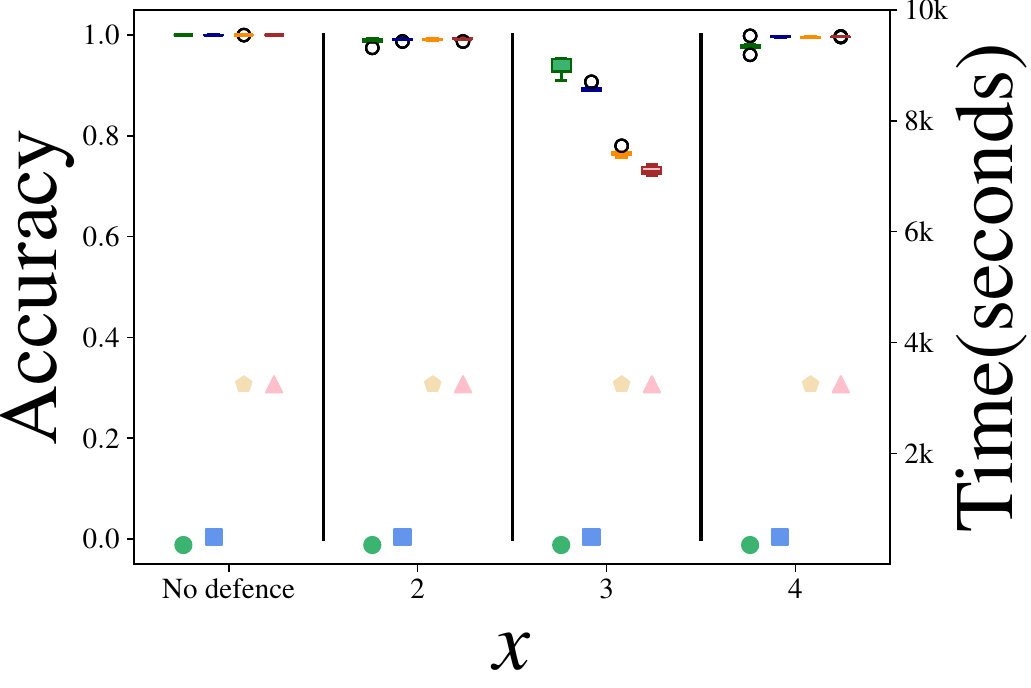}
		\end{minipage}
	}
        \subfigure[$|W|=5000$]
	{
 \label{fig:padding_seal_wiki_5000}
		\begin{minipage}{.30\linewidth}
			\centering
			\includegraphics[width=\linewidth]{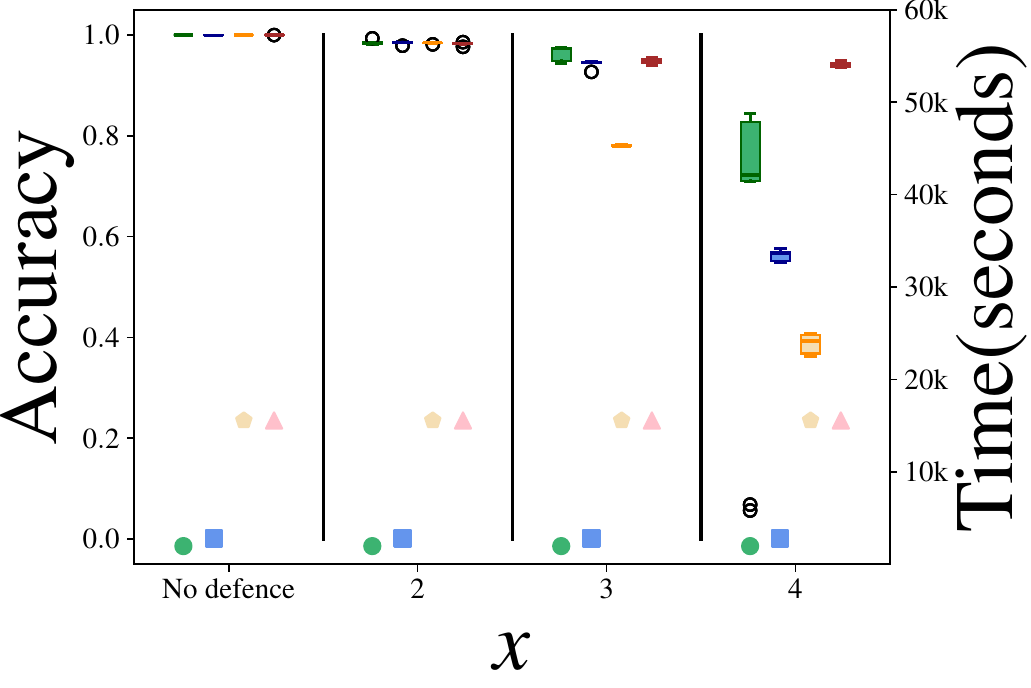}
		\end{minipage}
	}
	\caption{{Comparisons with RSA and IHOP against the padding in SEAL\cite{DBLP:conf/uss/DemertzisPPS20}, in Wikipedia.}}
	\label{fig:against_padding_seal_wiki}
\end{figure*}

\begin{figure}[!h]
	\centering
	\subfigure[Enron]
	{
 \label{fig:padding_seal_enron}
		\begin{minipage}{.45\linewidth}
			\centering
                \includegraphics[width=\linewidth]
                {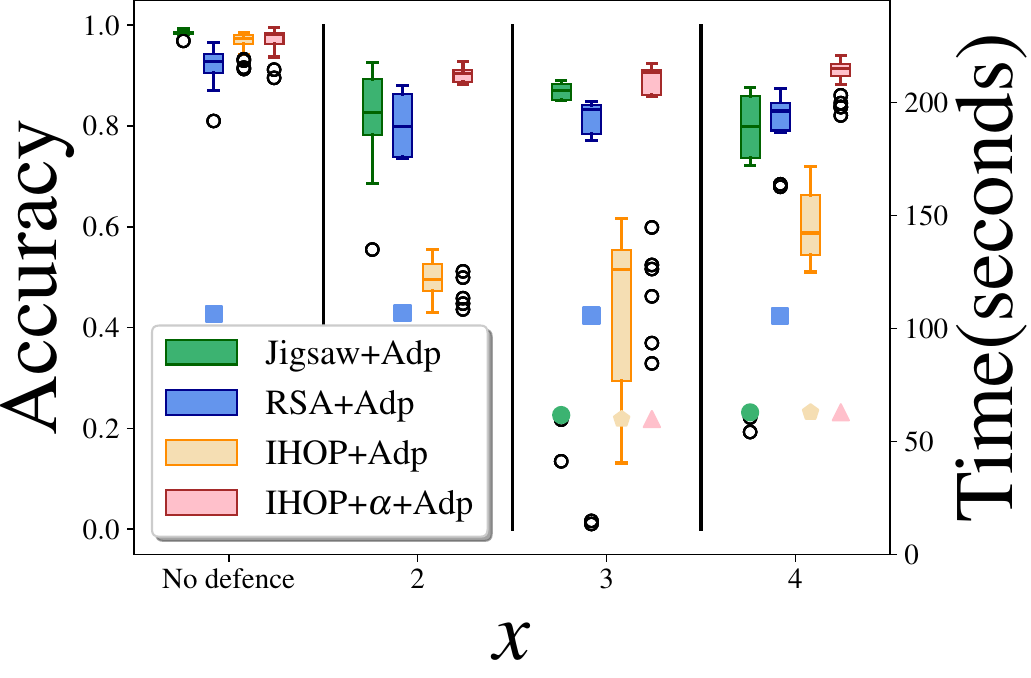}
		\end{minipage}
	}
	\subfigure[Lucene]
	{
 \label{fig:padding_seal_lucene}
		\begin{minipage}{.45\linewidth}
			\centering
			\includegraphics[width=\linewidth]{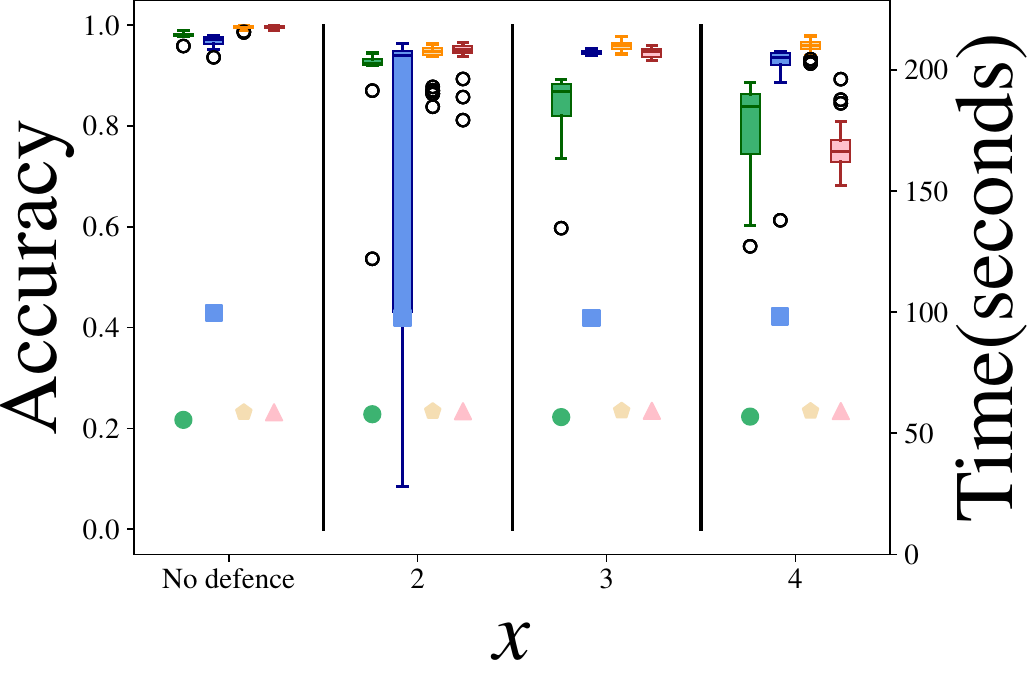}
		\end{minipage}
	}
	\caption{{Comparisons with RSA and IHOP against the padding in SEAL\cite{DBLP:conf/uss/DemertzisPPS20}, in Enron and Lucene.}}
	\label{fig:against_padding_seal_enron_lucene}
\end{figure}

\section{Results Against Other Padding Strategies}

\label{appendix:Results of Against Other Padding Strategy}

\textbf{Against the cluster-based padding}. Recall that the cluster-based padding\cite{DBLP:journals/corr/shielddb,DBLP:journals/iacr/BostF17} first divides all the keywords into clusters with each containing no less than $\alpha$ keywords. Then, this countermeasure pads each keyword to the largest volume in its cluster. 
To improve the padding efficiency, we sort all keywords in ascending order based on their volume and assign each continuous $\alpha$ keywords to the same cluster.
{We set $\alpha$ to 2, 4, and 8 and present the results for Enron and Lucene in Figure \ref{fig:against_padding_cluster_enron_lucene}, with evaluations under Wikipedia shown in Figure \ref{fig:against_padding_cluster_wiki}.
We set the $\alpha$ in Jigsaw to $0.1$ (considering the cluster-based padding that injects more noise to high-volume queries) and keep other parameters the same as in Section \ref{sec:Against Countermeasures}.}

{
We observe that the average accuracy of all the attacks exceeds $75\%$. 
But Enron consists of fewer distinctive queries than other datasets, leading to exceptional cases in the performance.  
As keywords with similar volumes are assigned to the same cluster, they become indistinguishable in terms of volume, which negatively impacts the accuracy of Jigsaw's first module. 
Beyond that, the cluster-based padding additionally brings instability in attacks' performance (see outliers in the figure). 
However, the average accuracy still remains practical.
}

\noindent\textbf{Against the Padding in SEAL}. We present the results against the padding in SEAL\cite{DBLP:conf/uss/DemertzisPPS20}, which pads the volume of keywords to the nearest power of an integer $x$. 
We note again that our attack only targets the padding strategy in SEAL.  
We set $x$ to 2, 3, and 4 for the padding and demonstrate the results in Figure \ref{fig:against_padding_seal_enron_lucene} (for Enron and Lucene) and Figure \ref{fig:against_padding_seal_wiki} (for Wikipedia), {and other parameters remain consistent with those previously configured.}   %
{
As $x$ grows, the average accuracy of Jigsaw maintains above $70\%$, although there are a few outliers in the results. 
The padding also affects the performance of RSA and IHOP, causing an unstable and dropping trend.

In summary, though the results contain outliers when against the padding in SEAL and the cluster-based padding, the tested attacks provide fine accuracy in most cases under our settings. 
}

\input{table_overhead}
{

\section{Overheads of Padding and Obfuscation}
\label{Appendix:overhead}
We demonstrate storage and communication overheads brought by the countermeasures in Table \ref{tab:table_overhead}.
The padding in CGPR\cite{cash2015leakage} significantly incurs the increase of communication costs, about 3x increase with $k=1500$ in Enron, and above 4x increase in Wikipedia with $k=150000$. 
The obfuscation in CLRZ\cite{DBLP:conf/infocom/ChenLRZ18} exhibits similar effects on communication overhead but remains a storage overhead of 1 because it does not infuse extra padded documents. 
The cluster-based padding\cite{DBLP:journals/corr/shielddb,DBLP:journals/iacr/BostF17} proves to be efficient in communication, albeit at the cost of a 1.5x increase in storage overhead with $\alpha=8$ in Wikipedia. 
The SEAL's padding \cite{DBLP:conf/uss/DemertzisPPS20} increases storage by a factor of $x$, as it pads the dataset to a total of $x\cdot|D|$ documents. 
It also yields approximately 2x increase in communication costs with $x=4$.

}

\end{appendices}

\end{document}

%% file: table_related_work.tex
\begin{table*}[!t]
\centering
    \begin{threeparttable} 
	    \caption{Comparisons of existing passive attacks\tnote{1}. 
     }
		\label{tab:tab_passive_attacks}
		\centering
        \setlength{\tabcolsep}{2mm}{\begin{tabular}{lcccccccc}
			\hline
			\multirow{2}{*}{Attack} & \multirow{2}{*}{Leakage} & \multicolumn{2}{c}{Known prior knowledge} & \multicolumn{2}{c}{Similar prior knowledge} &  \multirow{2}{*}{Accuracy} & \multirow{2}{*}{Padding}\tnote{2}& \multirow{2}{*}{Obfuscation}\tnote{3}\\ 
   
			\cline{3-6}
				& & Document & Query & Document & Frequency & &\\
			\hline
			IKK \cite{islam2012access} & ap & \fullcirc & \halfcirc & \emptycirc  & \emptycirc& $\sim 80\%$ & - &-\\
			Count \cite{cash2015leakage} & ap, vp & \fullcirc & \emptycirc & \emptycirc&\emptycirc&$\sim 90\%$  &- &-\\
   
			SubgraphID \cite{blackstone2020revisiting} & ap & \halfcirc & \emptycirc & \emptycirc & \emptycirc & $\sim 90\%$ &- &-\\
			LEAP \cite{ning2021leap} & ap & \halfcirc & \emptycirc & \emptycirc & \emptycirc &$\sim 100\%$ &- & -\\
   
   \hline
   \hline
            RSA \cite{damie2021highly} & ap & \emptycirc & \halfcirc & \fullcirc &\emptycirc &$\sim 85\%$ & $<20\%$&$<20\%$\\
   \hline
   \hline
               Freq \cite{liu2014search} & sp & \emptycirc & \emptycirc & \emptycirc & \fullcirc&$\sim 20\%$ & $\sim 20\%$&$\sim 20\%$\\
               SAP \cite{oya2021hiding} & vp,sp & \emptycirc & \emptycirc & \fullcirc & \fullcirc&$\sim 50\%$ &$\sim 30\%$ &$\sim 30\%$\\
               GraphM \cite{pouliot2016shadow} & ap & \emptycirc & \emptycirc & \fullcirc & \emptycirc&$\sim 70\%$ &$< 20\%$ & $< 20\%$\\
               IHOP \cite{oya2021ihop} & sp,ap & \emptycirc & \emptycirc & \fullcirc & \fullcirc&$\sim 90\%$ & $< 20\%$&$\sim 85\%$\\

               Jigsaw (Ours)\tnote{4}  & vp,sp,ap & \emptycirc & \emptycirc & \fullcirc& \halfcirc &$\sim 90\%$ & $\sim 60\%$& $\sim 85\%$\\

			\hline
		\end{tabular}}
            
		\begin{tablenotes}    
            \footnotesize               
            \item[1]``ap'' denotes the access pattern, ``vp'' denotes the volume pattern, and ``sp'' denotes the search pattern. The ``\fullcirc'' indicates that the attack needs nearly all known data or strongly relies on the corresponding similar data. The ``\halfcirc'' indicates that the attack needs partial known data or not particularly relies on similar data. The ``\emptycirc'' means that the attack does not need any known data or similar data. 
            The first four attacks are known-data attacks, and the last five are similar-data attacks. The RSA mainly relies on similar data but needs a few known queries to start the attack. We do not present the performance of the first four attacks against padding and obfuscation (denoting ``-'') since we mainly focus on similar-data attacks.
            \item[2]The performance against the padding of CGPR\cite{cash2015leakage} with $k=1,000$ on Enron. 
            {We utilize an adaptation (Appendix \ref{appendix:Modifications of the Similar Data}) for Jigsaw.} Padding is claimed to mitigate the listed known-data attacks effectively \cite{blackstone2020revisiting,ning2021leap}.  
            \item[3] The performance against the obfuscation of CLRZ\cite{DBLP:conf/infocom/ChenLRZ18} with TPR$=0.999$,  FPR$=0.05$ on Enron. {Note adaptations (Appendix \ref{appendix:Modifications of the Similar Data}) are used for IHOP and Jigsaw.}
            \item[4] Our attack can reach 90\% accuracy even if a defense hides the sp, and thus one may consider the sp an optional attack advantage. 
            
        \end{tablenotes}  
    \end{threeparttable} 
\end{table*}

%% file: alg1.tex
\begin{algorithm}[!t]
    \footnotesize
    \caption{Recover the top-$BaseRec$ distinctive queries.}
    \label{alg1}
    \SetKwFunction{FMain}{RecoverDQ}
    \SetKwProg{Bn}{procedure}{}{end}
    \Bn{\FMain{$Td_{r},V_{r},F_{r},W_{s},V_{s},F_{s},\alpha,BaseRec$}}{
        $Dis\gets\emptyset$; \Comment{$Dis$ maintains the differential distances of queries}\;
        \For {all $td_i\in Td_{r}$}{ \label{A1Caldiss}
            $d_{td_i}=\min\limits_{td_j\in Td_r\land j\ne i}\alpha\cdot|v_{td_i}-v_{td_j}|+(1-\alpha)|f_{td_i}-f_{td_j}|$\;
            append $(td_i,d_{td_i})$ to $Dis$\;
        } \label{A1Caldise}
        Sort $Dis$ in descending order according to $Dis.d_{td}$\;\label{A1sortdis}
        $Pred\gets\emptyset$; \Comment{$Pred$ stores the recoveries of distinctive queries}\;
        \For{$i\in [BaseRec]$}{\label{A1recs}
            $td_i,d_{td_i}=Dis[i]$\;
            $w=\arg\min\limits_{w_j\in W_s}\alpha\cdot|v_{td_i}-v_{w_j}|+(1-\alpha)|f_{td_i}-f_{w_j}|$\;
            append $(td_i,w)$ to $Pred$\;
        }\label{A1rece}
        \KwRet $Pred$\;
    }
    
    
    
\end{algorithm}

%% file: alg2.tex
\begin{algorithm}[!t]
    \footnotesize
    \caption{Verification by co-occurrence matrices.}
    \label{alg2}
    \SetKwFunction{FMain}{Verify}
    \SetKwProg{Bn}{procedure}{}{end}
    \Bn{\FMain{$Pred,C'_{r},C'_{s},BaseRec,ConfRec$}}{
        $Temp\_Pred \gets Pred$\;
        $Td'\gets Pred.td$\;
        $Revconf \gets\emptyset$\;\label{A2calconfs}
        \For {$i \in [|Td'|]$}{
            $revconf=||C'_{r}[i]-C'_{s}[i]||$\;
            append $(Td'[i],revconf)$ to $Revconf$\;
        }\label{A2calconfe}
        Sort $Revconf$ in descending order according to the $Revconf.revconf$\;\label{A2sort}
        \For{$i\in [BaseRec-ConfRec]$}{\label{A2res}
            $td,revconf=Revconf[i]$\;
            Remove the prediction of $td$ from $Temp\_Pred$\;
        }
        \KwRet $Temp\_Pred$\;\label{A2ree}
    }
\end{algorithm}

%% file: alg3.tex
\begin{algorithm}[!t]
    \footnotesize
    \caption{Dynamically recovering all queries.}
    \label{alg3}
    \SetKwFunction{FMain}{RecoverAll}
    \SetKwProg{Bn}{procedure}{}{end}
    \Bn{\FMain{$Pred,Td_{r},W_{s},C_{r},C_{s},RefSpeed$}}{
        $Final\_Pred\gets Pred$\;
        $unknownTd\gets Td_r - Final\_Pred.td$\;
        $unpairedW\gets W_s - Final\_Pred.w$\;
        Extract $C^s_{r}$ and $C^s_{s}$ from $C_{r}$ and $C_{s}$, respectively; \Comment{
        $C_r^s$ and $C_s^s$ is the co-occurrence matrix between $unknownTd$ and recovered queries, and between $unpairedW$ and paired keywords, respectively
        }\;
        \While{$unknownTd\neq\emptyset$}{
            $Temp\_Pred\gets\emptyset$\;
            \For{all $td\in unknownTd$}{ \label{A3calcers}
                $Cand\gets\emptyset$; \Comment{ $Cand$ stores candidate matches for $td$}\;
                \For{all $w\in unpairedW$}{
                    $score =-\ln( \beta||C^s_{r}[td]-C^s_{s}[w]||+(1-\beta)s(td,w))$\;
                    Add $(w,score)$ to $Cand$\; 
                }
                Sort $Cand$ in descending order according to the $Cand.score$\;
                $certainty = Cand[0].score-Cand[1].score$\;
                Add $(td,Cand[0].w,certainty)$ to $Temp\_Pred$\;
            }\label{A3calcere}
            \eIf {$|unknownTd|<RefSpeed$}{\label{A3addpres}
                Add all predictions in $Temp\_Pred$ to $Final\_Pred$\;
            }
            {
                Add the $RefSpeed$ predictions in $Temp\_Pred$ with largest $certainty$ to $Final\_Pred$\;
            }\label{A3addpree}
            Update $unknownTd$, $unpairedW$, $C^s_{r}$ and $C^s_{s}$\;\label{A3update}
        }
        \KwRet $Final\_Pred$\;
    }
\end{algorithm}

%% file: figures_in_experiments_distribution_enron_only.tex
\begin{figure*}[!ht]
	\centering
	
	\subfigure[HVHF]
	{
 \label{HVHF_3D_Enron}
		\begin{minipage}{.17\linewidth}
			\centering
			\includegraphics[width=\linewidth]{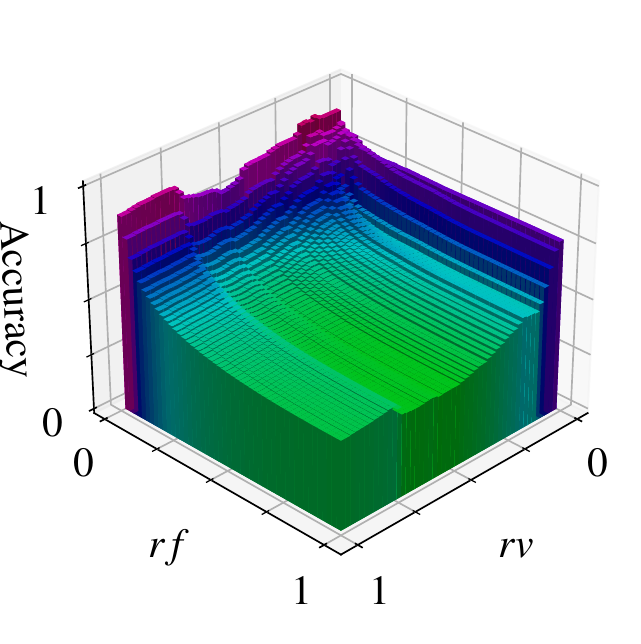}
		\end{minipage}
	}
	\subfigure[HVLF]
	{
 \label{HVLF_3D_Enron}
		\begin{minipage}{.17\linewidth}
			\centering
			\includegraphics[width=\linewidth]{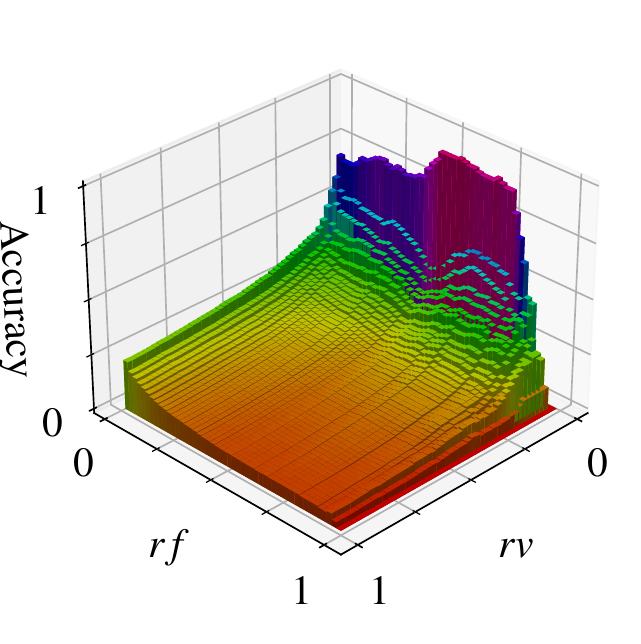}
		\end{minipage}
	}
        \subfigure[LVHF]
	{
 \label{LVHF_3D_Enron}
		\begin{minipage}{.17\linewidth}
			\centering
			\includegraphics[width=\linewidth]{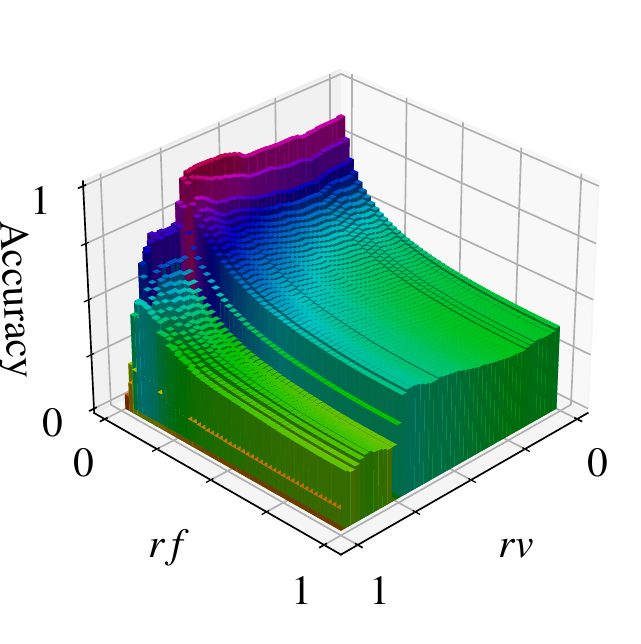}
		\end{minipage}
	}
        \subfigure[LVLF]
	{
 \label{LVLF_3D_Enron}
		\begin{minipage}{.17\linewidth}
			\centering
			\includegraphics[width=\linewidth]{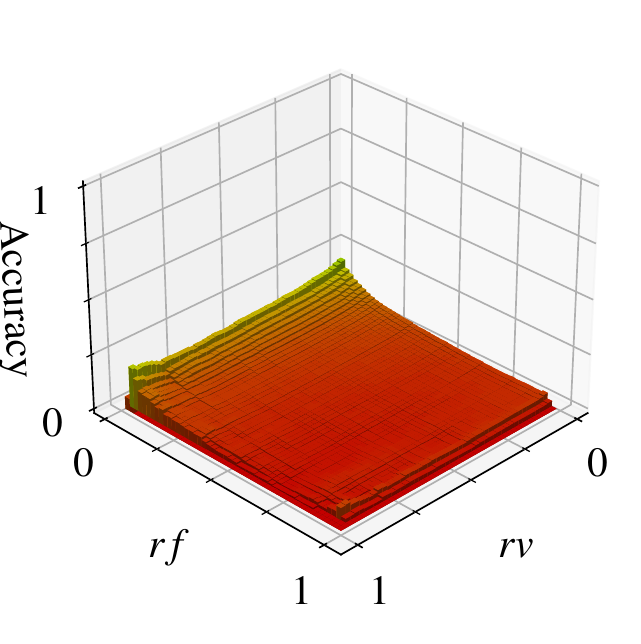}
		\end{minipage}
	}

	\caption{The accuracy of Algorithm \ref{alg1} in four quadrants with different $rv$ and $rf$, where we treat keywords with top-$rv\cdot l$ highest volume as high-volume keywords and treat keywords with top-$rf\cdot l$ highest frequency as high-frequency keywords (A larger $rv$ means more queries are considered as high-volume queries. Similarly, a larger $rf$ yields more queries that are categorized as high-frequency queries). }
	\label{fig:quadrants_enron_only}
\end{figure*}

	


\begin{figure*}[!t]
	\centering
	
	\subfigure[HVHF]
	{
 \label{HVHF_enron}
		\begin{minipage}{.17\linewidth}
			\centering
			\includegraphics[width=\linewidth]{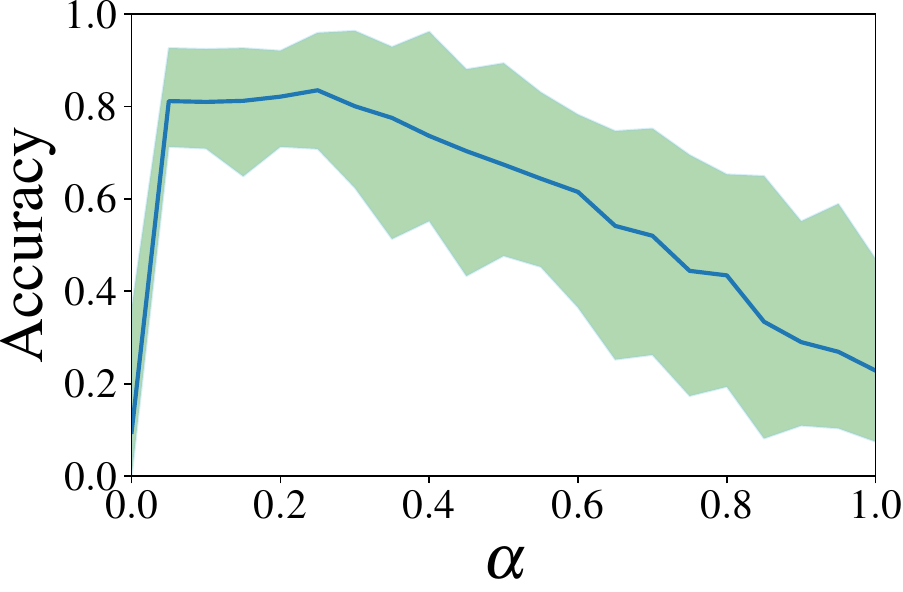}
		\end{minipage}
	}
	\subfigure[HVLF]
	{
 \label{HVLF_enron}
		\begin{minipage}{.17\linewidth}
			\centering
			\includegraphics[width=\linewidth]{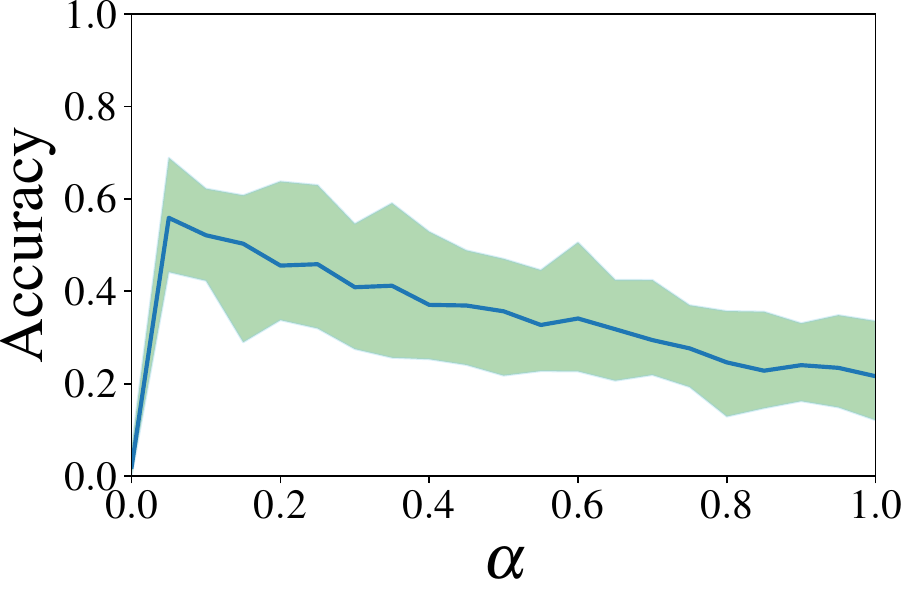}
		\end{minipage}
	}
        \subfigure[LVHF]
	{
 \label{LVHF_enron}
		\begin{minipage}{.17\linewidth}
			\centering
			\includegraphics[width=\linewidth]{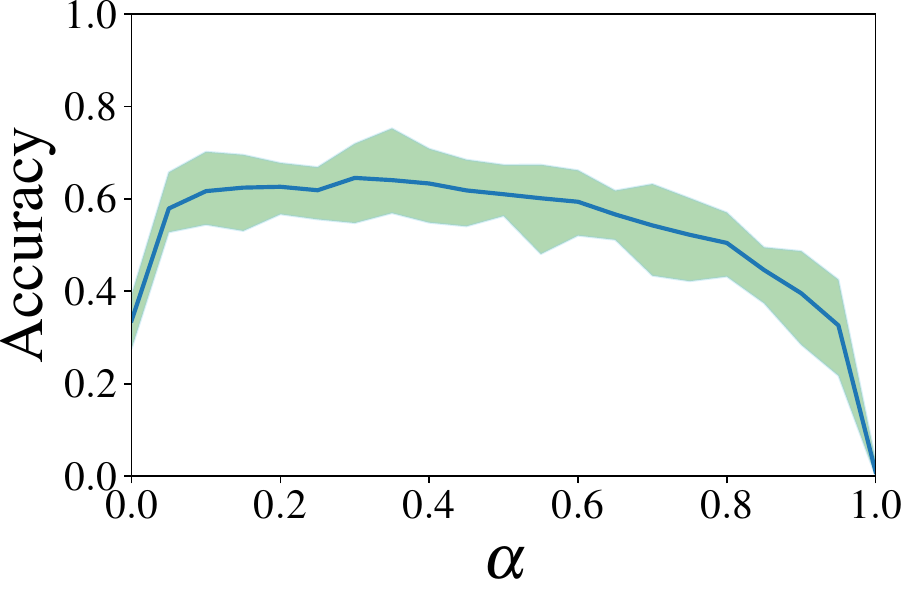}
		\end{minipage}
	}
        \subfigure[LVLF]
	{
 \label{LVLF_enron}
		\begin{minipage}{.17\linewidth}
			\centering
			\includegraphics[width=\linewidth]{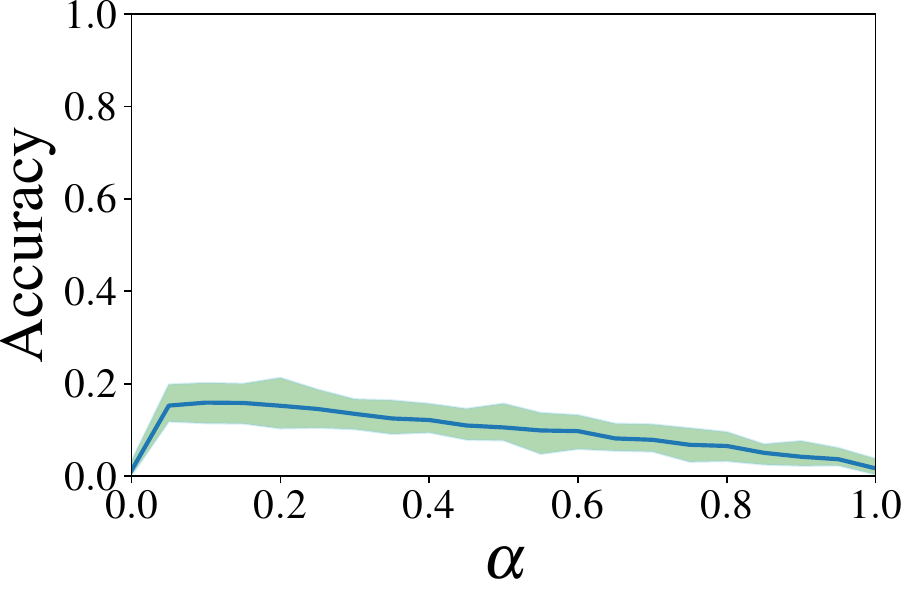}
		\end{minipage}
	}

	\caption{The accuracy of Algorithm \ref{alg1} on four quadrants with different $\alpha$, where $\alpha$ is the weight of volume and $(1-\alpha)$ is the weight of frequency in measurement.}
 
	\label{fig:quadrants_test_alpha_enron_only}
\end{figure*}

%% file: table_Rec_Ver_Num.tex
\begin{table*}[!h]
    \centering
    \begin{threeparttable} 
	    \caption{Results of Algorithm \ref{alg1} and Algorithm \ref{alg2} on Enron, Lucene, and Wikipedia.
     The $ConfRec$ in Algorithm 2 equals to $BaseRec\times 100\%$, $BaseRec\times 50\%$, and $BaseRec\times 20\%$ respectively.}
		\label{tab:tab_Rec_Ver_Num}
		\centering
        \setlength{\tabcolsep}{4mm}{\begin{tabular}{lcccc}
			\hline
			\multirow{2}{*}{Dataset} &\multirow{2}{*}{$BaseRec$} &\multicolumn{3}{c}{$ConfRec/BaseRec\times 100\%$ (accuracy/recovery rate/correctly recovered number)\tnote{1}}\\
            \cline{3-5}
            
            & &$100\%$\tnote{2}& $50\%$ & $20\%$\\
			\hline
            \hline
            \multirow{3}{*}{Enron} 
& 25  &$ 94.14 \%/ 23.25 \%/ 23.0$&$ 100.00 \%/ 3.46 \%/ 12.0$&$ 100.00 \%/ 0.98 \%/ 5.0$\\
& 100  &$ 77.33 \%/ 43.14 \%/ 57.6$&$ 91.53 \%/ 17.18 \%/ 39.5$&$ 96.95 \%/ 5.48 \%/ 19.0$\\
& 400  &$ 51.85 \%/ 80.10 \%/ 112.2$&$ 72.06 \%/ 46.41 \%/ 81.1$&$ 83.96 \%/ 17.37 \%/ 50.2$\\

		\hline
            \multirow{3}{*}{Lucene} 
& 25  &$ 99.36 \%/ 27.51 \%/ 24.6$&$ 99.98 \%/ 4.17 \%/ 12.0$&$ 100.00 \%/ 1.93 \%/ 5.0$\\
& 100  &$ 86.31 \%/ 48.64 \%/ 76.3$&$ 99.58 \%/ 25.55 \%/ 48.9$&$ 99.80 \%/ 5.85 \%/ 19.9$\\
& 400  &$ 63.07 \%/ 83.97 \%/ 147.7$&$ 82.85 \%/ 50.04 \%/ 103.6$&$ 96.55 \%/ 27.64 \%/ 67.1$\\

		\hline
            \multirow{3}{*}{Wikipedia} 
& 25  &$ 99.71 \%/ 11.89 \%/ 23.8$&$ 100.00 \%/ 2.58 \%/ 12.0$&$ 100.00 \%/ 0.83 \%/ 5.0$\\
& 100  &$ 91.57 \%/ 28.67 \%/ 80.9$&$ 99.61 \%/ 15.04 \%/ 47.7$&$ 100.00 \%/ 3.59 \%/ 19.9$\\
& 400  &$ 68.79 \%/ 58.17 \%/ 177.3$&$ 88.87 \%/ 34.47 \%/ 120.2$&$ 97.54 \%/ 15.94 \%/ 64.3$\\

		\hline
  
  \end{tabular}}

		\begin{tablenotes}    
            \footnotesize               
            \item[1]Each result is presented as (accuracy/recovery rate/correctly recovered number). The accuracy denotes the percentage of correctly recovered queries out of recovered queries. The recovery rate is the percentage of recovered queries out of all queries. The correctly recovered number is the number of correctly recovered and distinct queries.
            
            \item[2]This column shows the results of Algorithm \ref{alg1} as Algorithm \ref{alg2} does not remove any predictions.
            
        \end{tablenotes}  
    \end{threeparttable} 
\end{table*}

%% file: figures_in_test_beta.tex
\begin{figure}[!t]
	\centering

	\subfigure[Enron; With frequency]
	{
 \label{fig:test_beta_enron}
		\begin{minipage}{.45\linewidth}
			\centering
                \includegraphics[width=\linewidth]
                {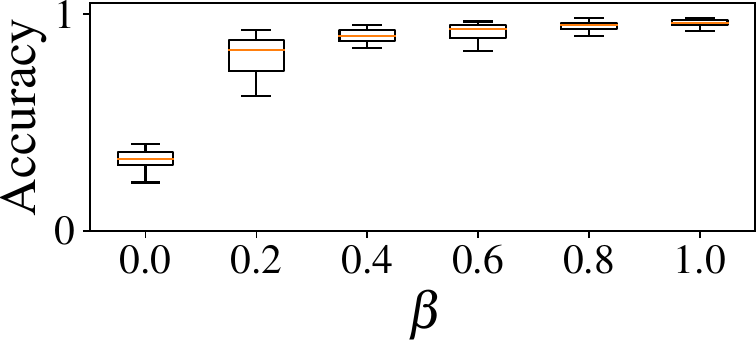}
		\end{minipage}
	}
        \subfigure[Enron; Without frequency]
	{
 \label{fig:test_beta_enron_without_f}
		\begin{minipage}{.45\linewidth}
			\centering
                \includegraphics[width=\linewidth]
                {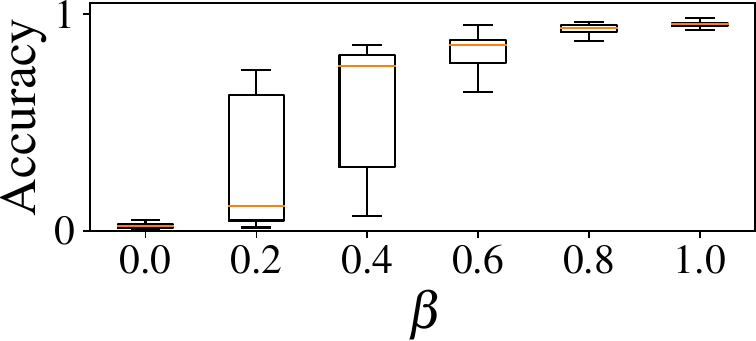}
		\end{minipage}
	}

	\subfigure[Lucene; With frequency]
	{
 \label{fig:test_beta_lucene}
		\begin{minipage}{.45\linewidth}
			\centering
			\includegraphics[width=\linewidth]{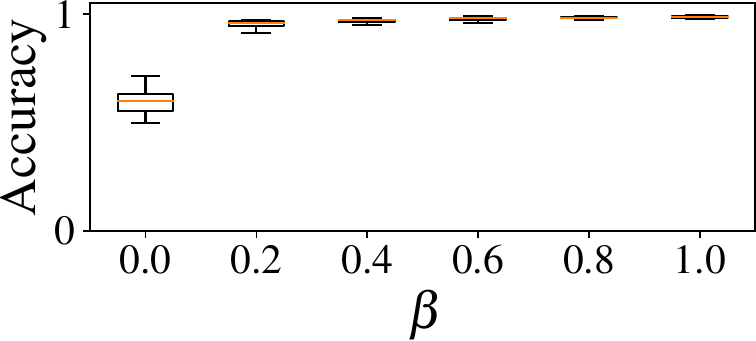}
		\end{minipage}
	}
        \subfigure[Lucene; Without frequency]
	{
 \label{fig:test_beta_lucene_enron_without_f}
		\begin{minipage}{.45\linewidth}
			\centering
			\includegraphics[width=\linewidth]{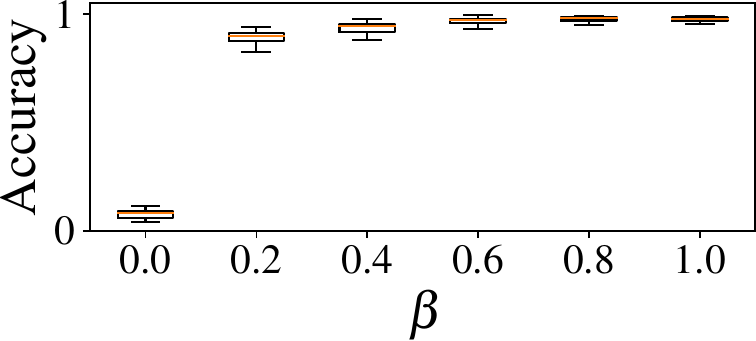}
		\end{minipage}
	}

        \subfigure[Wikipedia; With frequency]
	{
 \label{fig:test_beta_wiki}
		\begin{minipage}{.45\linewidth}
			\centering
			\includegraphics[width=\linewidth]{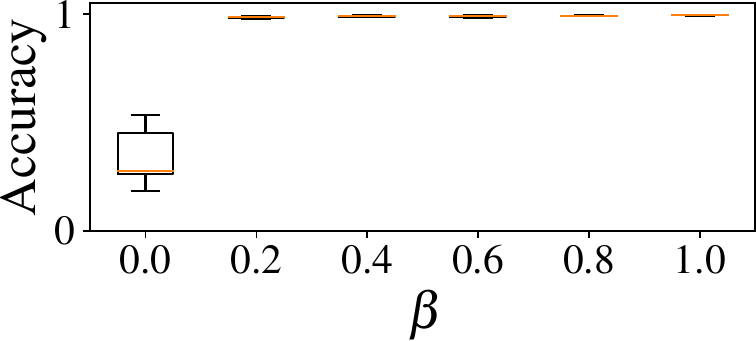}
		\end{minipage}
	}
        \subfigure[Wikipedia; Without frequency]
	{
 \label{fig:test_beta_wiki_enron_without_f}
		\begin{minipage}{.45\linewidth}
			\centering
			\includegraphics[width=\linewidth]{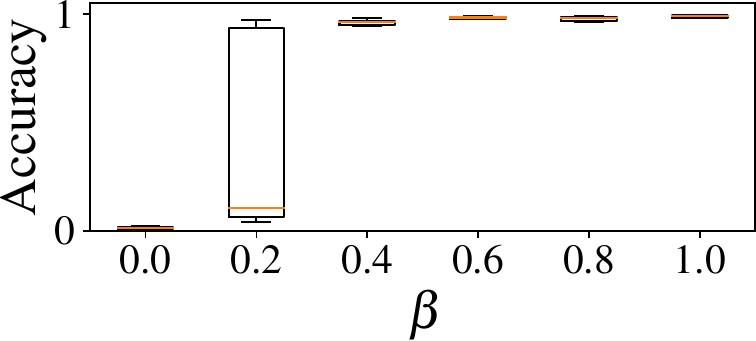}
		\end{minipage}
	}
 
	\caption{The accuracy of Jigsaw with different $\beta$, where $beta$ is the weight of co-occurrence information and $(1-\beta)$ is the weight of volume and frequency information in calculating $score$. The left and right columns display the results with and without frequency information.} 
	\label{fig:test_beta}
\end{figure}

%% file: table_durability.tex
\begin{table}[!ht]
    \centering
    \begin{threeparttable} 
	    \caption{Results of recovery accuracy with outdated frequency in different $\tau$, where $\tau$ is the time offsets between attacker's prior knowledge of the frequency and user's queries (measured in weeks for Enron and Lucene, and in months for Wikipedia). }
		\label{tab:table_durability}
		\centering
        \setlength{\tabcolsep}{1.8mm}{\begin{tabular}{lcccc}
			\hline
            Dataset & $\tau=10$w & $\tau=50$w & $\tau=100$w & $\tau=150$w\\
			\hline
            Enron & $0.9279$ & $0.9150$ & $0.8881$ & $0.8824$\\
            
            \hline
            Lucene & $0.9959$ & $0.9963$ & $0.9955$ & $0.9897$ \\
            \hline
            \hline
            Dataset & $\tau=2$m  & $\tau=10$m & $\tau=20$m & $\tau=30$m\\
			\hline
            Wikipedia & $0.9959$ & $0.9919$ & $0.9721$ & $0.9608$\\
            \hline
		\end{tabular}}
      
    \end{threeparttable} 
\end{table}

%% file: table_notations.tex
\begin{table}[t]
    \centering
    \caption{Summary of notations.}
    \label{tab:sum_of_nota}	
    \resizebox{\linewidth}{!}{
    \begin{tabular}{ll}
    \hline
    Notation& Description \\
    \hline
    \multicolumn{2}{c}{\textbf{Notations in SSE scheme and leakage}}\\
    
    \hline
    $D$ & User's document set $D=[d_1,d_2,\ldots,d_n]$.\\
    
    $Td^s$ & The list of all queries $Td^s=[td(x_1),td(x_2),\ldots,td(x_s)]$.\\
    
    $Td_r$ & The list of non-repeated queries $Td_r=[td_1,td_2,\ldots,td_l]$.\\
    
    $V_r$ & The volume of $Td_r$, $V_r=[v_{td_1},v_{td_2},\ldots,v_{td_l}]$.\\
    
    $F_r$ & The frequency of $Td_r$, $F_{r} = [ f_{td_1},f_{td_2},\ldots,f_{td_l}]$.\\
    
    $C_r$ & The co-occurrence matrix of $Td_r$.\\
    
    $D_s$ & A similar dataset used by the adversary.\\
    
    $W_s$ & Keyword universe generated from $D_s$.\\
    
    $V_s$ & The volume of $W_s$, $V_{s}=[ v_{w_1},v_{w_2},\ldots,v_{w_m}]$.\\
    
    $F_s$ & The frequency of $W_s$, $F_{s}=[ f_{w_1},f_{w_2},\ldots,f_{w_m}]$.\\
    
    $C_s$ & The co-occurrence matrix of $W_s$.\\
    \hline

    \multicolumn{2}{c}{\textbf{Notations in Jigsaw}}\\
    \hline
    $d_{td}$ &  The distance between $td$ and its nearest neighbor.\\
    
    $s(td,w)$ &  The distance between $td$ and keyword $w$.\\
    
    $revconf$ & The reversed confidence of a prediction.\\
    
    $score$ & The score of a prediction.\\
    
    $certainty$ & The largest score minus the second largest of predictions for $td$.\\
    
    $BaseRec$ & Recovered query number in our first module.\\
    
    $ConfRec$ & Recovered query number after our second module.\\
    
    $\alpha$ & The parameter controls the weight of volume and
    frequency.\\
    
    $\beta$ & The parameter in calculating $score$.\\
    
    $RefSpeed$ & Recovered query number in an iteration.\\
    \hline

    \multicolumn{2}{c}{\textbf{Notations in experiment}}\\
    \hline
     $\tau$ & The time offset between auxiliary frequency and users' query.\\
    
    $\eta$ & The observed user's query number in each time interval.\\
    \hline
    \end{tabular}
    }
\end{table}

%% file: figures_in_distribution.tex
\begin{figure}[!t]
	\centering
	\subfigure[Enron]
	{
 \label{fig:distribution_enron_2}
		\begin{minipage}{.26\linewidth}
			\centering
                \includegraphics[width=\linewidth]
                {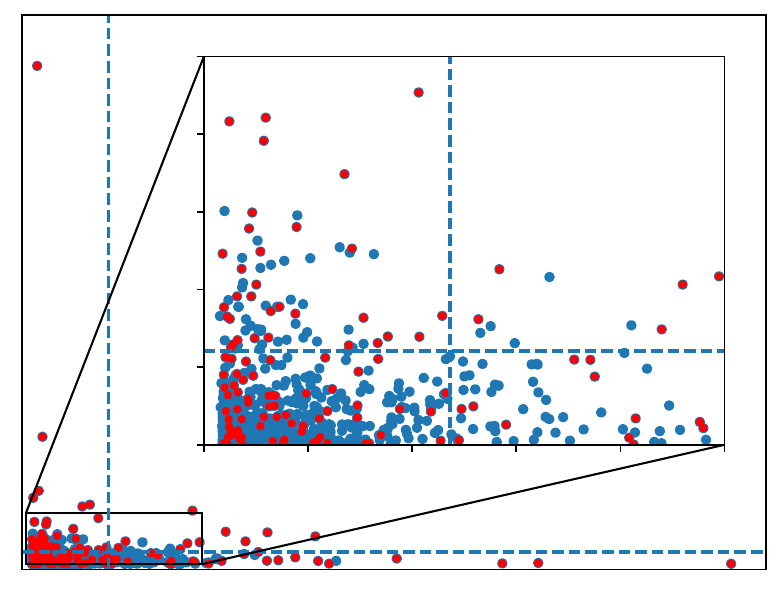}
		\end{minipage}
	}
	\subfigure[Lucene]
	{
 \label{fig:distribution_lucene_2}
		\begin{minipage}{.26\linewidth}
			\centering
			\includegraphics[width=\linewidth]{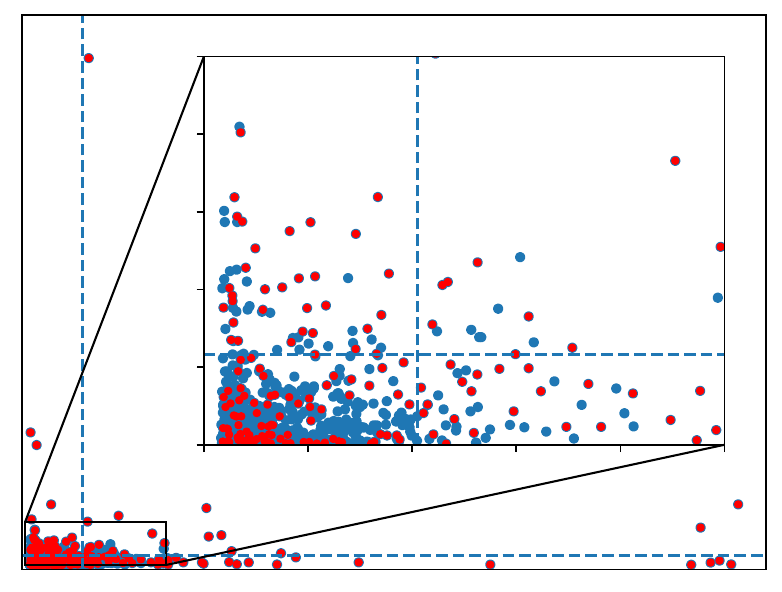}
		\end{minipage}
	}
        \subfigure[Wikipedia]
	{
 \label{fig:distribution_wiki_2}
		\begin{minipage}{.26\linewidth}
			\centering
			\includegraphics[width=\linewidth]{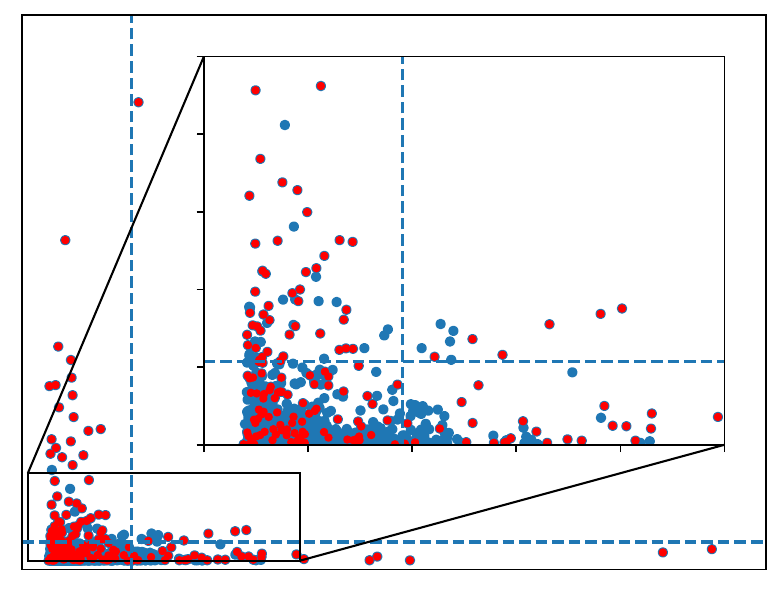}
		\end{minipage}
	}
	\caption{{The distribution of queries in normalized volume and frequency. The figures own the same format as Figure \ref{fig:distribution}.}}
	\label{fig:distribution_a}
\end{figure}

\begin{figure}[!t]
	\centering
	\subfigure[Enron]
	{
 \label{fig:distribution_enron}
		\begin{minipage}{.26\linewidth}
			\centering
                \includegraphics[width=\linewidth]
                {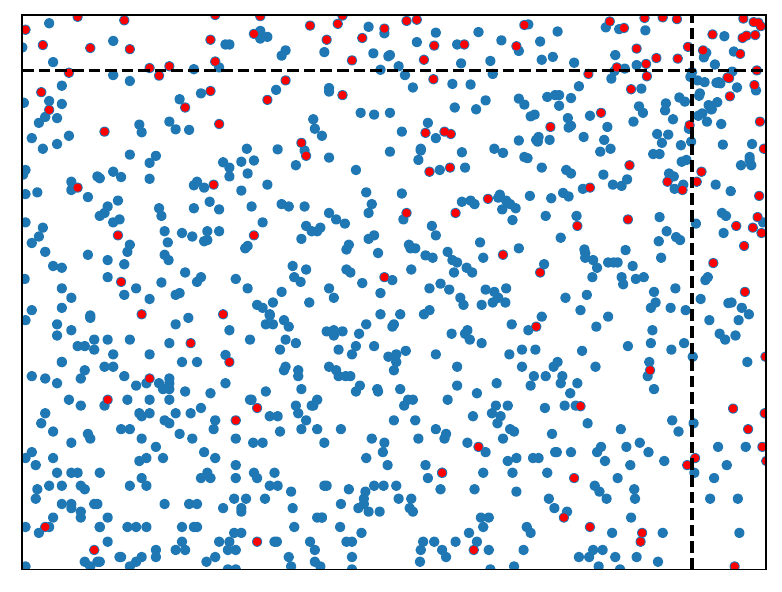}
		\end{minipage}
	}
	\subfigure[Lucene]
	{
 \label{fig:distribution_lucene}
		\begin{minipage}{.26\linewidth}
			\centering
			\includegraphics[width=\linewidth]{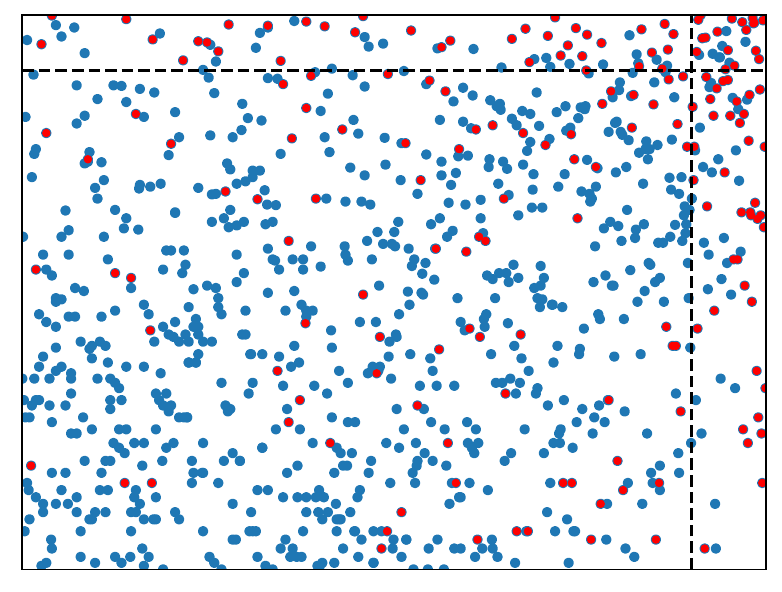}
		\end{minipage}
	}
        \subfigure[Wikipedia]
	{
 \label{fig:distribution_wiki}
		\begin{minipage}{.26\linewidth}
			\centering
			\includegraphics[width=\linewidth]{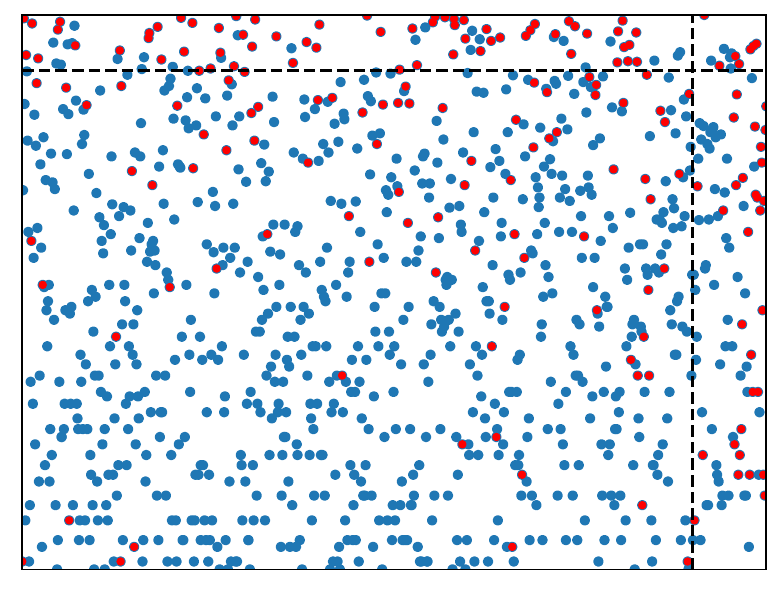}
		\end{minipage}
	}
	\caption{{The distribution of queries (showing their ranks based on volume and frequency). The dot and dashed lines share the same meaning as in Figure \ref{fig:distribution}. Those at further to the right indicate higher rankings in volume; those moving upwards represent elevated ranking in frequency.} } 
	\label{fig:distribution_b}
\end{figure}

%% file: figures_in_experiments_distribution_lucene_and_wiki.tex
\begin{figure}[!t]
	\centering

	\subfigure[HVHF]
	{
 \label{HVHF_3D_Lucene}
		\begin{minipage}{.20\linewidth}
			\centering
			\includegraphics[width=\linewidth]{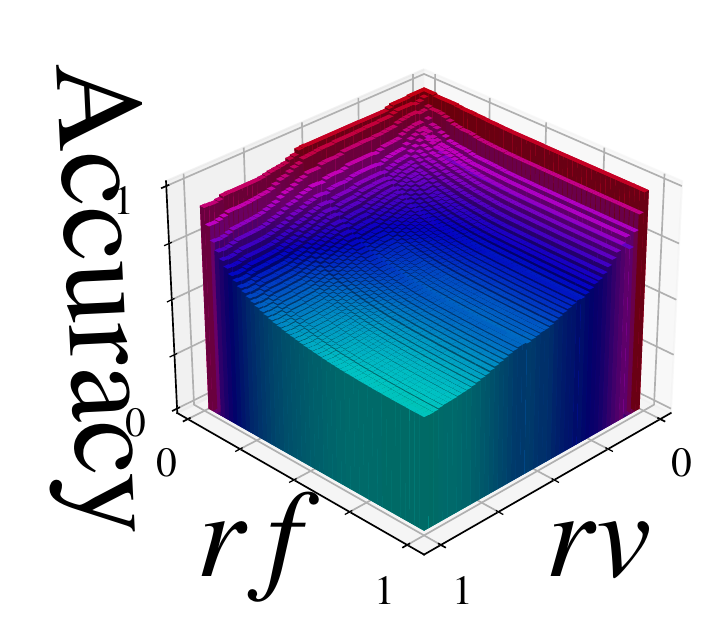}
		\end{minipage}
	}
	\subfigure[HVLF]
	{
 \label{HVLF_3D_Lucene}
		\begin{minipage}{.20\linewidth}
			\centering
			\includegraphics[width=\linewidth]{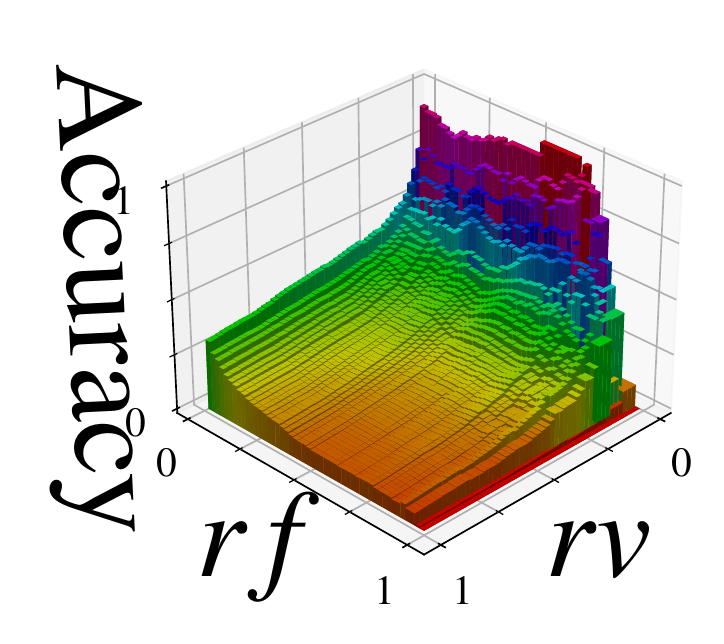}
		\end{minipage}
	}
        \subfigure[LVHF]
	{
 \label{LVHF_3D_Lucene}
		\begin{minipage}{.20\linewidth}
			\centering
			\includegraphics[width=\linewidth]{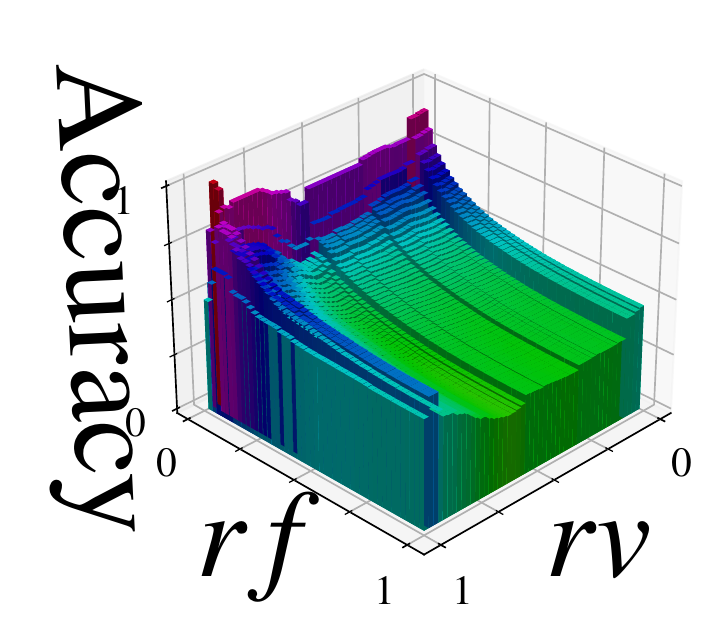}
		\end{minipage}
	}
        \subfigure[LVLF]
	{
 \label{LVLF_3D_Lucene}
		\begin{minipage}{.20\linewidth}
			\centering
			\includegraphics[width=\linewidth]{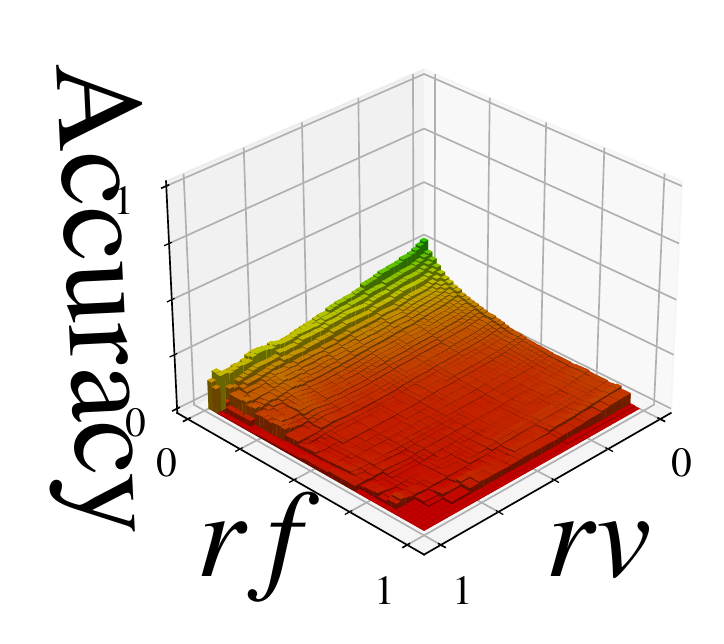}
		\end{minipage}
	}

	\subfigure[HVHF]
	{
 \label{HVHF_3D_Wiki}
		\begin{minipage}{.20\linewidth}
			\centering
			\includegraphics[width=\linewidth]{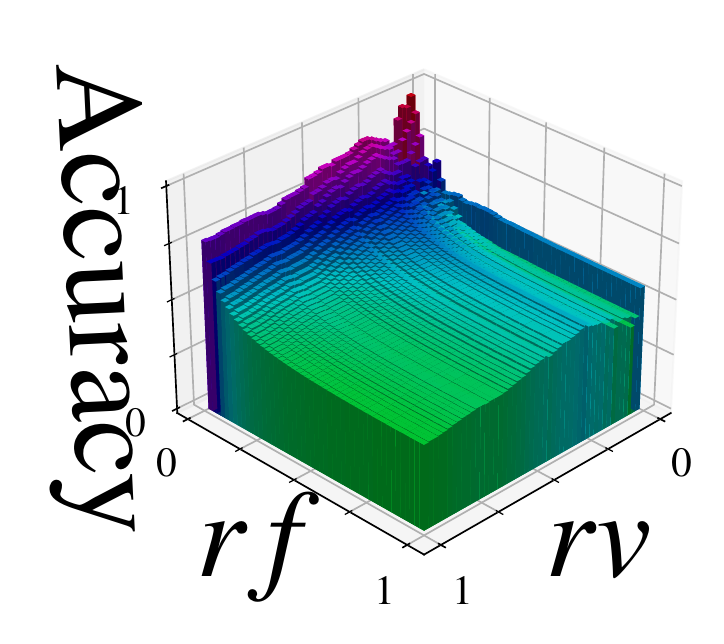}
		\end{minipage}
	}
	\subfigure[HVLF]
	{
 \label{HVLF_3D_Wiki}
		\begin{minipage}{.20\linewidth}
			\centering
			\includegraphics[width=\linewidth]{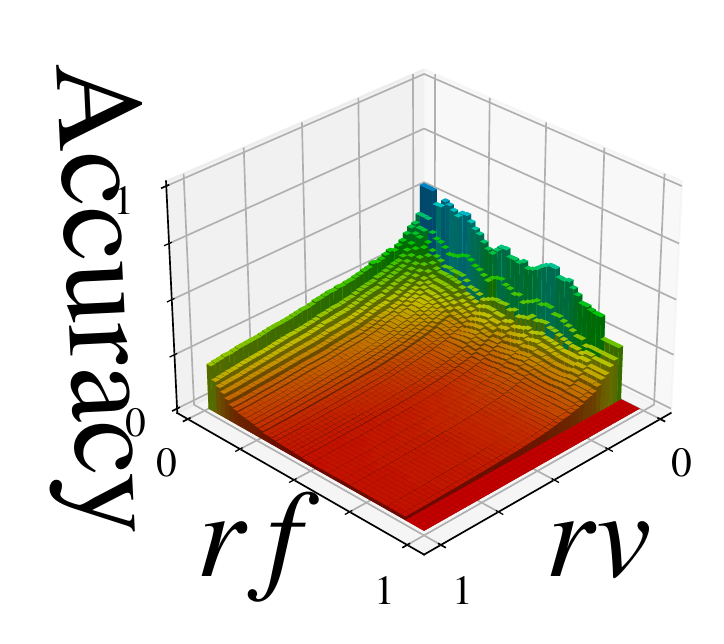}
		\end{minipage}
	}
        \subfigure[LVHF]
	{
 \label{LVHF_3D_Wiki}
		\begin{minipage}{.20\linewidth}
			\centering
			\includegraphics[width=\linewidth]{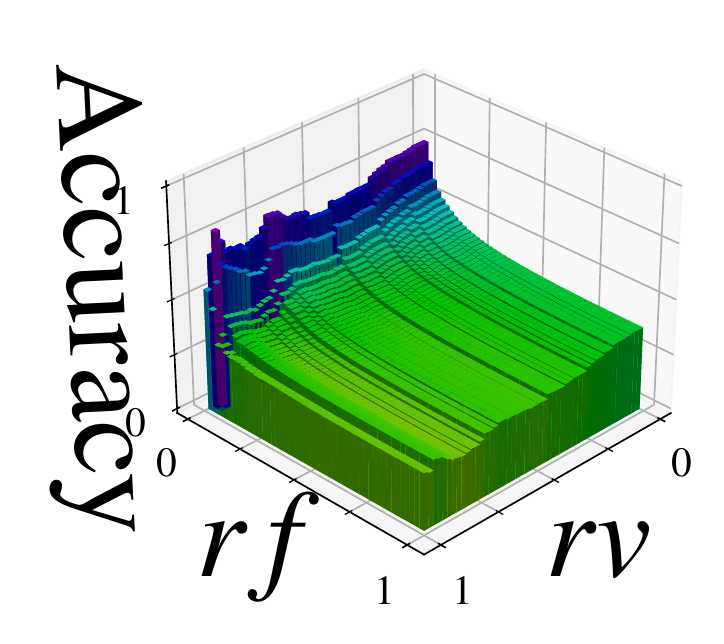}
		\end{minipage}
	}
        \subfigure[LVLF]
	{
 \label{LVLF_3D_Wiki}
		\begin{minipage}{.20\linewidth}
			\centering
			\includegraphics[width=\linewidth]{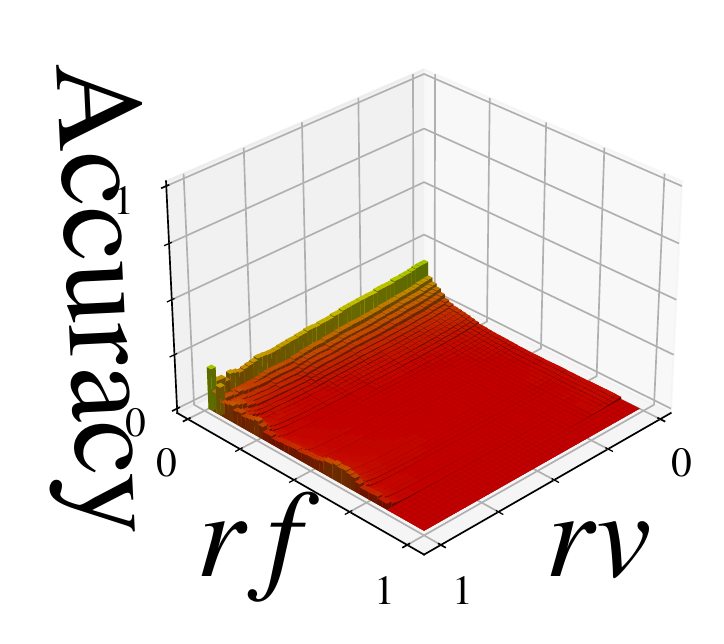}
		\end{minipage}
	}
	\caption{The accuracy of Algorithm \ref{alg1} in four quadrants with different $rv$ and $rf$. {Upper row - Lucene; lower row - Wikipedia}. }
	\label{fig:quadrants_lucene_wiki}
\end{figure}

\begin{figure}[!h]
	\centering

	\subfigure[HVHF]
	{
 \label{HVHF_lucene}
		\begin{minipage}{.20\linewidth}
			\centering
			\includegraphics[width=\linewidth]{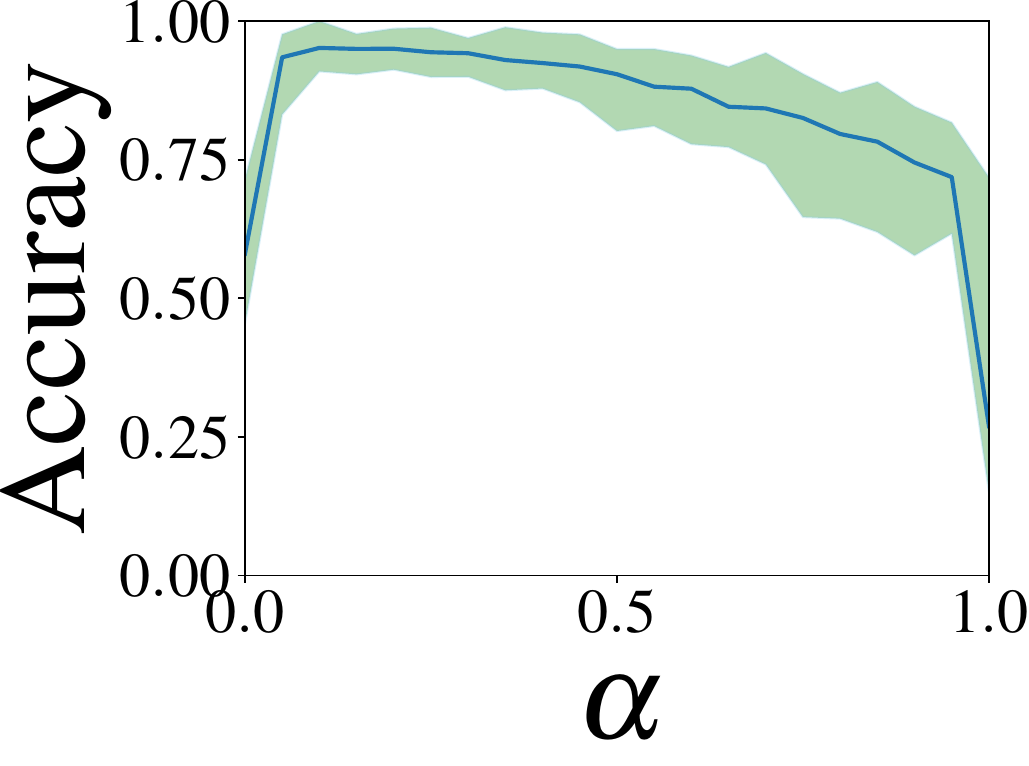}
		\end{minipage}
	}
	\subfigure[HVLF]
	{
 \label{HVLF_lucene}
		\begin{minipage}{.20\linewidth}
			\centering
			\includegraphics[width=\linewidth]{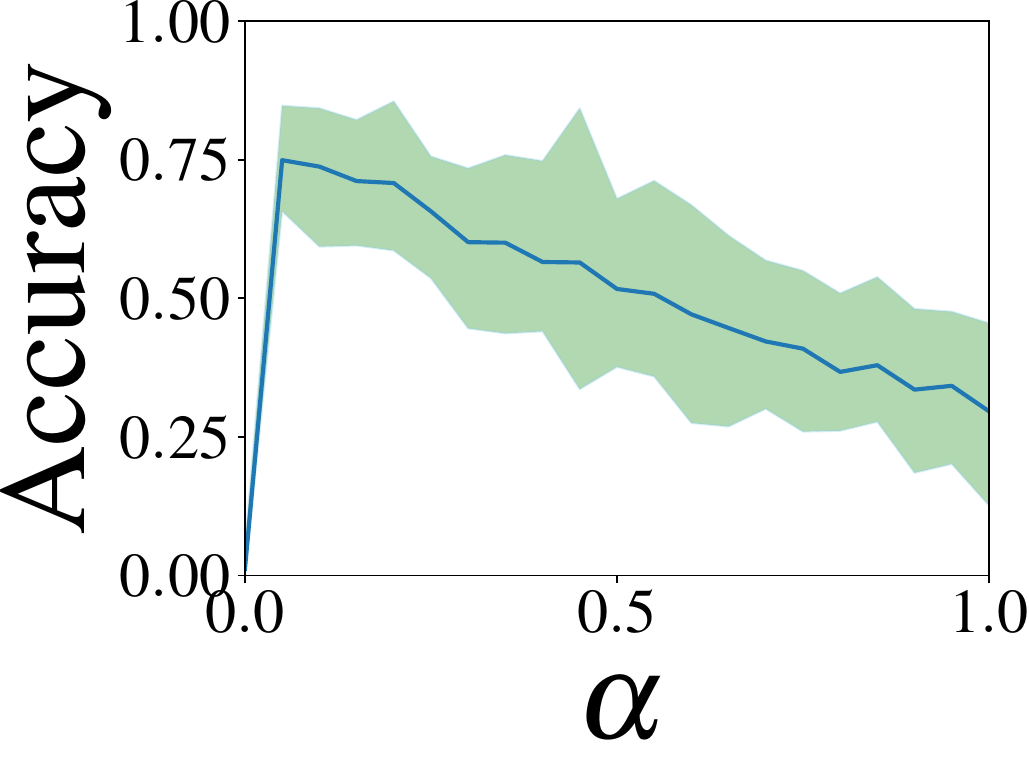}
		\end{minipage}
	}
        \subfigure[LVHF]
	{
 \label{LVHF_lucene}
		\begin{minipage}{.20\linewidth}
			\centering
			\includegraphics[width=\linewidth]{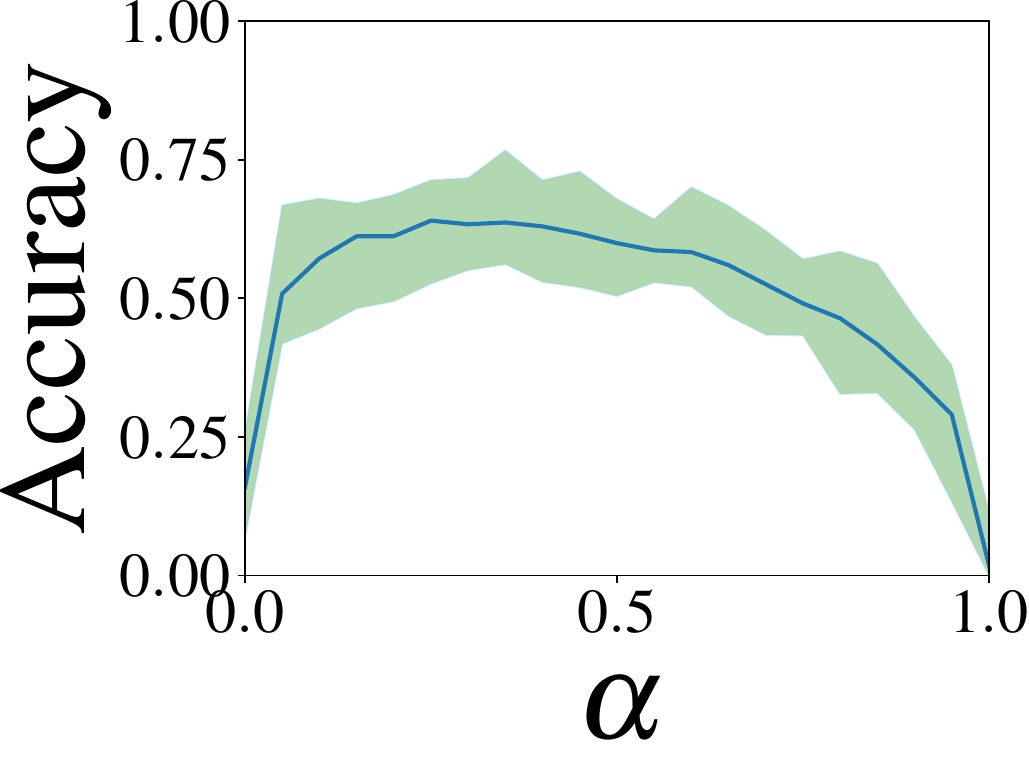}
		\end{minipage}
	}
        \subfigure[LVLF]
	{
 \label{LVLF_lucene}
		\begin{minipage}{.20\linewidth}
			\centering
			\includegraphics[width=\linewidth]{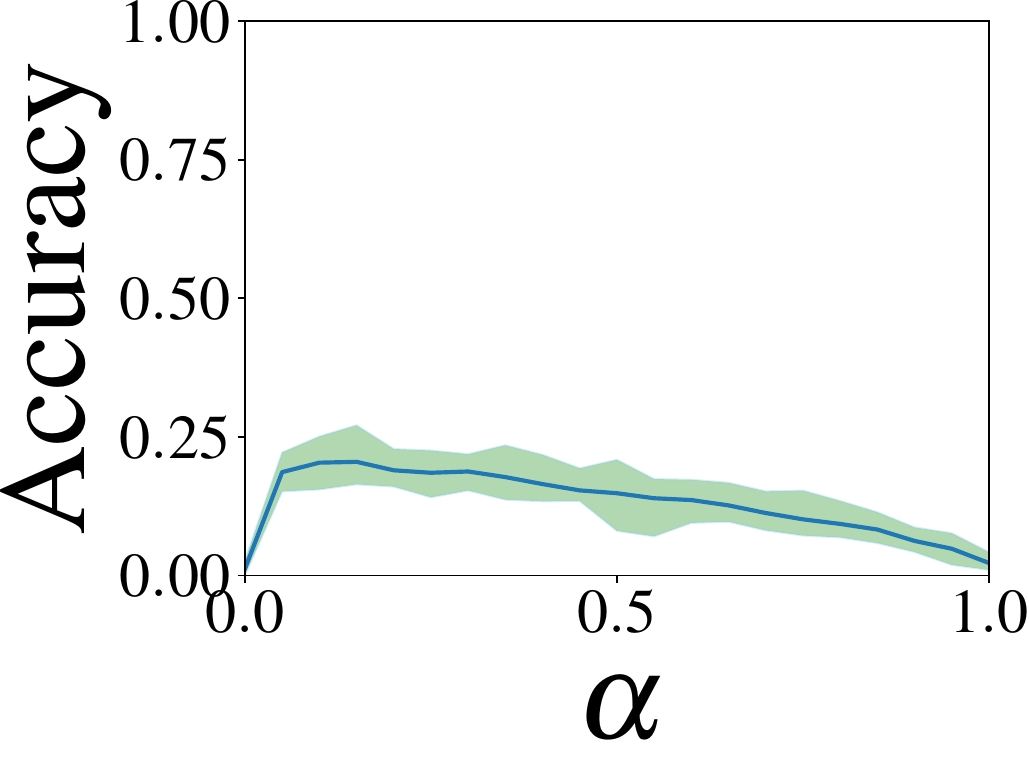}
		\end{minipage}
	}

	\subfigure[HVHF]
	{
 \label{HVHF_Wiki}
		\begin{minipage}{.20\linewidth}
			\centering
			\includegraphics[width=\linewidth]{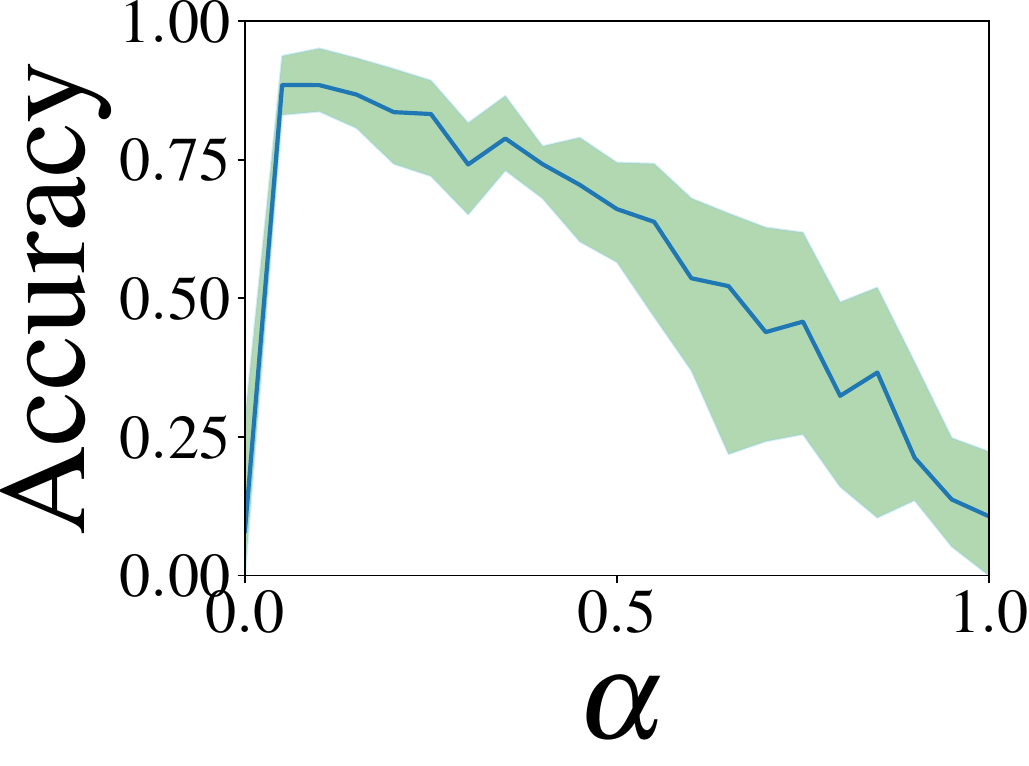}
		\end{minipage}
	}
	\subfigure[HVLF]
	{
 \label{HVLF_Wiki}
		\begin{minipage}{.20\linewidth}
			\centering
			\includegraphics[width=\linewidth]{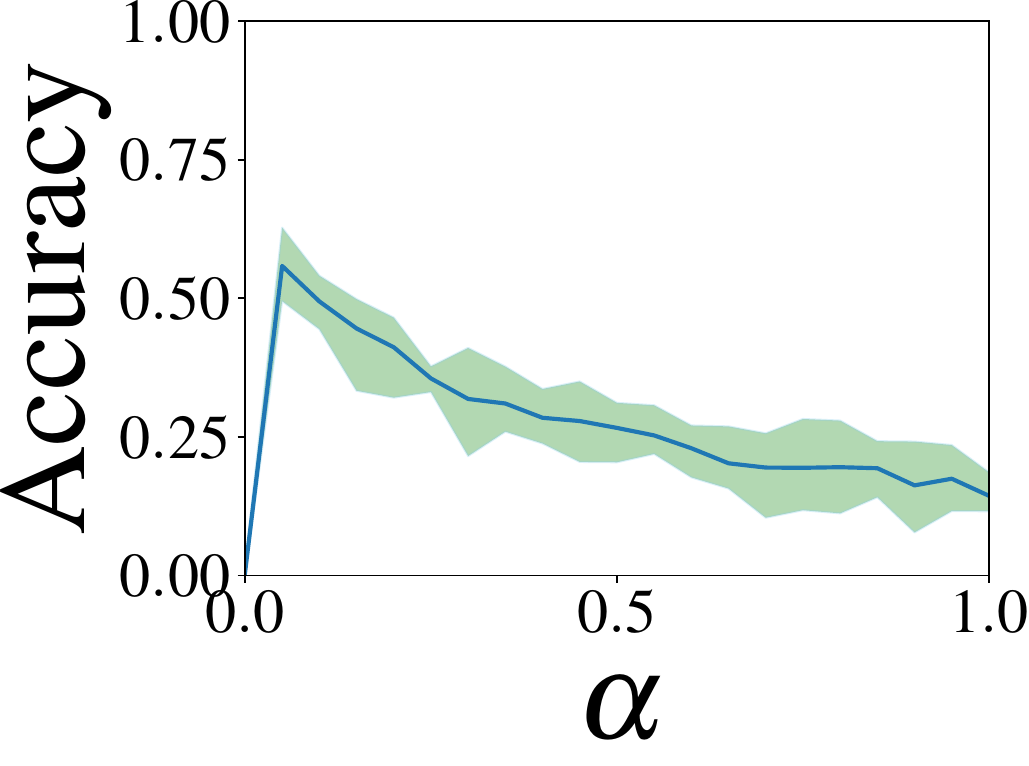}
		\end{minipage}
	}
        \subfigure[LVHF]
	{
 \label{LVHF_Wiki}
		\begin{minipage}{.20\linewidth}
			\centering
			\includegraphics[width=\linewidth]{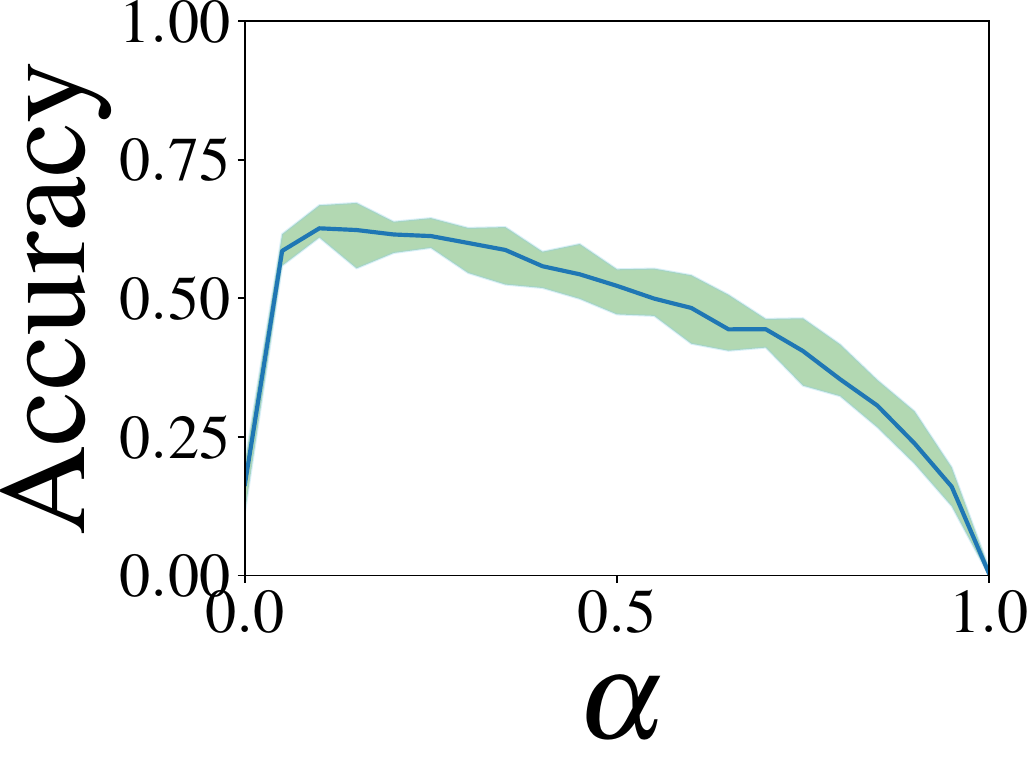}
		\end{minipage}
	}
        \subfigure[LVLF]
	{
 \label{LVLF_Wiki}
		\begin{minipage}{.20\linewidth}
			\centering
			\includegraphics[width=\linewidth]{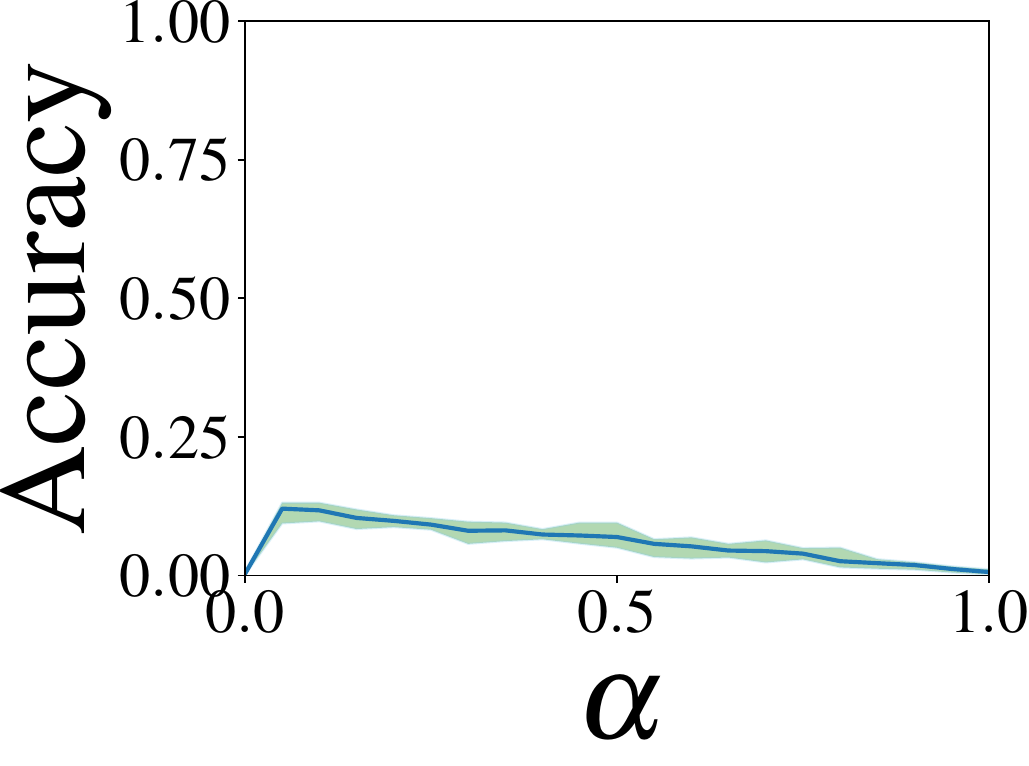}
		\end{minipage}
	}
	\caption{{The accuracy of Algorithm \ref{alg1} on four quadrants with different $\alpha$. Upper row - Lucene; lower row - Wikipedia.}
 }
	\label{fig:quadrants_test_alpha_lucene_wiki}
\end{figure}

%% file: table_RSA_with_different_query_numbers_wiki.tex
\begin{table*}[!t]
    \centering
    \begin{threeparttable} 
	    \caption{Comparison between Jiasaw and RSA with different known query numbers on Enron, Lucene, and Wikipedia.}
		\label{tab:table_RSA_kqn}
		\centering
        \setlength{\tabcolsep}{1.5mm}{\begin{tabular}{lccccccc}
			\hline
           Dataset & Attack(Known query percentage)& Jigsaw($0\%$) &  RSA($0.5\%$) & RSA($1\%$) & RSA($2.5\%$) & RSA($5\%$) \\
			\hline

            \hline
            \multirow{2}{*}{Enron} & Accuracy(Select $W$ with highest volume)& $0.942$ & $0.864$ & $0.919$ & $0.927$ & $0.922$\\
            
            & Accuracy(Randomly selected $W$)& $0.855$&$0.564$&$0.702$&$0.763$&$0.777$\\
            \hline

			\hline
            \multirow{2}{*}{Lucene} & Accuracy(Select $W$ with highest volume)& $0.973$ & $0.958$ & $0.972$ & $0.972$ & $0.974$\\
            
            & Accuracy(Randomly selected $W$)& $0.859$ &$0.663$&$0.800$&$0.826$&$0.838$\\
           
            \hline

			\hline
            \multirow{2}{*}{Wikipedia} & Accuracy(Select $W$ with highest volume)& $0.995$ & $0.990$ & $0.990$ & $0.990$ & $0.990$\\
            
            & Accuracy(Randomly selected $W$)& 0.992 &
            $0.987$ & $0.986$ & $0.987$ & $0.988$\\
           
            \hline
            
		\end{tabular}}
      
    \end{threeparttable} 
\end{table*}

%% file: figures_bef_and_after_adp.tex
\begin{figure}[!t]

	\centering
    
	\subfigure[CGPR]
	{
 \label{fig:padding_CGPR_adp}
		\begin{minipage}{.22\linewidth}
			\centering
                \includegraphics[width=\linewidth]
                {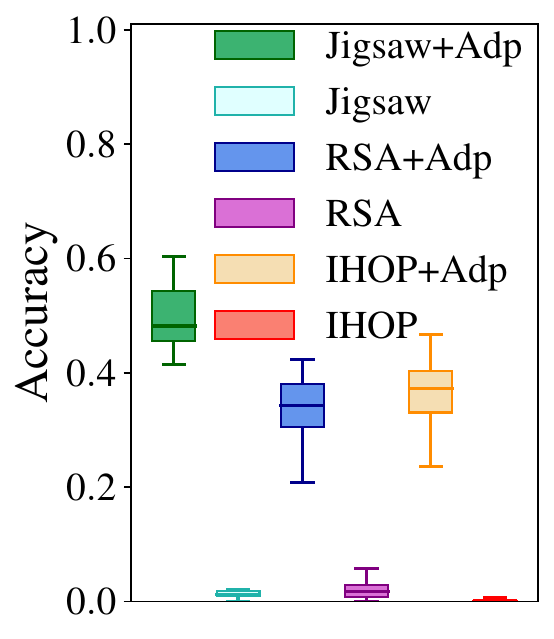}
		\end{minipage}
	}
  \subfigure[Cluster-based]
	{
 \label{fig:padding_cluster_adp}
		\begin{minipage}{.22\linewidth}
			\centering
                \includegraphics[width=\linewidth]
                {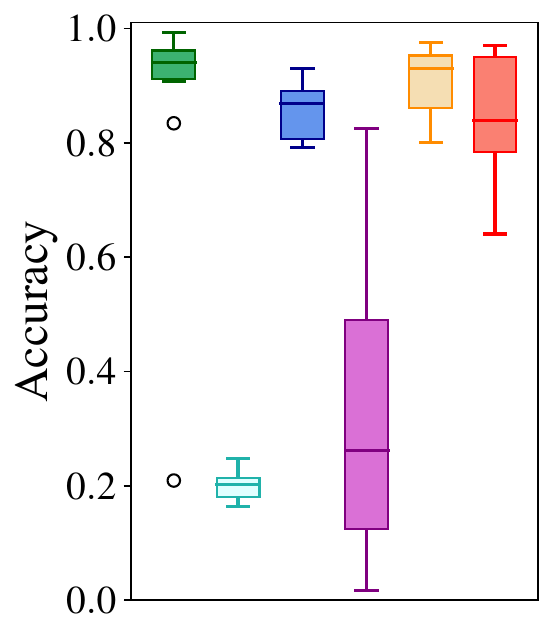}
		\end{minipage}
	}
  \subfigure[SEAL]
	{
 \label{fig:padding_SEAL_adp}
		\begin{minipage}{.22\linewidth}
			\centering
                \includegraphics[width=\linewidth]
                {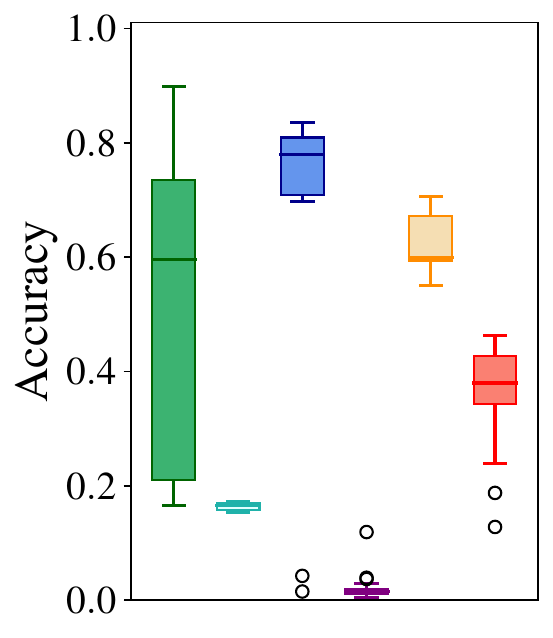}
		\end{minipage}
	}
\subfigure[Obfuscation]
	{
 \label{fig:obfuscation_adp}
		\begin{minipage}{.22\linewidth}
			\centering
                \includegraphics[width=\linewidth]
                {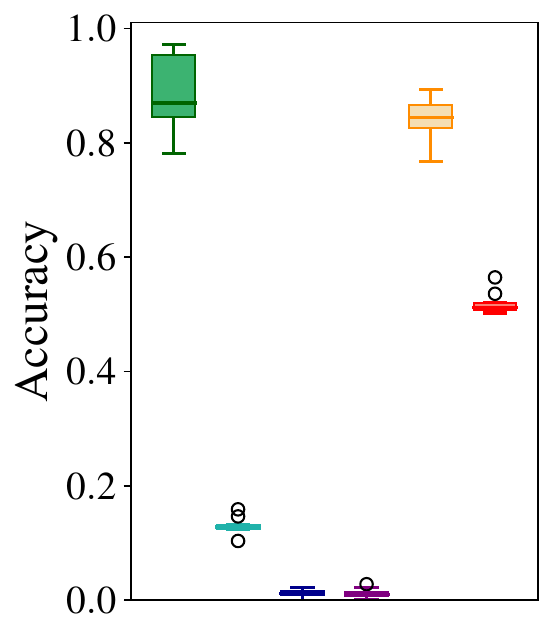}
		\end{minipage}
	}
	 \caption{The results of RSA and IHOP with (+Adp) / without the adaptations on similar data against the padding in CGPR\cite{cash2015leakage}, the obfuscation\cite{DBLP:conf/infocom/ChenLRZ18}, the cluster-based padding\cite{DBLP:journals/corr/shielddb,DBLP:journals/iacr/BostF17}, and the padding in SEAL\cite{DBLP:conf/uss/DemertzisPPS20}.}
 \label{fig:before and after adp}
	
\end{figure}

%% file: table_overhead.tex

      
{
\begin{table*}[!t]
    \centering
    \begin{threeparttable} 
	    \caption{Communication and storage overheads of the padding in CGPR\cite{cash2015leakage}, the obfuscation in CLRZ\cite{DBLP:conf/infocom/ChenLRZ18}, the cluster-based padding\cite{DBLP:journals/corr/shielddb,DBLP:journals/iacr/BostF17}(noted as Cluster), and the padding in SEAL\cite{DBLP:conf/uss/DemertzisPPS20}. For Enron and Lucene, we set the $|W|=1000$. For Wikipedia, $|W|\in\{1000,3000,5000\}$ (noted as Wiki1000,Wiki3000, and Wiki5000). The storage overhead (noted as Sto) is calculated as $(|Padded Documents|+|Original Documents|)/|Original Documents|$.
        The communication overhead (noted as Comm) is presented as $N_c/N$, where $N_{c}$ is the number of all returned documents with countermeasures and $N$ is the number without.}
		\label{tab:table_overhead}
		\centering
        \setlength{\tabcolsep}{2mm}{\begin{tabular}{l|ccc|cccc}
			\hline
            \multicolumn{1}{c}{}& & \multicolumn{1}{c}{Enron} & \multicolumn{1}{c}{Lucene} &  & \multicolumn{1}{c}{ Wiki1000}& Wiki3000 & Wiki5000 \\
			\hline
            \multirow{4}{*}{CGPR\cite{cash2015leakage}} & $k$ & Comm/Sto  & Comm/Sto  & $k$ & Comm/Sto & Comm/Sto & Comm/Sto\\
            \cline{2-8}
            & 500 &1.54/1.03 &1.12/1.01 &50000& 1.71/1.10 & 3.05/1.10 &4.09/1.10\\
            \cline{2-8}
            & 1000 &2.28/1.07 &1.29/1.03 &100000& 3.05/1.20 & 5.84/1.20 &7.94/1.20\\
            \cline{2-8}
            & 1500 &3.07/1.10 &1.46/1.04 &150000& 4.48/1.30&8.70/1.30 &11.86/1.30\\
            \hline
            \hline
            \multirow{4}{*}{CLRZ\cite{DBLP:conf/infocom/ChenLRZ18}} & FPR & Comm/Sto  & Comm/Sto  &FPR & Comm/Sto & Comm/Sto & Comm/Sto\\
            \cline{2-8}
            &0.01 &1.27/1.00 &1.16/1.00 &0.1 &2.39/1.00 &3.80/1.00 &4.87/1.00 \\
            \cline{2-8}
            &0.02 &1.55/1.00 &1.31/1.00 &0.2 &3.78/1.00 &6.63/1.00 &8.74/1.00 \\
            \cline{2-8}
            &0.05 &2.36/1.00 &1.79/1.00 &0.3 &5.17/1.00 &9.45/1.00 &12.61/1.00 \\
            \hline
            \hline
            \multirow{4}{*}{Cluster\cite{DBLP:journals/corr/shielddb,DBLP:journals/iacr/BostF17}} & $\alpha$ & Comm/Sto  & Comm/Sto  &$\alpha$ & Comm/Sto & Comm/Sto & Comm/Sto\\
           \cline{2-8}
            &2 &1.01/1.09 &1.01/1.17 &2 &1.00/1.16 &1.00/1.34 &1.00/1.18 \\
            \cline{2-8}
            &4 &1.03/1.17 &1.03/1.30 &4 &1.01/1.39 &1.01/1.48 &1.01/1.36 \\
            \cline{2-8}
            &8 &1.07/1.27 &1.07/1.41 &8 &1.03/1.49 &1.02/1.55 & 1.02/1.52\\
            \hline
            \hline
            \multirow{4}{*}{SEAL\cite{DBLP:conf/uss/DemertzisPPS20}} & $x$ & Comm/Sto  & Comm/Sto  &$x$ & Comm/Sto & Comm/Sto & Comm/Sto\\
            \cline{2-8}
            &2 &1.44/2.00 &1.46/2.00 &2 &1.44/2.00 &1.43/2.00 &1.43/2.00 \\
            \cline{2-8}
            &3 &1.84/3.00 &2.16/3.00 &3 &1.78/3.00 &1.79/3.00 &1.80/3.00 \\
            \cline{2-8}
            &4 &2.04/4.00 &2.31/4.00 &4 &2.26/4.00 &2.18/4.00 &2.15/4.00 \\
            \hline
		\end{tabular}}
      
    \end{threeparttable} 
\end{table*}
}